\documentclass{aa}
\usepackage{graphicx}
\usepackage{natbib}
\usepackage{txfonts}

\def\lesssim{\mathrel{\hbox{\rlap{\hbox{\lower4pt\hbox{$\sim$}}}\hbox{$<$}}}}
\def\grtsim{\mathrel{\hbox{\rlap{\hbox{\lower4pt\hbox{$\sim$}}}\hbox{$>$}}}}
\def\chemlk#1 {\stackrel{#1}{\longrightarrow}}
\def\kms{km~s$^{-1}$}
\def\vlsr{$V_{\rm LSR}$}
\def\hcop{HCO$^+$}
\def\hthcop{H$^{13}$CO$^+$}
\def\hthcn{H$^{13}$CN}
\def\hnthc{HN$^{13}$C}
\def\pow#1#2{#1$\times$10$^{#2}$}
\def\scm{cm$^{-2}$}
\def\ccm{cm$^{-3}$}
\def\tkin{$T_{\rm kin}$}
\def\txc{$T_{\rm ex}$}
\def\tmb{$T_{\rm mb}$}

\def\nhh{$n$(H$_2$)}
\def\gtsim{{_>\atop{^\sim}}}

\usepackage{ulem}
\def\newnewnew#1{#1}
\def\newnewold#1{}
\def\newnew#1{#1}
\def\newold#1{}
\def\new#1{#1}
\def\old#1{}

\begin{document}

\title{Dense molecular gas towards W49A: A template for extragalactic starbursts?}
\titlerunning{Dense molecular gas towards W49A}

\author{H. Roberts \inst{1} \and 
        F. F. S. van der Tak \inst{2,3} \and
        G. A. Fuller\inst{4} \and 
        R. Plume\inst{5} \and
        E. Bayet\inst{6}}

\institute{Astrophysics Research Centre, School of Mathematics and Physics, Queens University Belfast, Belfast, BT7 1NN, UK\\
\email{h.roberts@qub.ac.uk}
\and SRON Netherlands Institute for Space Research, PO Box 800, 9700 AV Groningen, The Netherlands
\and Kapteyn Astronomical Institute, PO Box 800, 9700 AV Groningen, The Netherlands
\and Jodrell Bank Centre for Astrophysics, School of Physics and Astronomy, University of Manchester, Manchester, M13 9PL, UK
\and Department of Physics and Astronomy, University of Calgary, Calgary, T2N 1N4, AB, Canada
\and Department of Physics and Astronomy, University College London, Gower Street, London WC1E 6BT, UK
             }

\date{Received 17 May 2010; accepted 21 October 2010}

\abstract
{The HCN, HCO$^+$, and HNC molecules are commonly used as tracers of dense star-forming gas in external galaxies, but such observations are spatially unresolved. Reliably inferring the properties of galactic nuclei and disks requires detailed studies of sources whose structure is spatially resolved.}
{To understand the origin of extragalactic molecular line emission, we compare the spatial distributions and abundance ratios of HCN, HCO$^+$, and HNC in W49A, the most massive and luminous star-forming region in the Galactic disk.}
{Maps of a 2$'$ (6.6\,pc) field at 14$''$ (0.83\,pc) resolution of the J=4--3 transitions of HCN, H$^{13}$CN, HC$^{15}$N, HCO$^+$, H$^{13}$CO$^+$, HC$^{18}$O$^+$, and HNC are combined with supplementary observations of the J=5--4 transition of DCN and the J=3--2 transitions of HCN and H$^{13}$CO$^+$. Most of the data are from HARP/ACSIS, with supplementary data from JCMT Receiver~A and the SCUBA archive. We use maps of the integrated intensity and line-profiles to pick out regions of the source to study in more detail. We compare column densities and abundance ratios towards these regions with each other and with predictions from gas-phase chemical models.}
{The kinematics of the molecular gas in W49A appears complex, with a mixture of infall and outflow motions. 
\newnewnew{Both the line profiles and comparison of the main and rarer species show that the main species are optically thick.}
Two ``clumps'' of infalling gas that we look at in more detail appear to be at $\sim$40~K, compared to $\geq$100~K at the source centre, and may be $\sim$10$\times$ denser than the rest of the outer cloud. The chemical modelling suggests that the HCN/HNC ratio probes the current gas temperature, while the HCN/HCO$^+$ ratio and the deuterium fractionation were set during an earlier, colder phase of evolution.}
{The similarity in the derived physical conditions in W49A and those inferred for the molecular gas in external galaxies suggest W49A is an appropriate analogue of an extragalactic star forming region. Our data show that the use of HCN/HNC/\hcop\ line ratios as proxies for the abundance ratios is incorrect for W49A \newnewold{because the lines are optically thick}, suggesting that using these line ratios as abundance ratios in galactic nuclei is invalid too. On the other hand, our observed isotopic line ratios such as H$^{13}$CN/H$^{13}$CO$^+$ approach our modeled abundance ratios quite well in W49A. Second, the 4--3 lines of HCN and \hcop\ are much better tracers of the dense star-forming gas in W49A than the 1--0 lines, confirming similar indications for galactic nuclei. Finally, our observed HCN/HNC and HCN/\hcop\ ratios in W49A are inconsistent with homogeneous PDR or XDR models, indicating that irradiation does not strongly affect the gas chemistry in W49A. Overall, the W49A region appears to be a useful template for starburst galaxies.}

\keywords{ISM: molecules -- ISM: clouds -- ISM: abundances -- stars: formation}

\maketitle

\section{Introduction}\label{s:intro}

In recent years, submillimeter line emission from a variety of molecules has been detected from external galaxies in the early and local Universe (\citealt{kramer08}). The first detection of HCO$^+$ at high redshift, the J=1--0 transition towards the Cloverleaf quasar at z=2.56 \citep{riechers06}, was quickly followed up by higher-$J$ HCN, HNC and CN lines towards the ultra-luminous quasar APM~08279+5255 at z=3.911 \citep{wagg05,guelin07}.  Whereas low-$J$ CO lines are used to trace the total molecular gas mass in Galactic nuclei, HCN and HCO$^+$ lines are often used to trace the higher-density gas. This dense gas is much more closely related to star formation activity than CO, as shown by the tighter correlation of $L$(HCN 1-0) with $L$(far-IR) \citep{gao}. However, factors other than the mass of dense gas presumably affect the HCN emission, and high-$J$ CO lines may trace the star forming activity as well as or even better than HCN 1-0 \citep{krumholz,narayanan,bayet_CO}.

The HCN / HCO$^+$ ratio is also used to trace the chemical state of the dense gas, in particular its irradiation by UV light from a starburst or by X-rays from an AGN.  For example, \citet{kohno} used the HCN/HCO$^+$ ratio in nearby Seyfert galaxies to distinguish between AGNs and starbursts; \citet{greve} find that excited HCN transitions trace dense gas in the two starburst galaxies, Arp~220 and NGC~6240, while sub-thermally excited HCO$^+$ lines trace gas which is 10--100~times less dense; the HCN/HCO$^+$ ratios observed by \citet{aalto2007} towards NGC~4418 suggest emission from dense, warm gas with a PDR component, but with no X-ray chemistry.
Another commonly used line ratio is HCN / HNC, which is thought to trace the gas temperature (e.g., \citealt{baan08}).

The angular resolution of current instrumentation at submm wavelengths ranges from $\sim$30$''$ for single-dish telescopes to $\sim$1$''$ for interferometers. 
For nearby galaxies \newnew{($D\sim$10--100\,Mpc)}, single-dish observations thus probe \newnew{scales of 1.5--15\,kpc}, while interferometers probe \newnew{scales of 50--500\,pc}.
In the case of high-redshift galaxies, even interferometers probe size scales of $\sim$kpc. \newnew{The goal of this work is to improve understanding of such observations by using a local analogue. In particular, }
the use of line ratios to infer physical conditions depends on the assumption that the gas is roughly homogeneous on these scales, so that the lines trace the same gas.
Recent authors have already questioned this assumption \citep{bayet08,bayet09}, and
the aim of this paper is to see just how far this assumption of homogeneity is from reality, using the W49A region as a template for extragalactic regions with high star formation rates, and to explore how the observational limitations can be worked around.
However, the 6.6\,pc size of the images used here falls short of the size scales resolved in distant extragalactic systems by a factor of $\gtsim$10.

\begin{figure*}
\rotatebox{270}{
\resizebox{!}{\hsize}{\includegraphics*[2cm,2cm][11cm,24cm]{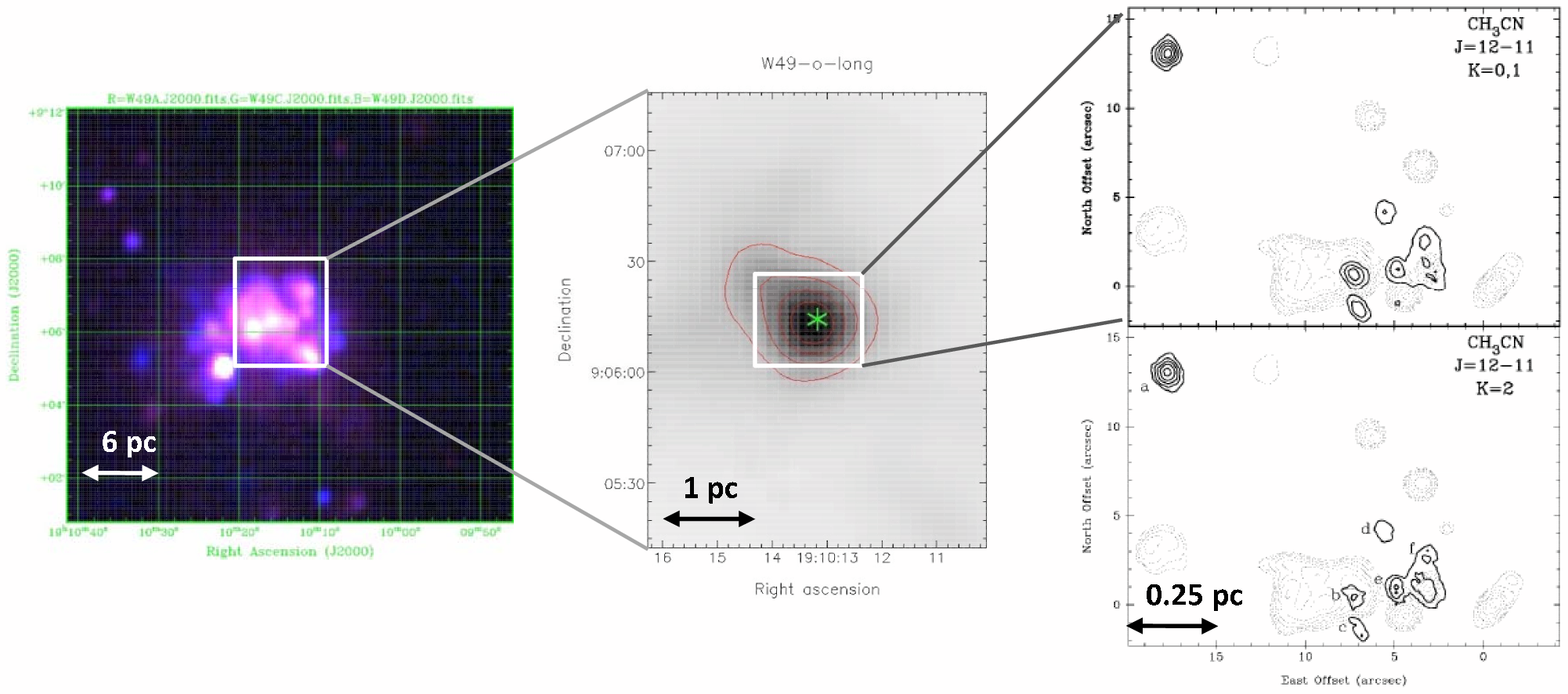}}
}\caption{Images of the W49A region at different wavelengths and spatial scales. Left: Three colour MSX image of W49, showing the clustered nature of sources in the region: 8.28$\mu$m (red); 12.13$\mu$m (green); 14.65$\mu$m (blue). Centre: Dust continuum emission at 850$\mu$m (SCUBA archive data; see text). Contours are at 20, 40, 60 and 80\% of the peak intensity. This map is $\sim$1.5$\times$2$'$ in size, slightly smaller than our HARP maps. Right: BIMA observations of hot cores, traced by CH$_3$CN \citep{wilner01}. We note that the central pixels of the HARP maps cover a cluster of several hot cores.}\label{f:spatial}
\end{figure*} 

With a luminosity in excess of 10$^7$~L$_{\sun}$ \citep{ward-th} and a mass $>$10$^7$~M$_{\sun}$, W49A is one of the most active star-forming regions in the Galactic disk, at a distance of 11.4~kpc \citep{gwinn92}.
The region is composed of a number of optically obscured, compact HII regions \citep{wilner01} surrounded by a molecular cloud with a mass $>$10$^5$~M$_{\sun}$ \citep{mufson,simon}. 
Spectrally, W49A is extremely complex, containing numerous features contributed by W49A itself, as well as additional clouds associated with the Sagittarius spiral arm which crosses the line-of-sight twice \citep{plume04}. Fig.~\ref{f:spatial} shows the structure of W49A at infra-red vs sub-mm wavelengths and at spatial scales from $\sim$30 to $<$1~pc. 
 
\citet{welch87} estimate the cloud density to be $>$3$\times$10$^{5}$~cm$^{-3}$ within the inner 1 pc (the core region) and $\sim$10$^4$~cm$^{-3}$ in the outer parts of the cloud. Based on observations of the CO J=7--6 transition, \citet{jaffe} estimate kinetic temperatures of $>$70~K in the centre, falling below 50~K in the outer parts of the cloud. From sub-millimetre continuum maps of the dust emission towards W49A, \citet{ward-th} find a dust temperature of 50~K and a dust mass 2400~M$_{\sun}$ over an effective emitting region of $\sim$2~pc. From near-infra-red IRAS data they infer an additional population of small, hot ($T\sim$350\,K) dust grains or Polycyclic Aromatic Hydrocarbons (PAHs). 

Before W49A can be used as an extragalactic template, a better understanding of its physical and chemical properties is needed.
Rotational transitions of interstellar molecules help reaching this goal in two ways: they probe the current density, temperature, and optical depth of the gas and they measure the abundances of molecular species, which can be compared with  chemical models to study the evolution of the chemical inventory and the physical conditions.
Two specific tracers of the chemical state of molecular gas are the HCN/HNC ratio and the deuteration fraction (D/H ratio) of the gas.

The HCN/HNC abundance ratio has long been known to vary with temperature \citep{goldsmith81,goldsmith86}. Although HNC and HCN are assumed to be formed at the same rate (through dissociative recombination of the HCNH$^+$ ion), \newnewnew{the reactions of HNC with O and with H, which destroy HNC,} have energy barriers which make them very slow at low temperatures. As a consequence, the HCN/HNC ratio increases with increasing \tkin\ \newnewnew{\citep{schilke92}}. 

Observations of molecular D/H ratios (i.e. the abundance of a chemical species where one or more D atoms have been substituted for the H atoms, relative to the abundance of the ``normal'' hydrogen-bearing species) also probe the kinetic temperature, as well as the chemical history, of the gas. Although the cosmic D/H ratio is only 3$\times$10$^{-5}$ \citep{linsky}, chemical fractionation concentrates deuterium in molecules at temperatures $<$50~K \citep{watson,mbh}. \citet{hatchell_dcn} measured DCN/HCN ratios towards a number of hot molecular cores (\tkin\ $\geq$ 100~K), associated with massive protostars, and found ratios  of a few times 10$^{-3}$. These results, and other molecular D/H ratios in warm gas, can be explained by assuming that the deuterium fractionation was set at temperatures $<$50~K and preserved in the icy mantles of interstellar grains, until the region was heated and the ices evaporated into the gas phase \citep[see also][]{loinardD2CO,pariseDmeth,vastel_d2s}. \citet{rodgers&millar} showed that deuterium fractionation of neutral species can survive for 10$^4$--10$^5$~years in hot gas, post-evaporation. Deuterated molecular ions (e.g.\ DCO$^+$), on the other hand, are destroyed much more rapidly when the gas is warmed to $>$30~K \citep[e.g.][]{roberts07} and are not expected to be abundant in protostellar regions.

\section{Observations}\label{s:obs}

This work is based on data obtained as part of the Spectral Legacy Survey \citep[SLS;][]{plume_sls}, being conducted at the James Clerk Maxwell Telescope (JCMT) on Mauna Kea, Hawai'i. The SLS is performing spectral imaging of $2' \times 2'$ fields towards four targets representing different star-forming environments. Once completed, the spectral range covered by the SLS will be 330--362~GHz. This paper presents observations of HCN, HNC, HCO$^+$, and isotopomers towards W49A (Table~\ref{t:summary}). For this data set, the ``central position'' refers to R.A. = 19:10:13.4; Dec.\ = 09:06:14 in J2000 co-ordinates. 

The SLS uses the 16-pixel HARP receiver (325--375~GHz) and the ACSIS correlator \citep{buckle09}. The observations were carried out using the HARP4\_mc jiggle position switch mode, which produces maps sampled every 7.5$''$, and the spectra were calibrated by observing at an off-position 14$'$ to the northeast of the source.  The angular resolution of the JCMT is $\sim$15$''$ \newnewnew{at 345\,GHz}, which is $\sim$0.8\,pc at the distance of W49. The spectral resolution of these observations is $\sim$0.8~km~s$^{-1}$ and the beam efficiency is 0.6 \citep{buckle09}. Pointing was checked every hour and the pointing is estimated to be accurate to 1.5$''$. 

\begin{table}\caption{Summary of the observations.}\label{t:summary}
\begin{tabular}{llccc}
\hline \hline
\noalign{\smallskip}
Line & Transition & Frequency & $t_{\rm int}$ $^a$ & rms ($T_{\rm mb}$) $^b$ \\
     &            & GHz      &  s                 & K \\ 
\noalign{\smallskip}
\hline
\noalign{\smallskip}
HC$^{18}$O$^+$ & J=4--3& 340.631 & 540 & 0.08 \\
HC$^{15}$N & J=4--3    & 344.200 & 120 & 0.19 \\
H$^{13}$CN & J=4--3    & 345.340 & 480 & 0.09 \\
H$^{13}$CO$^+$ & J=4--3& 346.998 & 480 & 0.12 \\
HCN & J=4--3           & 354.505 & 480 & 0.07 \\
HCO$^+$ & J=4--3       & 356.734 & 240 & 0.12 \\
DCO$^+$ & J=5--4       & 360.170 & 480 & 0.07 \\
DCN & J=5--4           & 362.046 & 480 & 0.06 \\
HNC & J=4--3           & 362.630 & 583 & 0.03 \\
\noalign{\smallskip}
\hline  
\noalign{\smallskip}
H$^{13}$CO$^+$ & J=3--2& 260.255 & 38 & 0.22 \\ 
HCN & J=3--2           & 265.886 & 19 & 0.52 \\
\noalign{\smallskip}
\hline
\noalign{\smallskip}
\multicolumn{5}{l}{$^a$: Integration time per pixel} \\
\multicolumn{5}{l}{$^b$: Noise level measured over the central 1$'\times$1.5$'$} \\
\end{tabular}
\end{table}

The data were reduced using the ORAC Data Reduction pipeline (ORAC-DR). This automatically processes time-series ACSIS data, applying reduction recipes which are determined by the incoming data type, and, for HARP, outputting gridded data cubes. For SLS data, these recipes include: checking for consistency between the calculated $T_{\rm sys}$ during the observations and the measured r.m.s.\ noise; checking for variations in the r.m.s.\ noise measured by each receptor across the map; checking for r.m.s.\ uniformity across the spectral range; removing baselines from every spectrum; and co-adding cubes taken at the same position and frequency. If any receptors are particularly noisy, or suffer from bad baselines, they are masked out of the time series data and not used in creating the final cubes. 

Supplementary observations of selected molecular transitions have also been imaged by the SLS team using the JCMT Receiver~A (211--276~GHz; beamwidth 20$''$). These include the HCN and H$^{13}$CO$^+$ J=3--2 lines, which are presented below. Half-beamwidth spaced raster maps were made, covering approximately the same field of view as HARP.  The 230~GHz data were reduced and calibrated with standard Starlink procedures. Linear baselines were removed and the data were Hanning smoothed. 15 pixel x 15 pixel spectral grids of the data were generated for each sight-line such that each pixel corresponds to an (8$\times$8)$''$ portion of sky. 

We also made SCUBA maps of W49A, based on data from the JCMT archive. These data were reduced using the ORAC-DR package which calculates the gain from photometry calibrations and the optical depth from sky-dip measurements. The 850\,$\mu$m emission map is shown in Fig.~\ref{f:spatial}. 
\newnewnew{Finally we make use of maps of W49A in the HCN, HNC and \hcop\ $J$=1--0 lines, made with the IRAM 30m telescope by Peng \& Wyrowski (priv. comm.).}

\section{Results}\label{s:res}

\subsection{Spatial distributions}\label{s:spatial}

\begin{figure*}
\includegraphics[height=20cm]{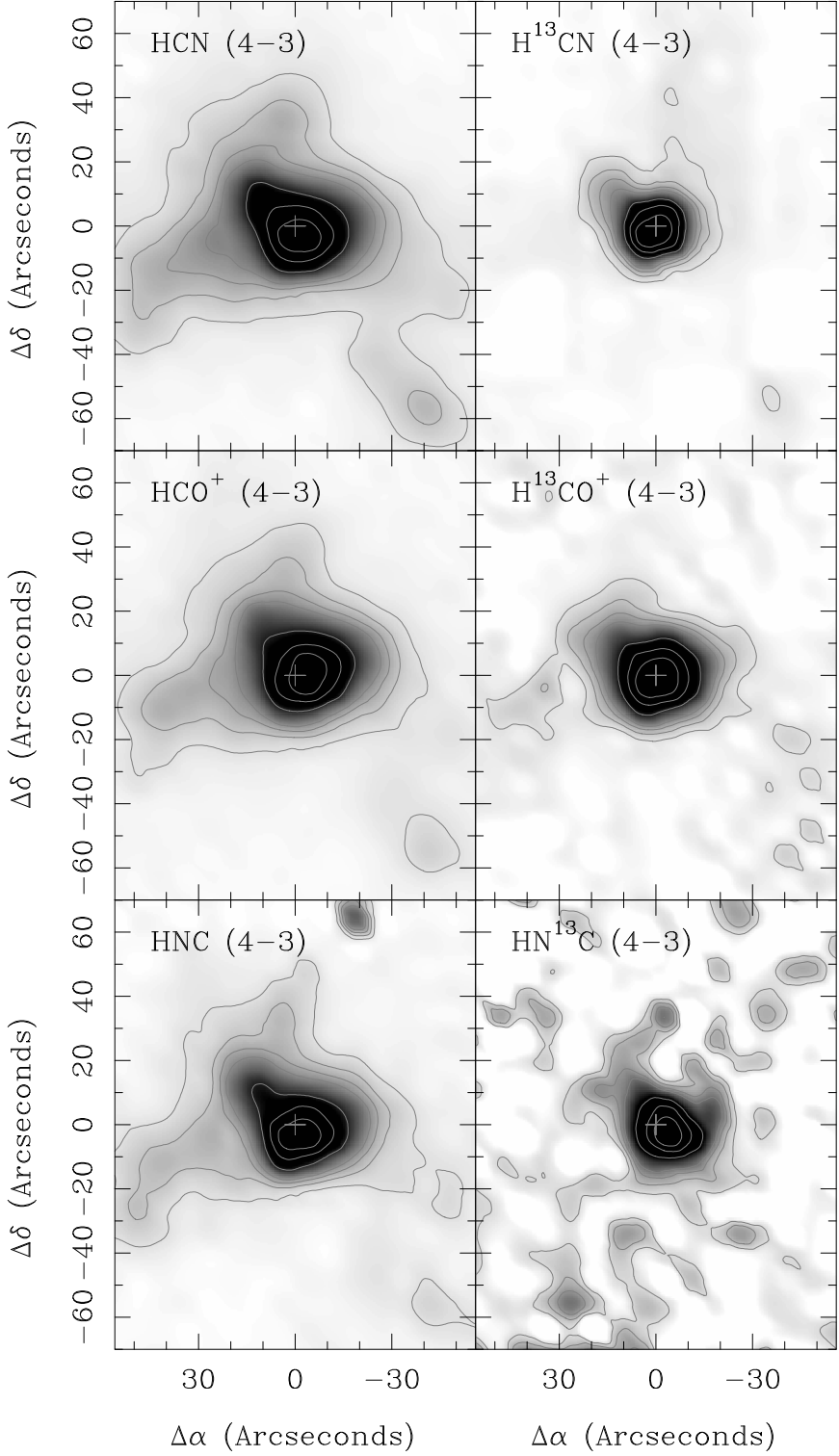}
\caption{Maps of the integrated intensity ($\int{T_A^*dV}$) across W49: HCN 4--3 (top-left); H$^{13}$CN 4--3 (top-right); HCO$^+$ 4--3 (middle-left); H$^{13}$CO$^+$ 4--3 (middle-right); HNC 4--3 (bottom-left); and HN$^{13}$C 4--3 (bottom-right) line intensities. The main lines are summed over 150 km s$^{-1}$, while the $^{13}$C-substituted lines are summed over 60 km s$^{-1}$. The contours show 5, 10, 15, 20, 40 ... 80\% of the peak integrated intensity, which is 161 K km s$^{-1}$ for HCN, 72 K km s$^{-1}$ for H$^{13}$CN, 319 K km s$^{-1}$ for HCO$^+$, and 34 K km s$^{-1}$ for H$^{13}$CO$^+$, 52 for HNC, and 4 for HN$^{13}$C.}
\label{f:W49main}
\end{figure*}

\begin{figure*}
\includegraphics[angle=-90,width=\textwidth]{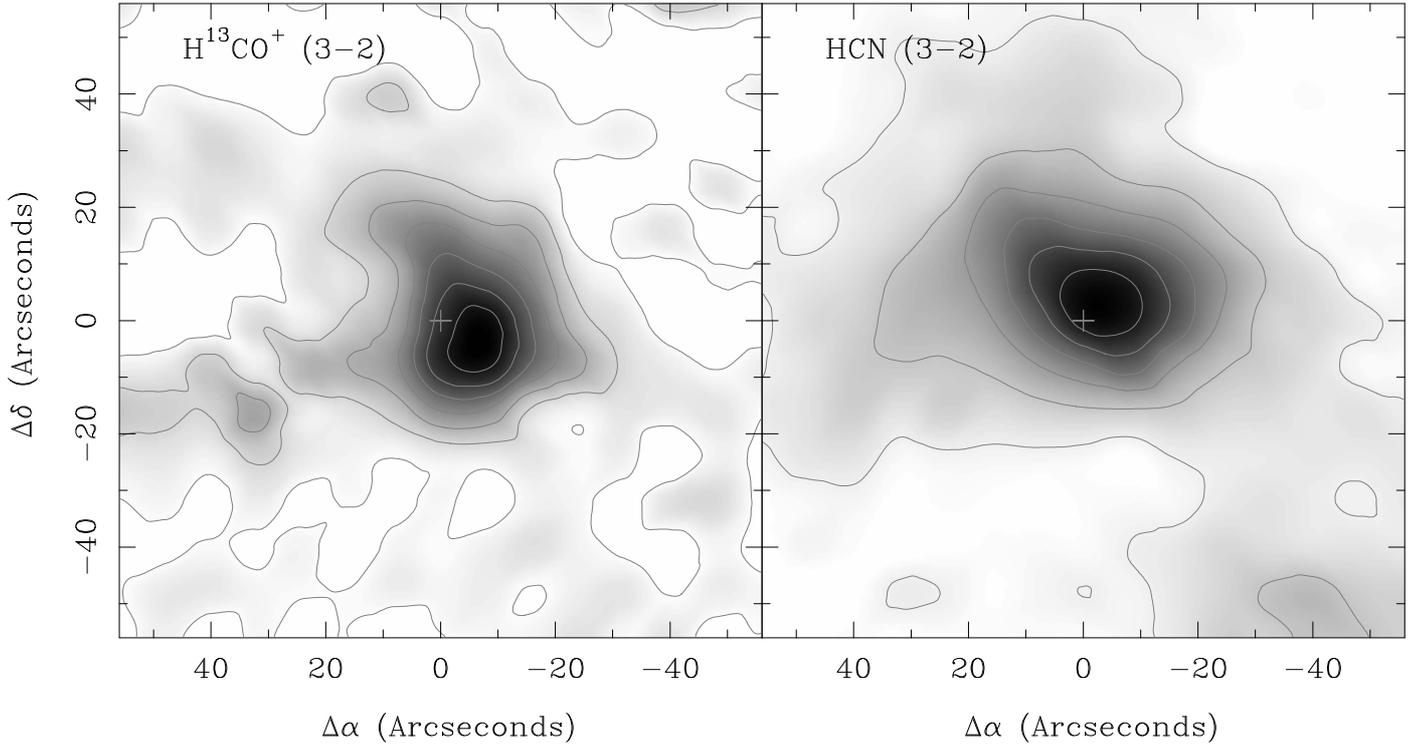}
\caption{Maps of the integrated intensity ($\int{T_A^*dV}$) across W49: H$^{13}$CO$^+$ 3-2 (left) and HCN 3-2 (right).  The contours show 0 to 100\% of the peak integrated intensity (in steps of 15\%), which is 24 K km s$^{-1}$ for H$^{13}$CO$^+$ and 172 for HCN.}
\label{f:W49_HCN32}
\end{figure*}

\begin{table}
  \caption{Observed line ratios}
\label{t:ratios}
\begin{tabular}{lcc} \hline \hline \noalign{\smallskip}
Species     & \multicolumn{2}{c}{Ratio} \\
            & J=1--0\tablefootmark{a} & J=4--3 \\ 
\noalign{\smallskip} \hline \noalign{\smallskip}
HCN/HCO$^+$ & 0.9   & 0.56 \\
HCN/HNC     & 1.0   & 3.7\\ 
\noalign{\smallskip} \hline \noalign{\smallskip}
 Species & Ratio J=4--3/J=1--0 \\
\noalign{\smallskip} \hline \noalign{\smallskip}
HCN & 0.73 \\
HCO$^+$ & 1.25\\
HNC & 0.22 \\
\noalign{\smallskip} \hline
\end{tabular}
\tablefoot{
Ratios of velocity-integrated line intensities between species in the J=1--0 and J=4--3 transitions (upper table) and between the J=4--3 and J=1--0 transition for each of the main species (lower table). The integrated line intensities have been    measured in a 67$''$ diameter aperture centred on the peak of the emission. \\
\tablefoottext{a}{IRAM 30m data from Peng \& Wyrowski, priv. comm.}
}
\end{table}

Figures~\ref{f:W49main}~and~\ref{f:W49_HCN32} show maps of the integrated intensities of the HCN, HCO$^+$, HNC, H$^{13}$CN, H$^{13}$CO$^+$ and HN$^{13}$C J=4--3 and the HCN J=3--2 lines around W49A. The same source structure, extended in the east-west direction, is seen in both the dust continuum emission (Fig.~\ref{f:spatial}) and the HCN and HCO$^+$ lines, and also in HNC but less pronounced.

Emission from the HCN J=3--2 line is the most extended, being seen across most of the (100$\times$100)$''$ map. The HCN and HCO$^+$ J=4--3 emission extends over (82.5$\times$60)$''$, based on the 10\% contour, and there is a slightly fainter clump of emission to the south-west, seen in both lines. The H$^{13}$CO$^+$ J=4--3 emission extends over (46$\times$38)$''$, while H$^{13}$CN 4--3 is slightly more compact. The weaker isotopic lines are seen only towards the central pixels of the map: HC$^{15}$N and HC$^{18}$O J=4--3 are detected in a (23$\times$23)$''$ region and DCN J=5--4 over (37$\times$23)$''$. The DCO$^+$ J=5--4 line was not detected.
\newnewnew{All sizes refer to the 10\% intensity level.}

\begin{figure*}
\includegraphics[angle=-90,width=\textwidth]{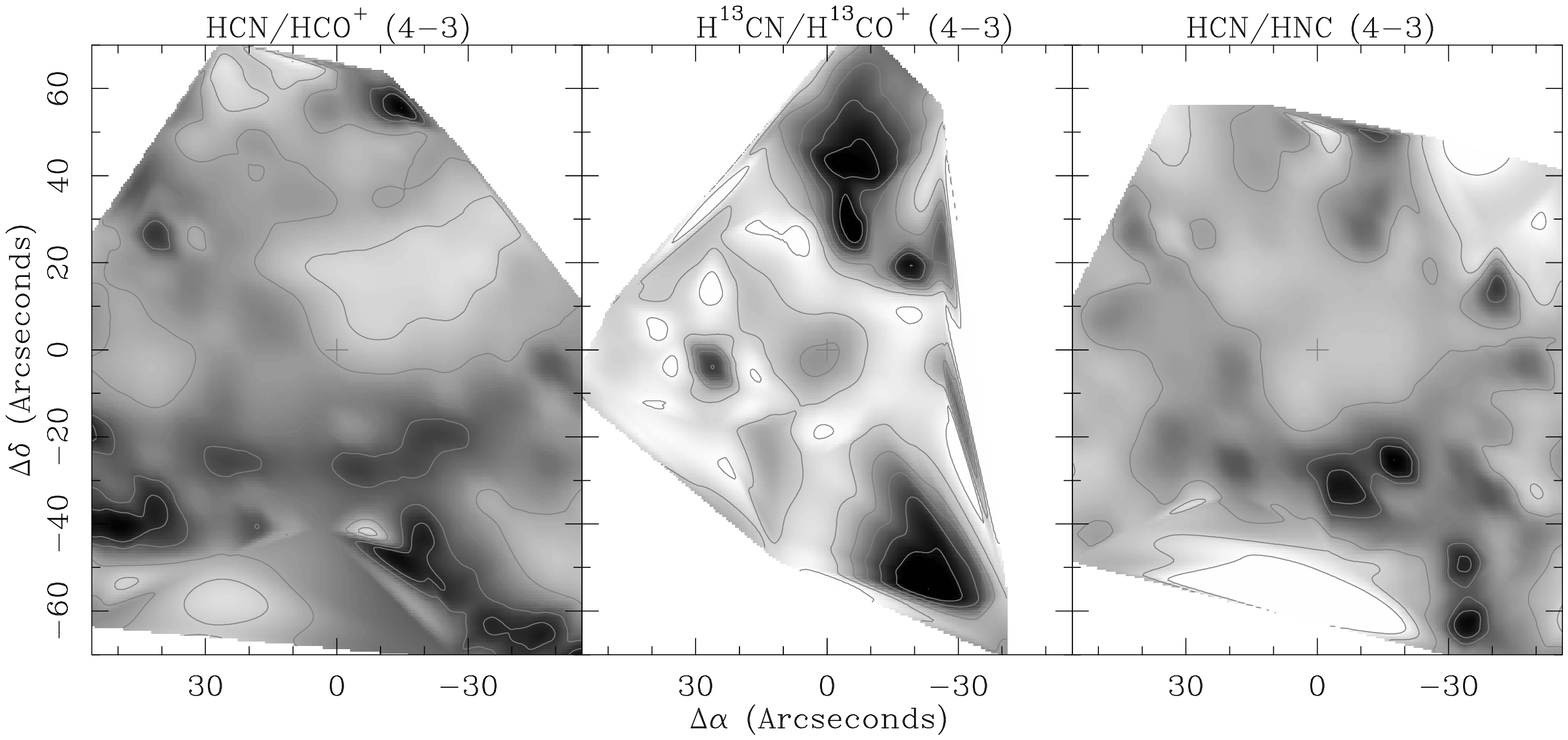}
\caption{Maps of the HCN/HCO$^+$, H$^{13}$CN/H$^{13}$CO$^+$, and HCN/HNC J = 4-3 integrated intensity ($\int{T_A^*dV}$) ratios across W49A (greyscale and contours).  Contours show 10 to 100\% of the peak values (in steps of 15\%) which are 1.4 for  HCN/HCO$^+$, 4.9 for H$^{13}$CN/H$^{13}$CO$^+$, and 10.3 for HCN/HNC.  A cutoff integrated intensity was applied to each transition before calculating the ratio to avoid divide-by-zero errors in low signal regions of the maps.  The cutoff values are 1.5 K km s$^{-1}$ for HCN, 0.5 K km s$^{-1}$ for HNC, 3.0 K km s$^{-1}$ for HCO$^+$, 0.7 K km s$^{-1}$ for H$^{13}$CN, and 0.3 K km s$^{-1}$ for H$^{13}$CO$^+$.}
\label{f:ratios}
\end{figure*}

Observations of external galaxies often have limited signal-to-noise and therefore consider ratios of velocity-integrated line intensities. Figure~\ref{f:ratios} shows the HCN/HCO$^+$, \hthcn/\hthcop\ and HCN/HNC ratios across the W49A map, based on the integrated intensities of the J=4--3 lines shown in Fig.~\ref{f:W49main}, \new{with cutoffs applied to only bring out significant structure}. 
Table~\ref{t:ratios} gives the values of these ratios integrated over our maps.
Over most of the map, the HCN 4--3 line is weaker than the \hcop\ line, giving integrated intensity ratios $<$1. The situation is reversed for the $^{13}$C-substituted species: \hthcn\ is generally stronger than \hthcop.
\newnew{This reversal indicates a high optical depth of the main isotopic HCN and \hcop\ lines, as also suggested by the limited dynamic range of the maps in Fig.~\ref{f:ratios}}
\newnewnew{and the fact that the HCN/\hthcn, HNC/\hnthc\ and \hcop/\hthcop\ intensity ratios are 5--10. These ratios are wel below the isotopic abundance ratio of 60, which indicates line optical depths of $\sim$10, assuming that the $^{13}$C--substituted lines are optically thin.}

\subsection{Line profiles}\label{s:profile}

\begin{figure}
\rotatebox{270}{
\resizebox{!}{\hsize}{\includegraphics*[2cm,2cm][19cm,27.2cm]{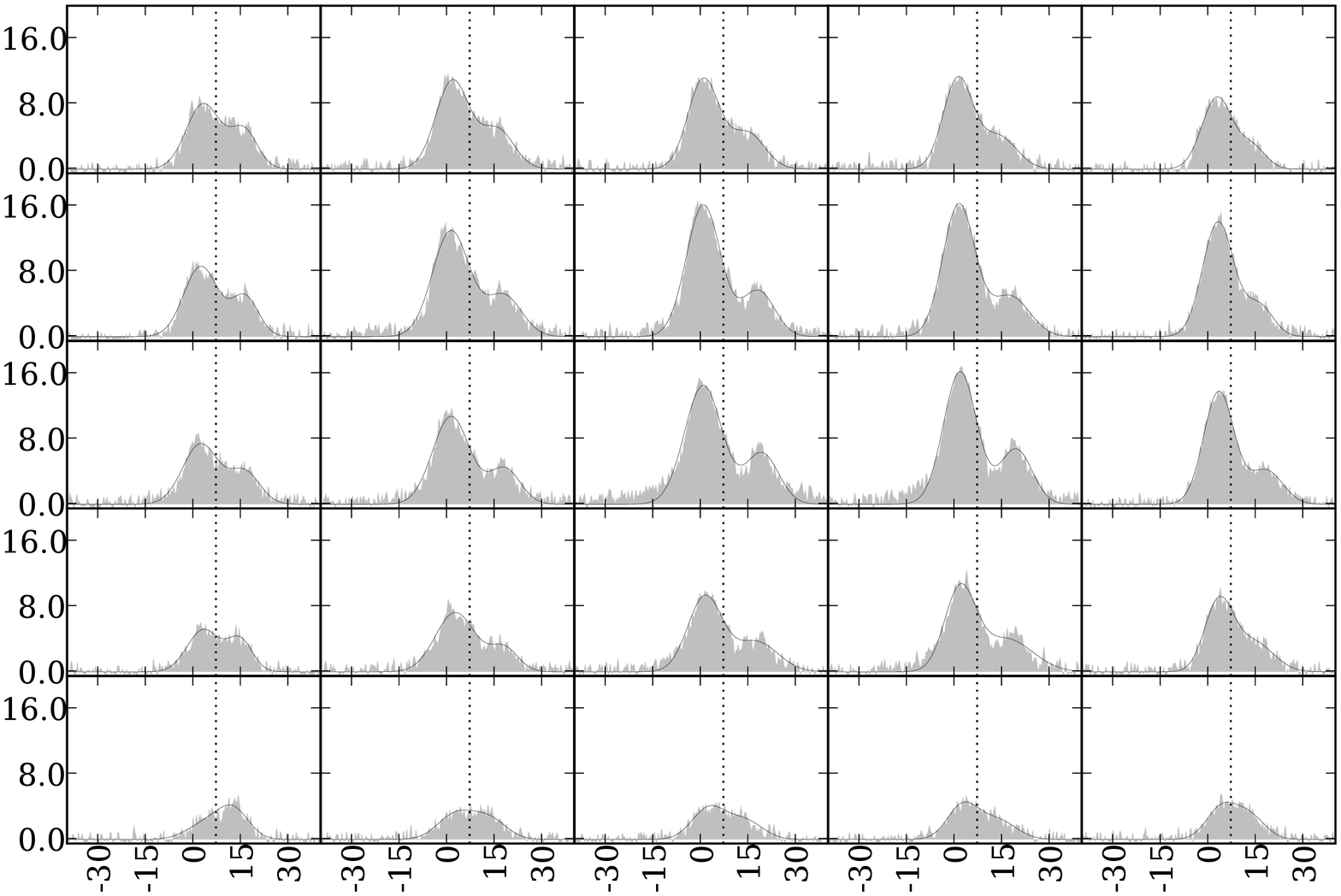}}}
\rotatebox{270}{
\resizebox{!}{\hsize}{\includegraphics*[2cm,2cm][19cm,27.2cm]{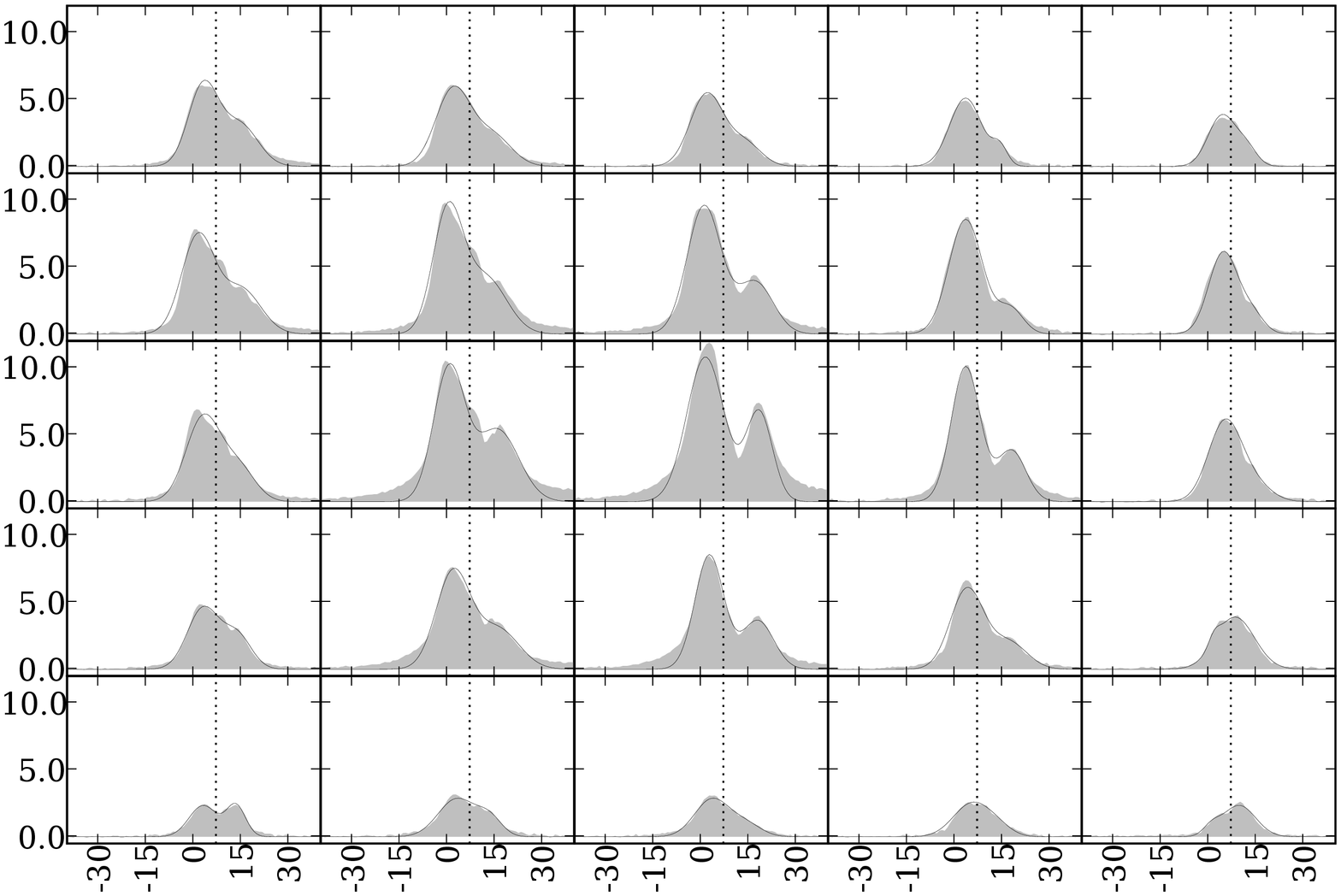}}}
\rotatebox{270}{
\resizebox{!}{\hsize}{\includegraphics*[2cm,2cm][19cm,27.2cm]{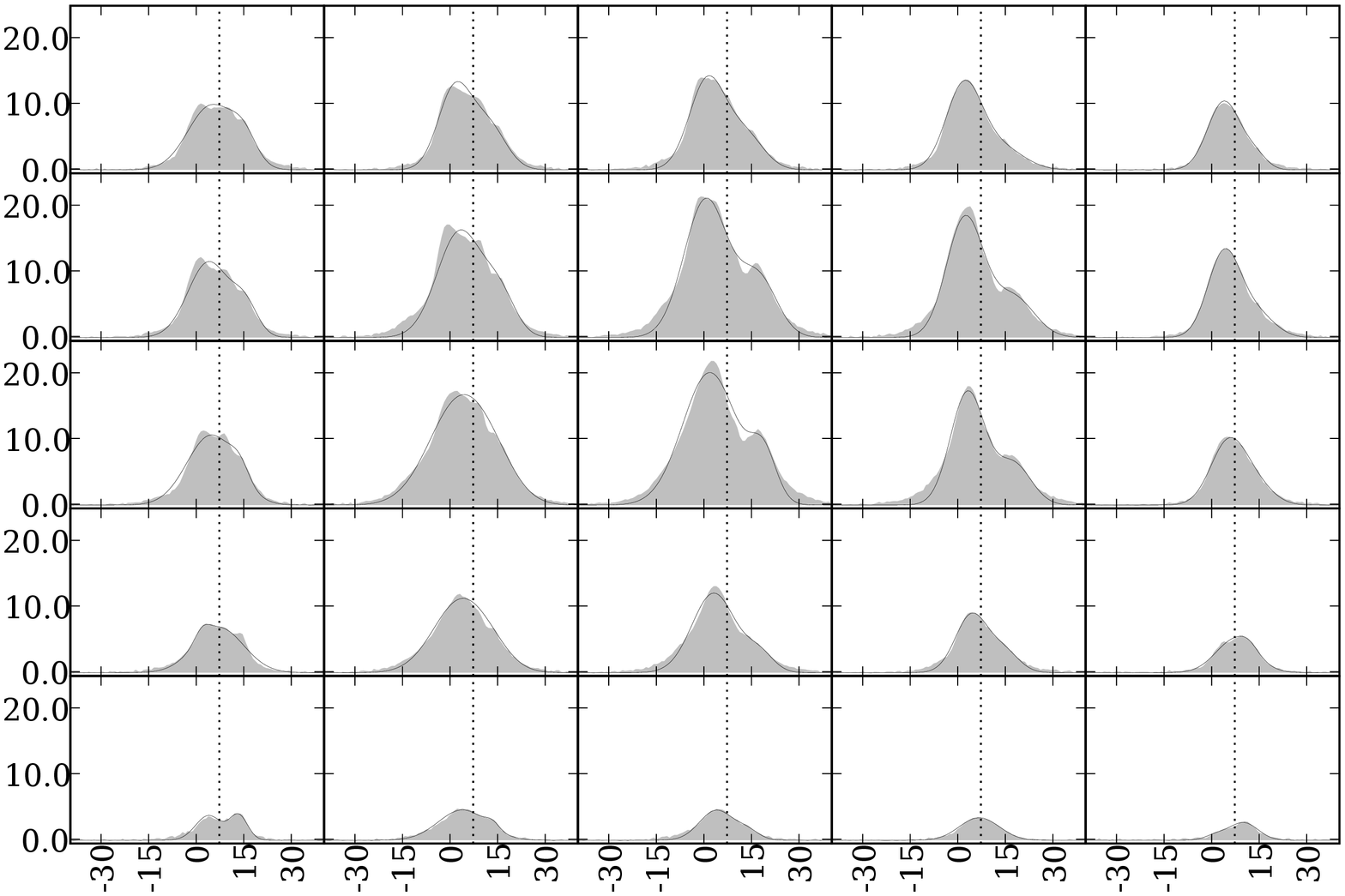}}}
\caption{Observed spectral lines (shaded grey) and Gaussian fit profiles (black lines) towards a 37.5$'' \times$ 37.5$''$ grid centred at R.A. = 19:10:13.4; Dec: 09:06:14. The X-axis is velocity relative to the line rest frequency (km~s$^{-1}$); the Y-axis is \tmb\ (K). HCN J=3--2 (top); HCN J=4--3 (middle); HCO$^+$ J=4--3 (bottom). The dashed vertical line on each plot is at the expected source velocity of 7~km~s$^{-1}$.}\label{f:W49_map_centre}
\end{figure}

\begin{figure}
\rotatebox{270}{
\resizebox{!}{\hsize}{\includegraphics*[2cm,-8cm][21cm,36cm]{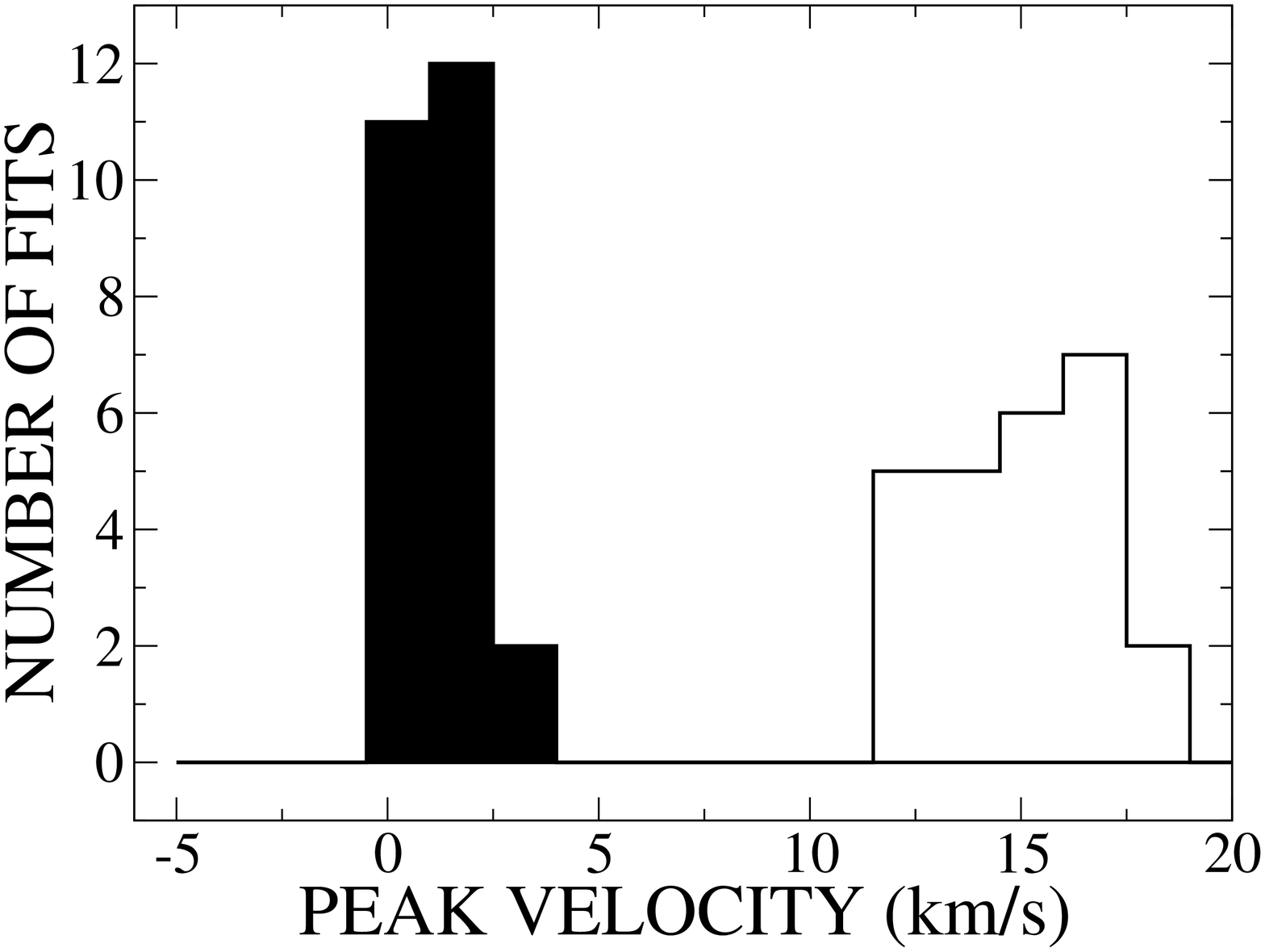}}}
\rotatebox{270}{
\resizebox{!}{\hsize}{\includegraphics*[2cm,-8cm][21cm,36cm]{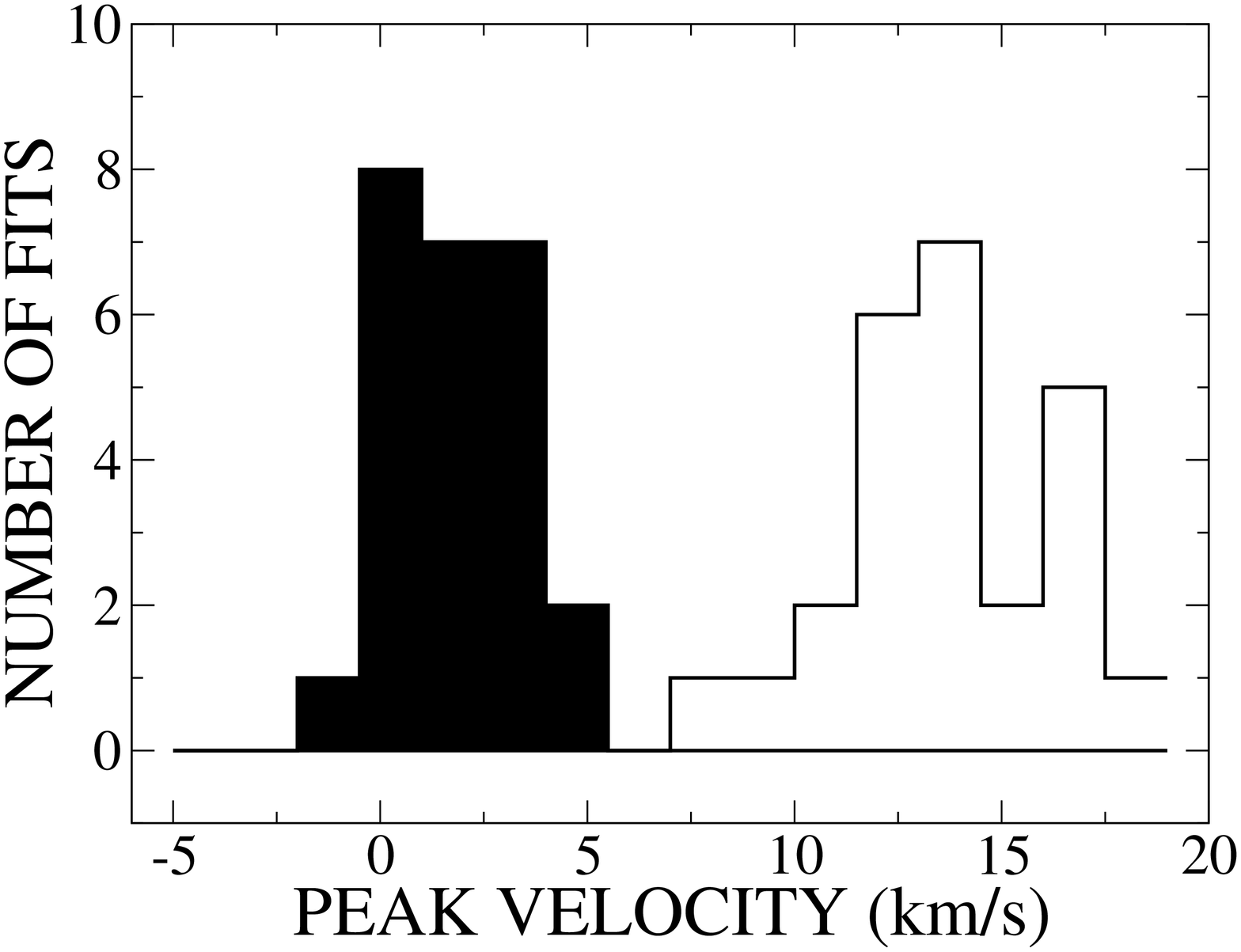}}}
\rotatebox{270}{
\resizebox{!}{\hsize}{\includegraphics*[2cm,-8cm][21cm,36cm]{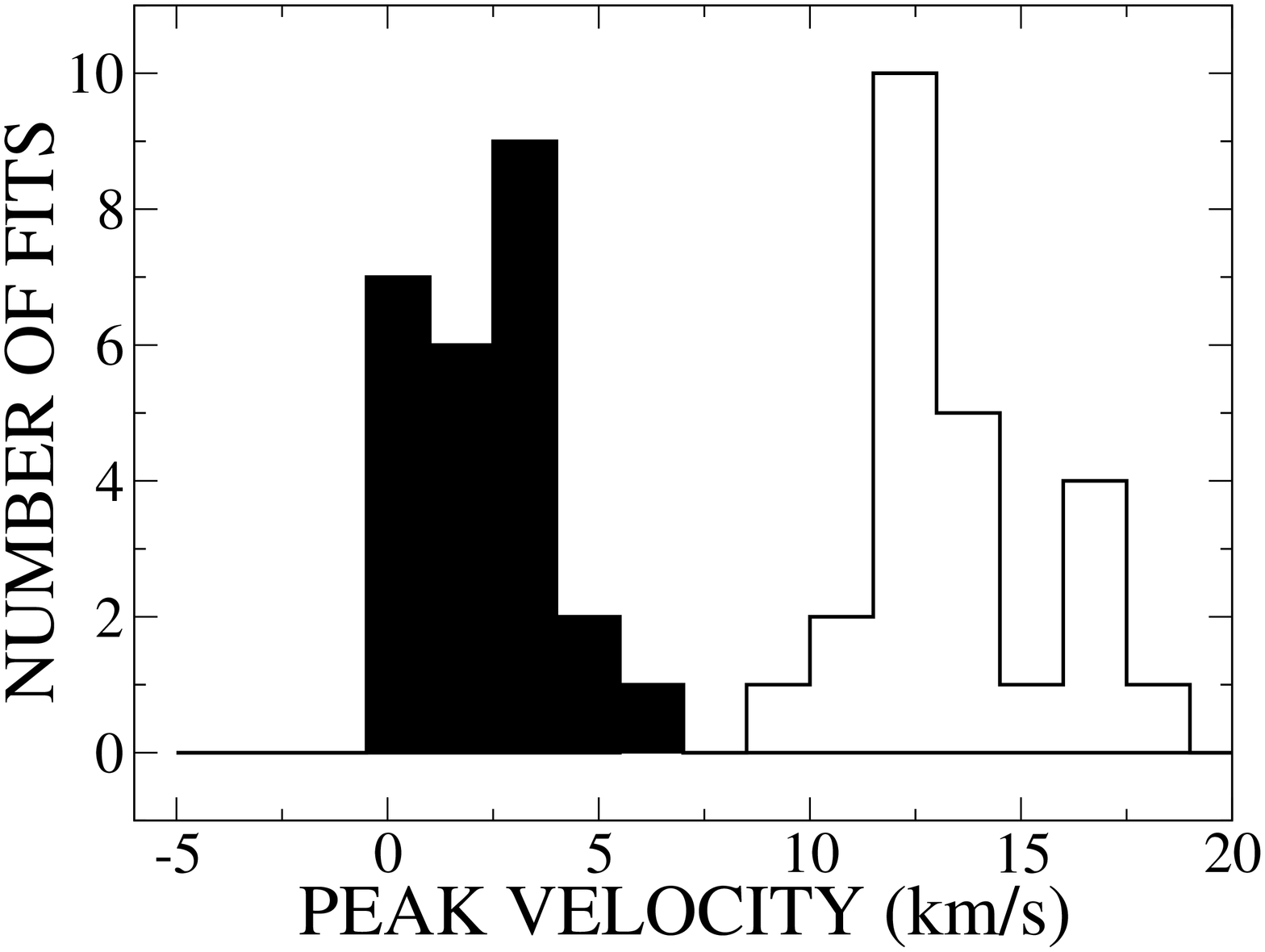}}}
\caption{Histograms of the peak velocity for the spectral lines observed towards the centre of W49A (see Fig.~\ref{f:W49_map_centre}) \newnewold{illustrating fit components} \newnewnew{measured with double Gaussian fits to the line profiles} at each map position. The number of lower velocity components is black and the number of higher velocity components is white. From top to bottom: HCN J=3--2; HCN J=4--3; HCO$^+$ J=4--3.}\label{f:W49_vel_centre}
\end{figure}

Figure~\ref{f:W49_map_centre} shows the HCN and HCO$^+$ line profiles observed towards 5$\times$5 pixels (37.5$\times$37.5~arcsec) centred at R.A. = 19:10:13.4; Dec.\ = 09:06:14, where the molecular emission peaks. Many of the profiles are double peaked and the rest are asymmetric. All profiles are best fitted by two Gaussians; the combined fit profiles are also shown. Fig.~\ref{f:W49_vel_centre} shows histograms of the number of spectra in each map as a function of the peak velocity of each Gaussian fit. 
\new{The bin width of 3\,\kms\ was chosen to be large enough to obtain significant signal in each bin, yet small enough to be able to locate the peaks of the distribution.}
For all three species, the lower-velocity component of each line peaks between 0 and $+$5~km~s$^{-1}$ while the higher-velocity component peaks between $+$12 and $+$15~km~s$^{-1}$.
On the other hand, our observations of the rarer isotopologues show single-peaked line profiles (Figs~\ref{f:W49_map_centre_HCNiso} and~\ref{f:W49_map_centre_HCOiso}) which peak close to the source velocity of $+$7$-$8\,\kms\ \citep{welch87,jaffe}. 
Thus, the HCN and HCO$^+$ lines are self-absorbed and highly optically thick. 

In the central region of W49A, the blue-shifted peaks of the main isotopic lines are brighter than the red-shifted peaks by 50--100\%, which is a classic sign of infall \citep[e.g.][]{zhou}. 
Infall motion has been observed toward similar regions: \new{on large scales toward} W43 \citep{motte03} and \new{on smaller scales toward} W51 \citep{zhang97}. 
In the case of W49A, \citet{dickel} suggest large scale free-fall collapse of the molecular cloud towards a central HII region based on HCO$^+$ $J$=1--0 and 3--2 line profiles.
However, \citet{serabyn93} find double-peaked line profiles in both CS and C$^{34}$S and argue that the starburst in W49A is due to the collision of two clouds.
Also, our maps show that although ``infall-type'' line profiles are seen in the east-west direction, to the north and south there are regions where the red-shifted emission is stronger than the blue, suggesting outflowing motions in this direction \newnewnew{(see also \S\,\ref{s:Nclump})}. 
Finally, the clump of emission in the south-west corner is characterised by single-peaked lines close to the expected source velocity. 
We conclude that the gas kinematics in W49A is complex, with infall and outflow motions probably occurring simultaneously.
\new{Imaging of W49A at high spatial and spectral resolution is necessary for a better understanding of its gas kinematics.}

\begin{figure*}
\includegraphics[width=\textwidth]{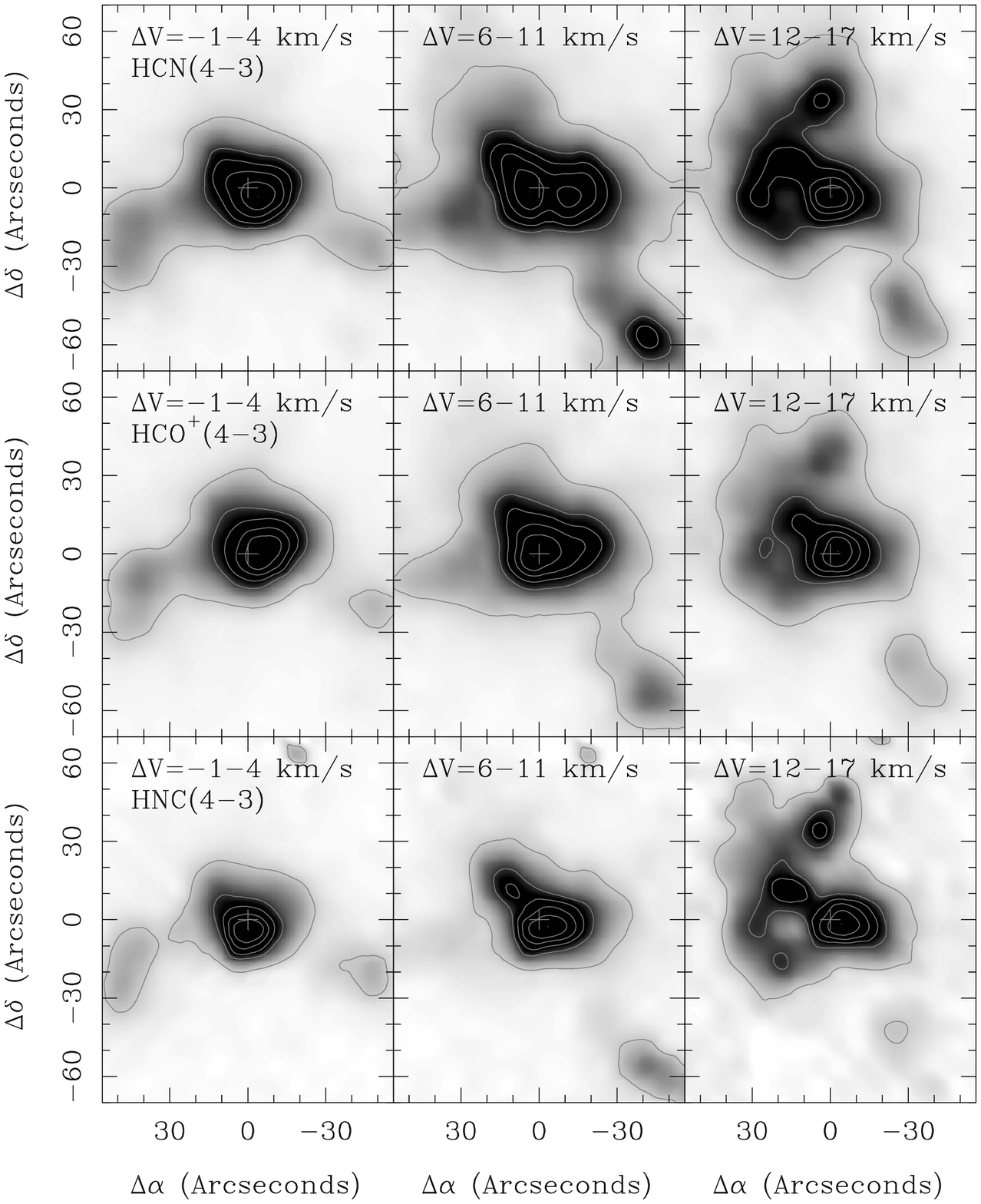}
\caption{Maps of the integrated intensity ($\int{T_A^*dV}$) across W49A over 5 km s$^{-1}$ velocity ranges. Left column: $-$1  to $+$4 km\,s$^{-1}$ (the blue-shifted peak); middle column: $+$6 to $+$11 km\,s$^{-1}$ (the expected source velocity); right column: $+$12 to $+$17 km\,s$^{-1}$ (the red-shifted peak). Top row: HCN J=4-3; centre row: HCO$^+$ J=4-3; bottom row: HNC J=4-3. The contours show 10 to 100\% of the peak integrated intensity in steps of 15\%. The peak intensities are 33, 18, and 16 K\,km\,s$^{-1}$, for the blue-shifted, central velocity, and red-shifted HCN J=4–3 emission, respectively; 63, 43, and 33 K\,km\,s$^{-1}$ for HCO$^+$; and 16, 12,and 5 K\,km\,s$^{-1}$  for HNC.}
\label{f:W49_line_map}
\end{figure*}

In order to investigate the velocity profiles across the maps, Fig.~\ref{f:W49_line_map} shows integrated intensity maps at the velocities of the HCN, HCO$^+$, and HNC J=4--3 fit components as well as at the expected source velocity.  While the spatial distributions are significantly different at the different velocity ranges, the maps of all molecules over the same velocity range are very similar. The blue-shifted emission is stronger at the east and west edges of the map, while the red-shifted emission is more extended in the north-south direction.
\new{The maps in Fig.~\ref{f:W49_line_map} also suggest the existence of unresolved substructure: Possible effects of this substructure will be discussed in \S~\ref{s:column}.}

Based on Fig.~\ref{f:W49_line_map}, we select three regions of the map away from the source centre to study in more detail: the ``South-West clump'' (centred at R.A.=19:10:10.6; Dec=09:05:18) is clearly seen at the source velocity and in the red-shifted component, but is much fainter in the blue-shifted component; the ``Northern clump'' (R.A.=19:10:13.6; Dec.=09:06:48) is a peak in the integrated intensity seen only in the red-shifted component; while the ``Eastern tail'' (R.A.=19:10:16.6; Dec.=09:05:48) is much stronger at the velocity of the blue-shifted component. 

\subsection{Column densities}\label{s:column}

\begin{figure}
\rotatebox{270}{
\resizebox{!}{\hsize}{\includegraphics*[2cm,1cm][19cm,28cm]{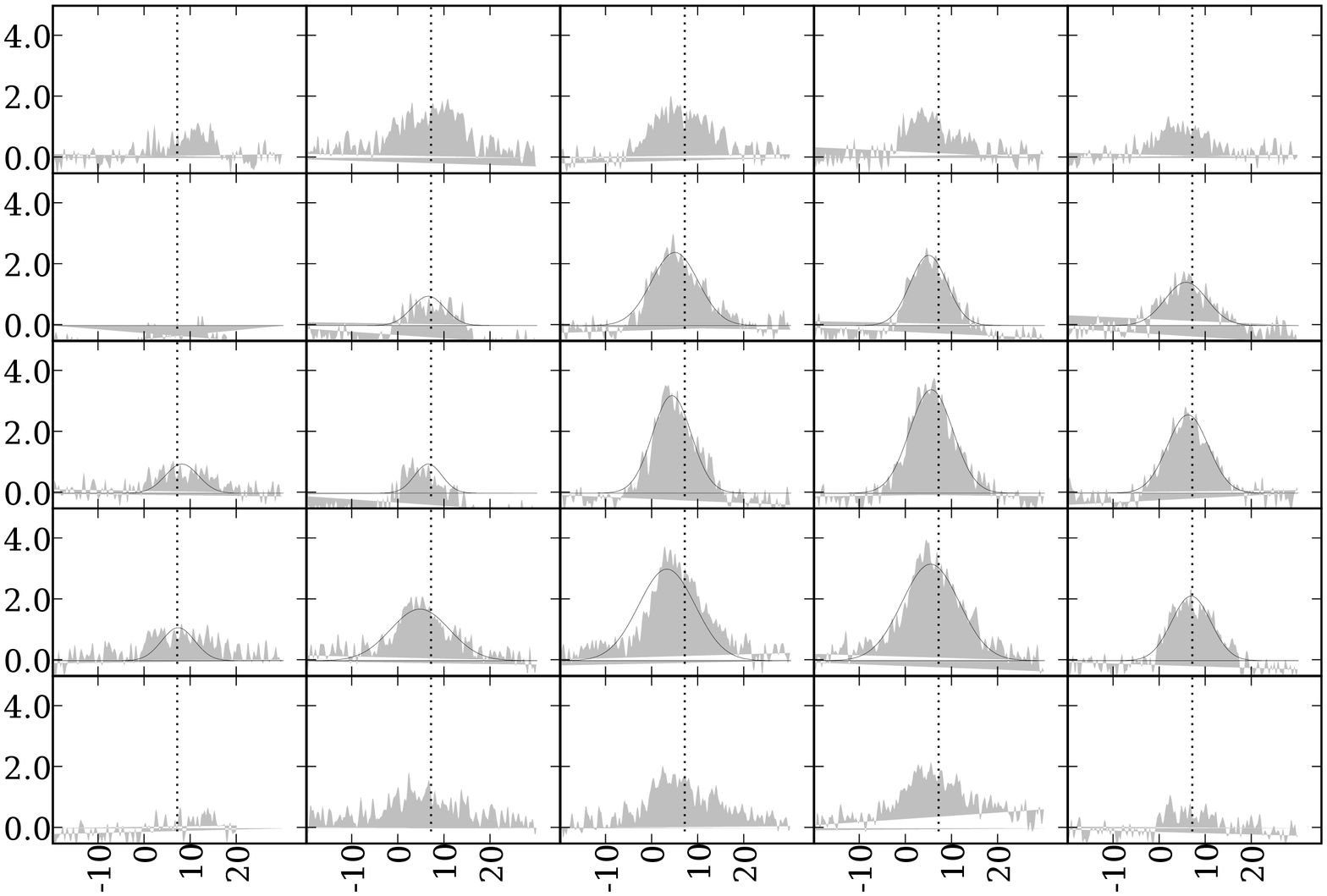}}}
\rotatebox{270}{
\resizebox{!}{\hsize}{\includegraphics*[2cm,1cm][19cm,28cm]{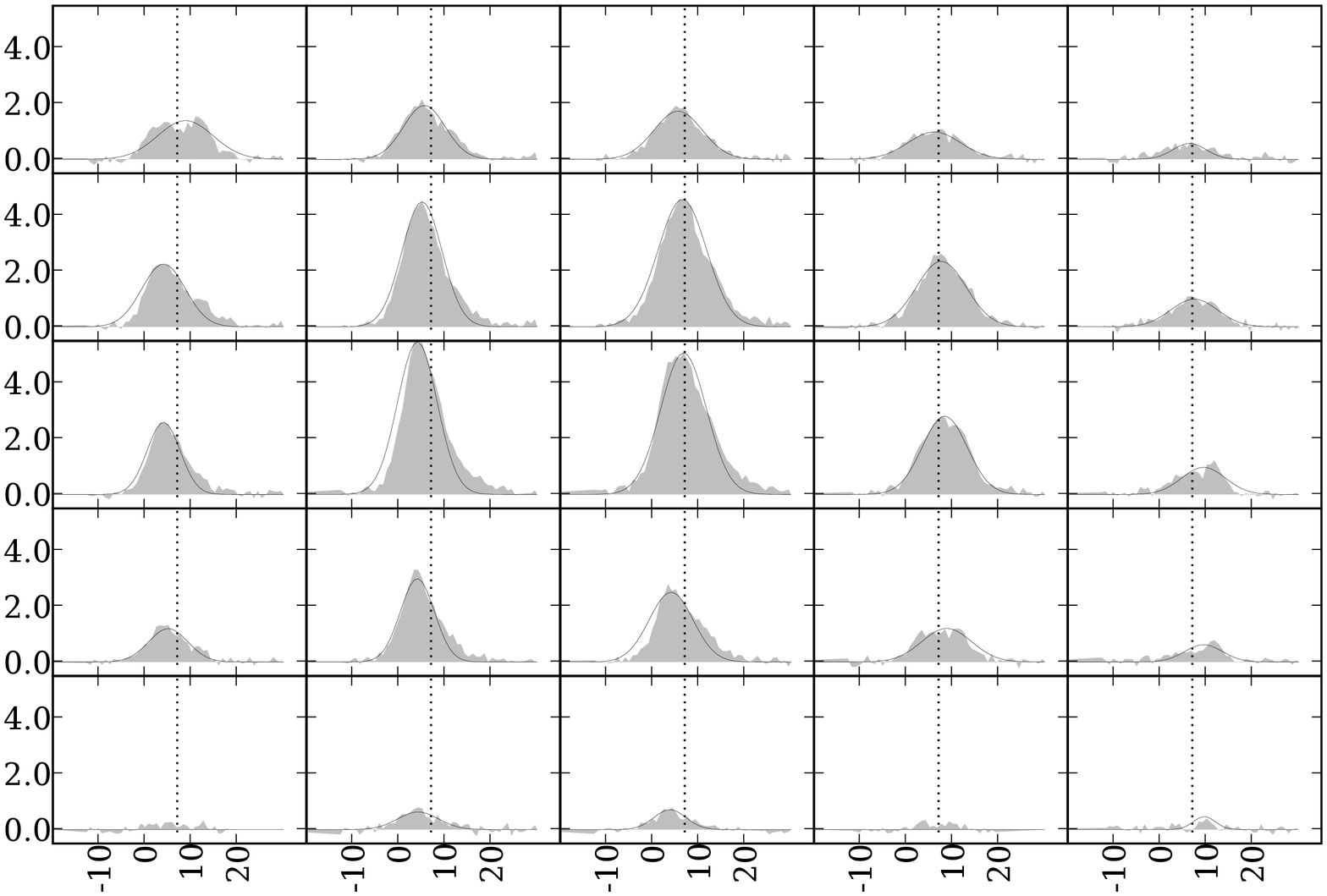}}}
\rotatebox{270}{
\resizebox{!}{\hsize}{\includegraphics*[2cm,1cm][19cm,28cm]{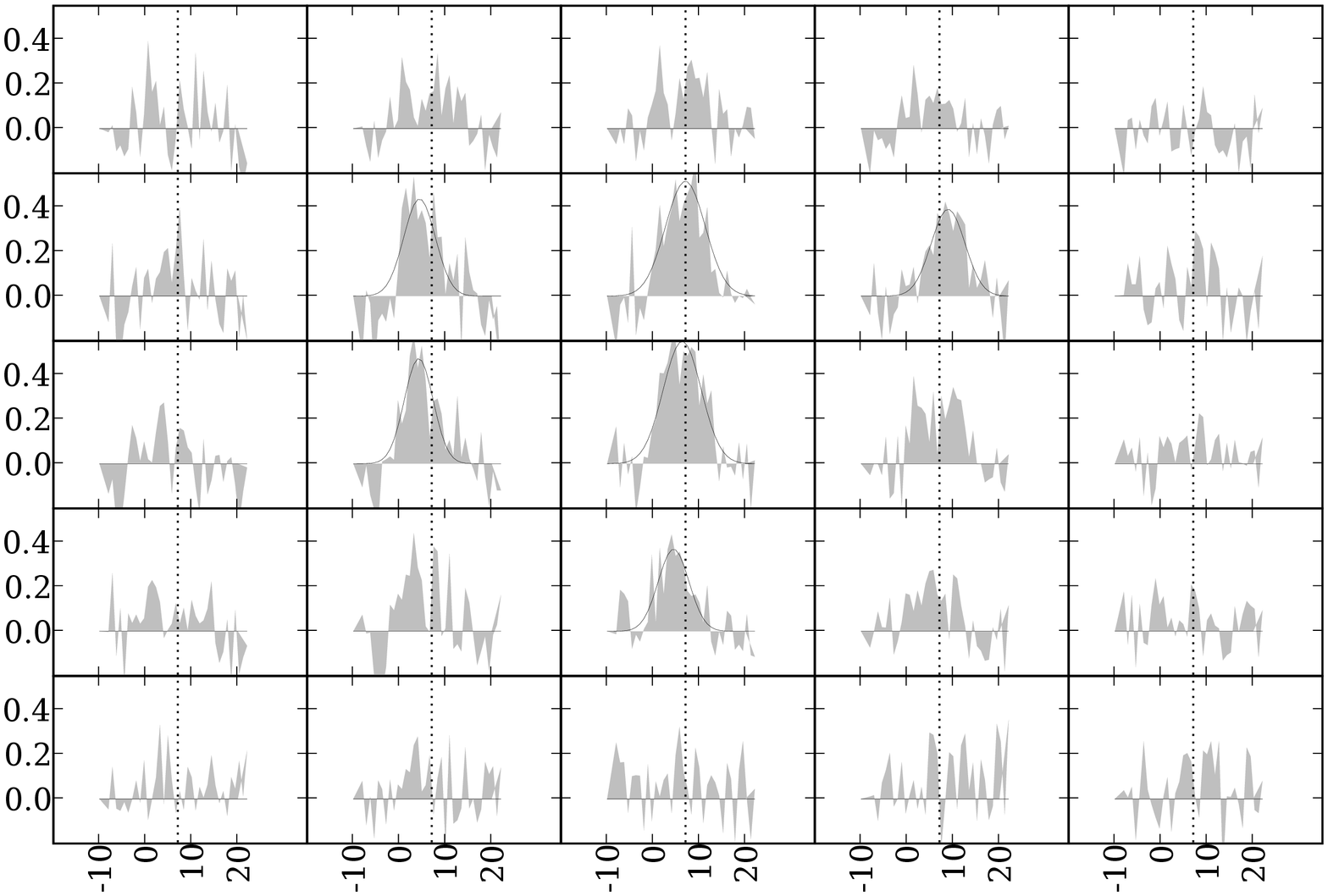}}}
\caption{As for Fig.~\ref{f:W49_map_centre} but, top to bottom: H$^{13}$CO$^+$ J=3--2; H$^{13}$CO$^+$ J=4--3; HC$^{18}$O$^+$ J=4--3. }\label{f:W49_map_centre_HCOiso}
\end{figure}

\begin{figure}
\rotatebox{270}{
\resizebox{!}{\hsize}{\includegraphics*[2cm,1cm][19cm,28cm]{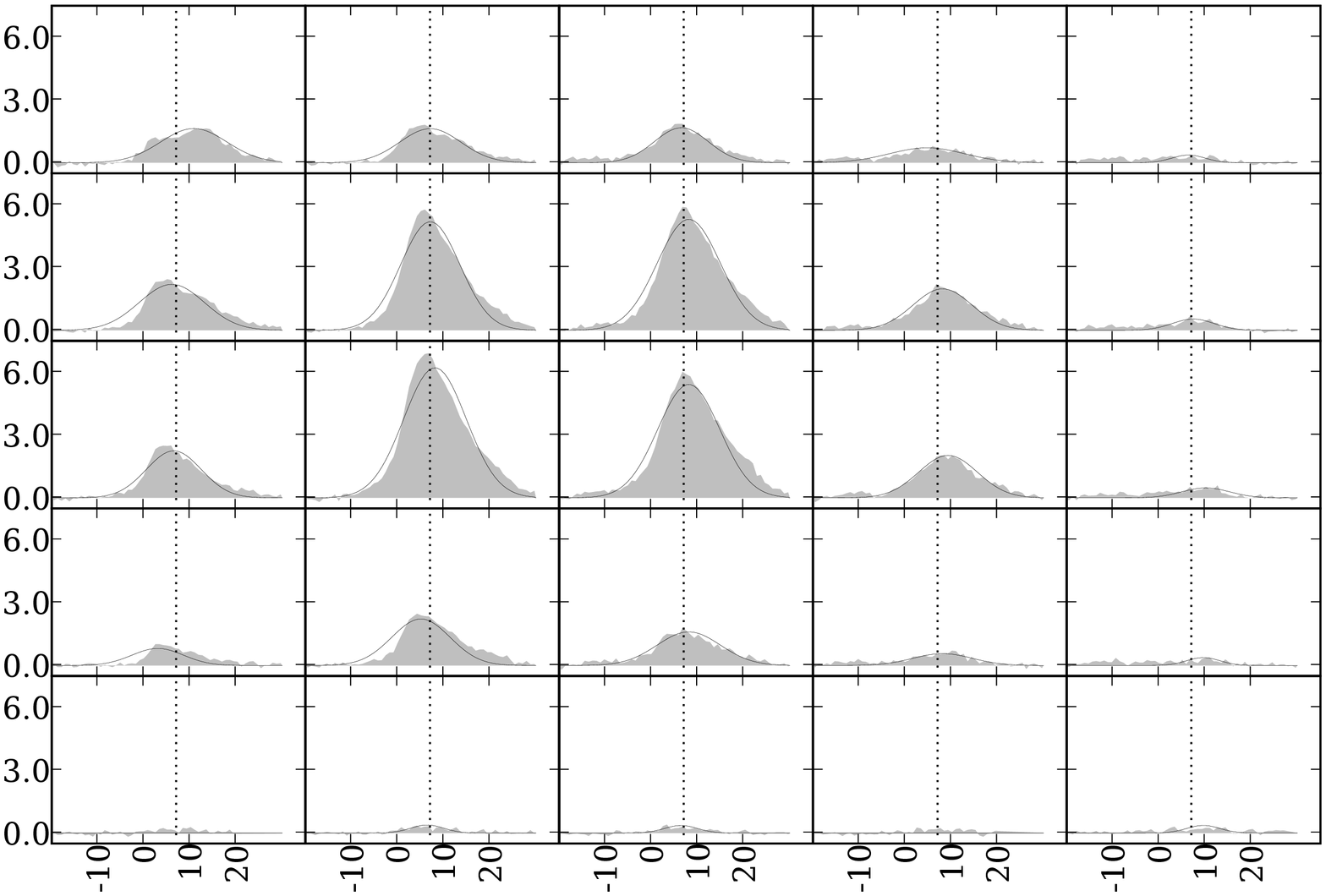}}}
\rotatebox{270}{
\resizebox{!}{\hsize}{\includegraphics*[2cm,1cm][19cm,28cm]{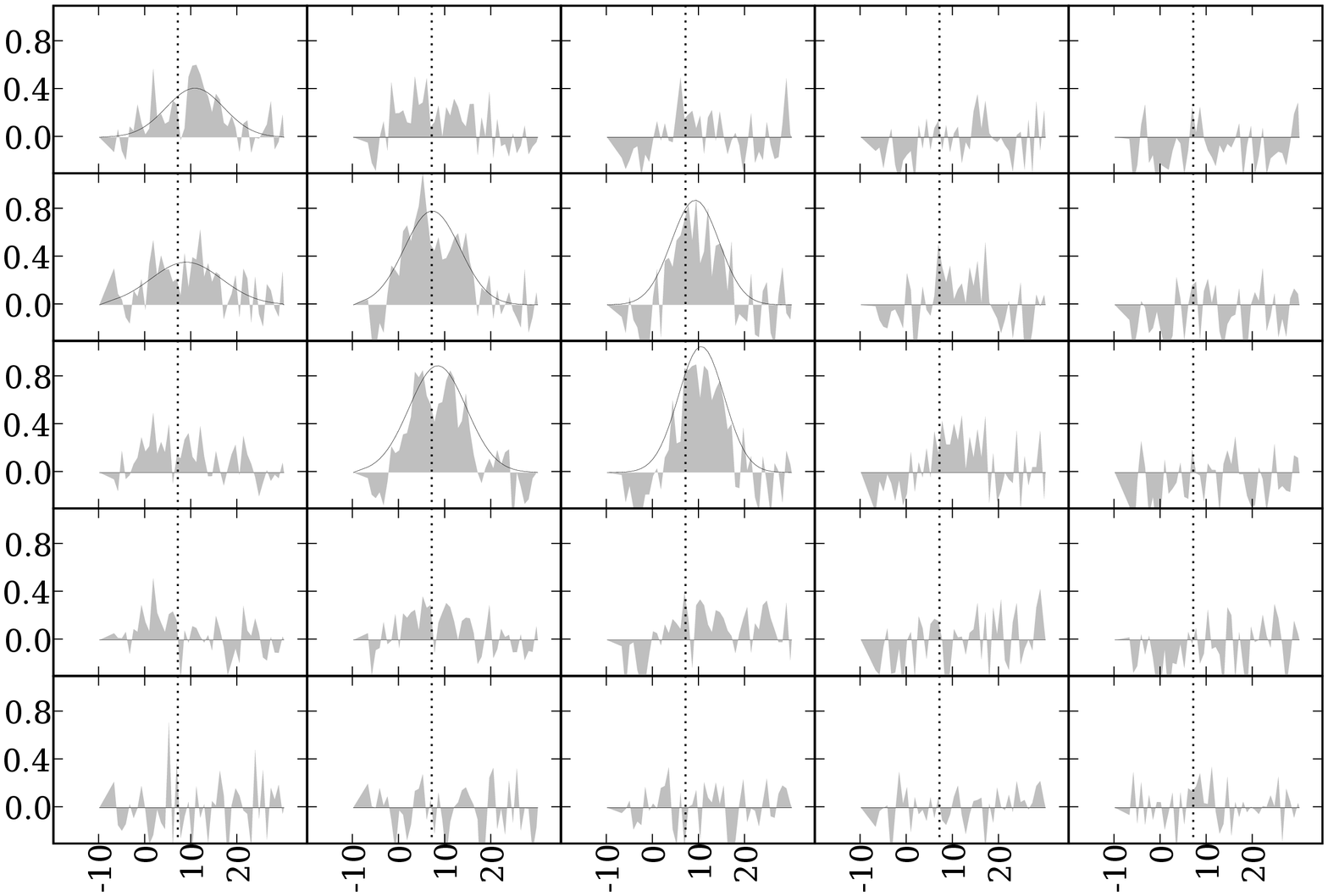}}}
\rotatebox{270}{
\resizebox{!}{\hsize}{\includegraphics*[2cm,1cm][19cm,28cm]{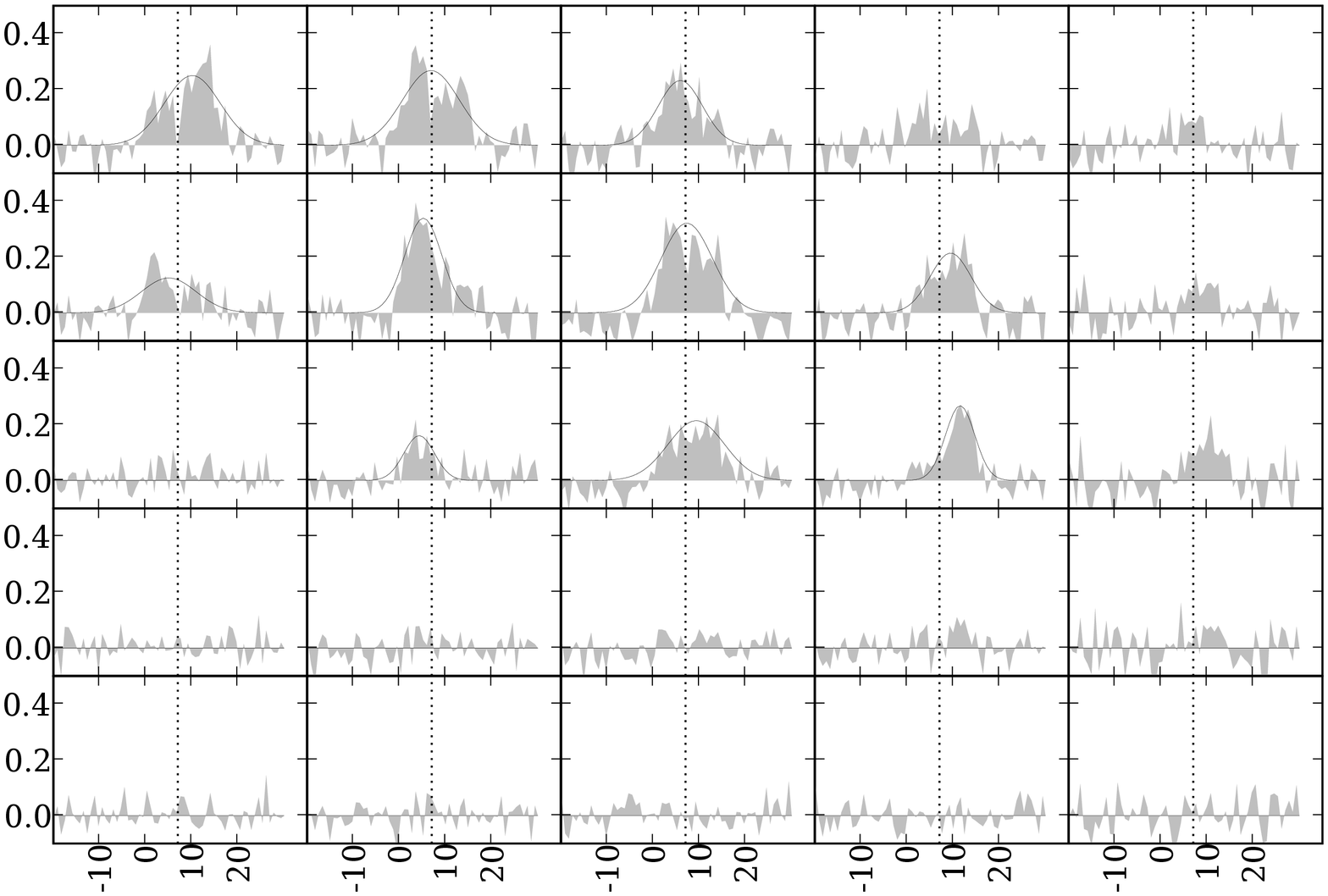}}}
\caption{As for Fig.~\ref{f:W49_map_centre} but, top to bottom: H$^{13}$CN J=4--3; HC$^{15}$N J=4--3; DCN$^+$ J=5--4.}\label{f:W49_map_centre_HCNiso}
\end{figure}

\begin{figure}
\rotatebox{270}{
\resizebox{!}{\hsize}{\includegraphics*[2cm,-5cm][21cm,33cm]{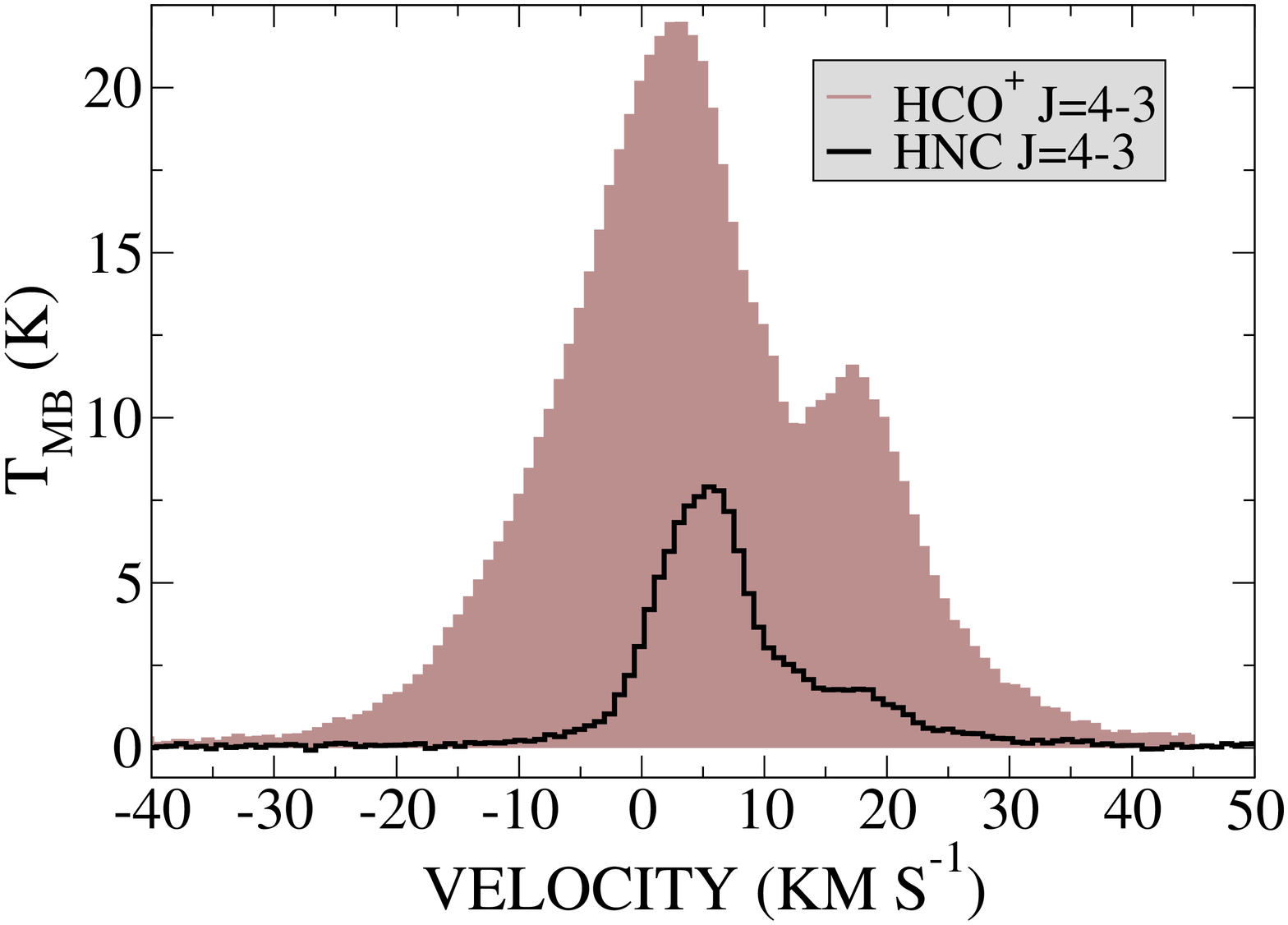}}}
\rotatebox{270}{
\resizebox{!}{\hsize}{\includegraphics*[2cm,-5cm][21cm,33cm]{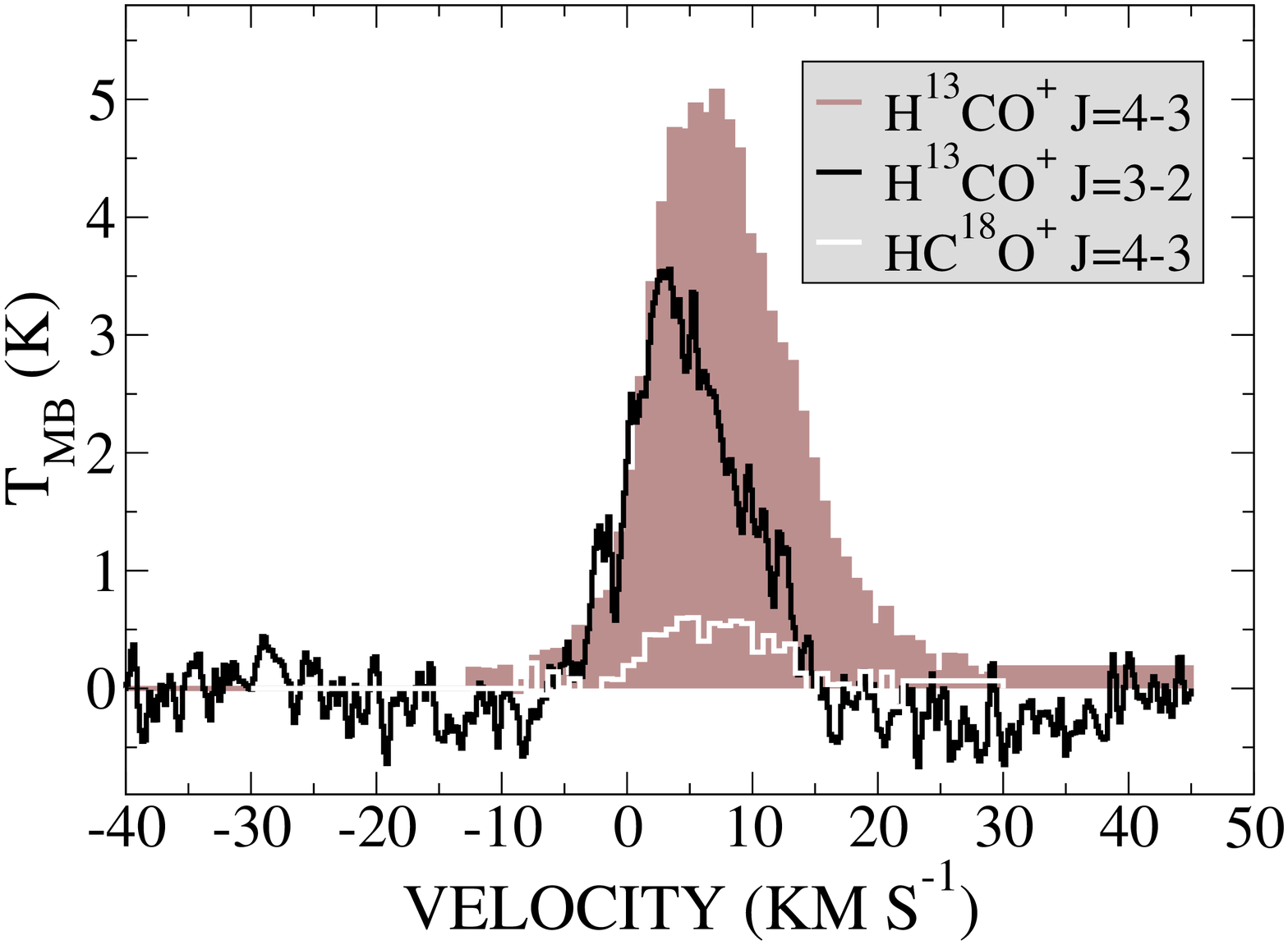}}}
\rotatebox{270}{
\resizebox{!}{\hsize}{\includegraphics*[2cm,-5cm][21cm,33cm]{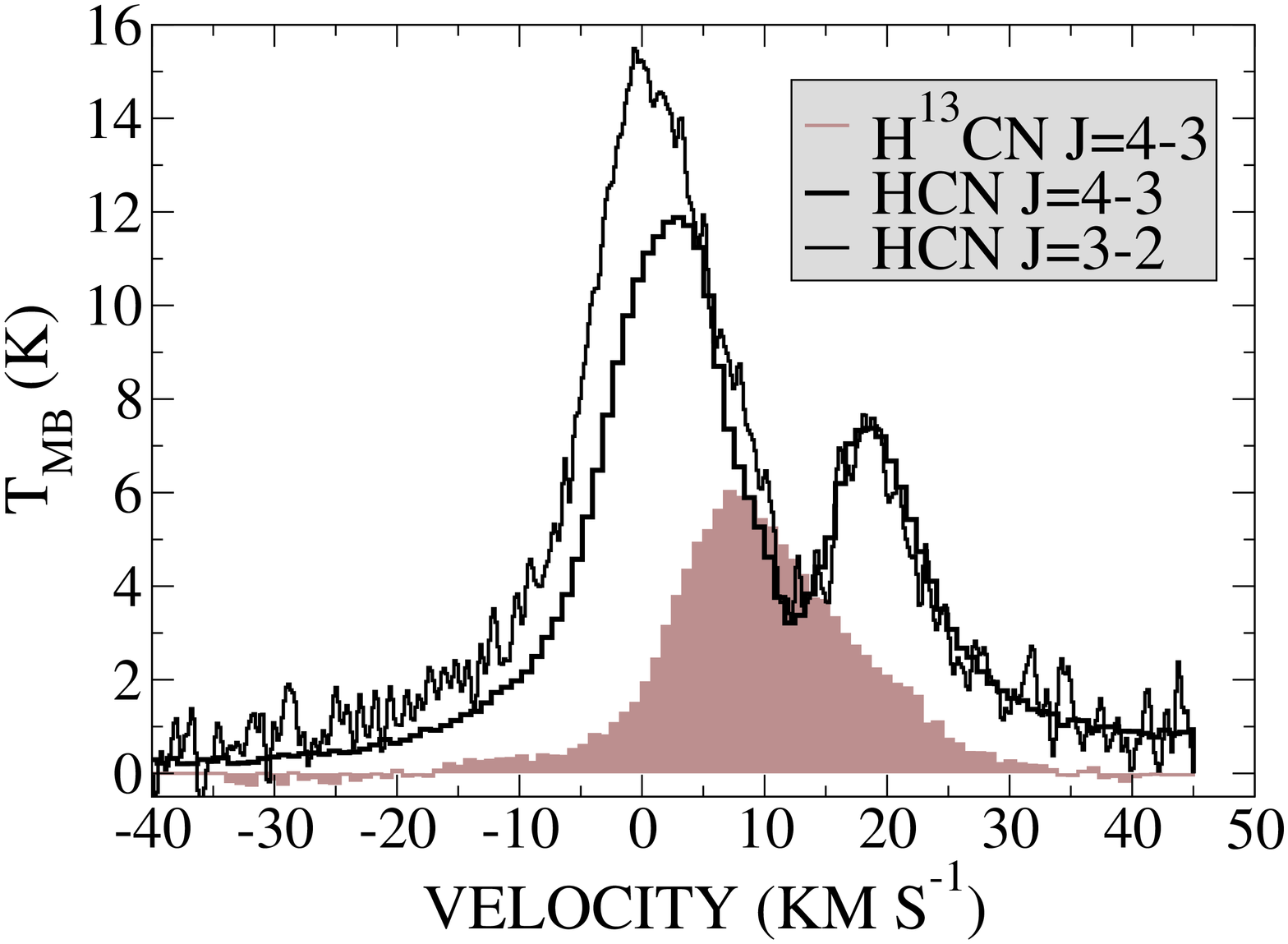}}}
\rotatebox{270}{
\resizebox{!}{\hsize}{\includegraphics*[2cm,-5cm][21cm,33cm]{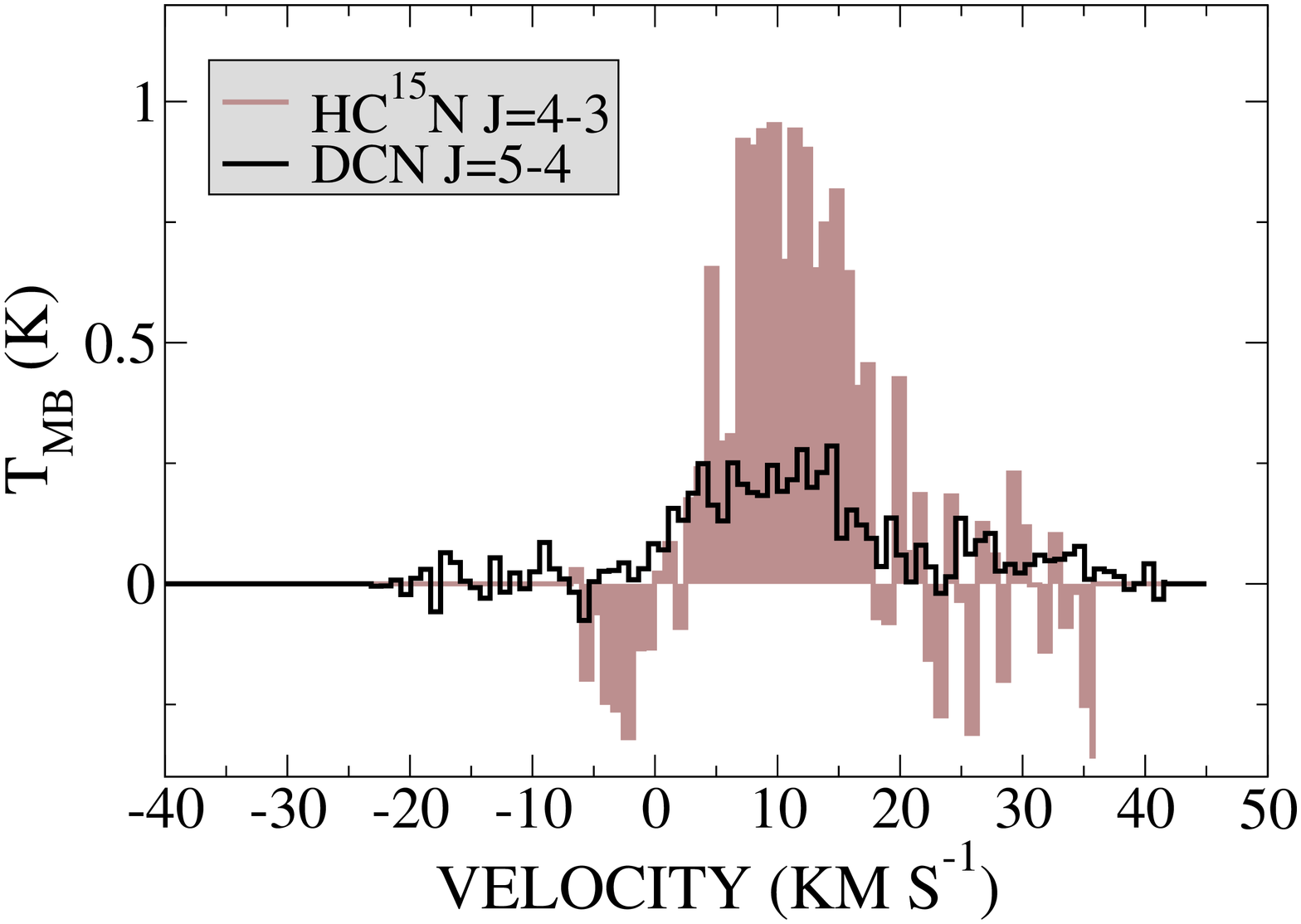}}}
\caption{Line profiles towards the central position (R.A.=19:10:13.4; Dec.=9:06:14). Note the difference in beam size between the 4-3 and 3-2 transitions. }\label{f:centre_line_shapes}
\end{figure}

To derive the column densities of HCN, HNC and HCO$^+$, we use the lines of their isotopically substituted species, which have a lower optical depth than the main species.
Figs.~\ref{f:W49_map_centre_HCOiso} and \ref{f:W49_map_centre_HCNiso} show the observed profiles of these lines, along with single Gaussian fits. Although the $^{13}$C-substituted lines are fitted reasonably well, these lines are broader than the weaker isotopologues and show somewhat asymmetric profiles, so their optical depths may not be negligible. This is shown in more detail in Fig.~\ref{f:centre_line_shapes}, which compares the line profiles of the HCO$^+$, HCN, and HNC (and isotopologue) spectra towards the central pixel of the W49A map. The $^{13}$C-substituted lines do peak at $\sim$8~km~s$^{-1}$, close to the source velocity, but, as well as being asymmetric, the H$^{13}$CN J=4--3 line FWHM is $\sim$1.4 times that of the HC$^{15}$N J=4--3 line.  For this reason we have, wherever possible, based our calculations of the HCO$^+$ and HCN column densities and the DCN/HCN ratios on the HC$^{15}$N and HC$^{18}$O$^+$ lines. 

We use the RADEX program \citep{vdt_radex} to estimate column densities from the observed widths and peak intensities of the Gaussian fits to the line emission. 
This program performs statistical equilibrium calculations including collisional and radiative processes, and treats optical depth effects with an escape probability formalism.
\new{The program assumes that the optical depth is independent of velocity, which is appropriate since we are modeling the velocity-integrated line intensities.}
Molecular input data for HCN and \hcop\ are taken from the LAMDA database \citep{schoeier}; we assume that collisional rates do not change upon isotopic substitution which is probably valid given the identical electronic structure. \newnewold{Since collision data for HNC do not exist,} 
We adopt the HCN collision rates for HNC, which assumption is a potential source of error since the two species have somewhat different dipole moments (3.0 vs 3.3~D).
\newnewnew{Very recently, \citet{sarrasin:hcn-he} and \citet{dumouchel:hcn-he} have presented collision data for HCN and HNC with He; the impact of these data on the present work will be analyzed in a forthcoming paper.}

Since the emission is spatially extended, we assume a beam filling factor of unity for all lines. 
\new{Due to the presence of substructure within the JCMT beam (e.g., \citealt{dickel99}), this assumption provides lower limits on the line optical depths and molecular column densities. The measured line widths provide an estimate of these limits.
For example, the width of the H$^{13}$CN 4--3 line toward the source centre,
which is 1.4$\times$ larger than that of the HC$^{15}$N line,
suggests a 3--5$\times$ higher optical depth than the value of $\tau=0.2$ estimated with RADEX.
However, since much of our analysis is based on line ratios, the optical depths of individual lines are not a major concern. Section~\ref{s:disc} provides further discussion on the effect of unresolved substructure.}

For HCN and \hcop, the intensities of two transitions are known, which allows an estimate of the ambient physical conditions (kinetic temperature and H$_2$ volume density). In the case of HNC, only one transition was observed, and we have adopted the temperature and density found for HCN and \hcop.

\subsubsection{The source centre}\label{s:centre}

We calculate the column density of the central source adopting \tkin = 100\,K
and \nhh = \pow{2}{6} cm$^{-3}$ (Tables 3 and 4).
Using these physical conditions, \txc\ (\hthcop\ $J$=3--2) $\sim$ \tkin,
and the HCO$^+$ column density determinations from \hthcop\ $J$=3--2,
\hthcop\ $J$=4--3, and HC$^{18}$O$^+$ $J$=4--3 towards the source centre all
agree to within a factor of 2, with the column density determined from the
$J$=4--3 and $J$=3--2 transitions becoming more discrepant at lower density.  
Over the range \tkin = 75 -- 200\,K and
\nhh = $5\times10^5-5\times10^6$ cm$^{-3}$ (\S~1), the derived column
density determinations from the H$^{13}$CN $J$=4--3 transitions agree very well
with those from the HC$^{15}$N $J$=4--3 transitions, assuming isotopic ratios
appropriate for the local ISM \citep{wilson_rood}.  The calculated HCN and
DCN column densities fall by a factor of 2 as the kinetic temperature
increases from 75 to 200 K or by almost an order of magnitude if the density
increases from $5\times10^5$ cm$^{-3}$ to $5\times10^6$ cm$^{-3}$.  Over this
range of temperature and density, the calculated excitation temperatures for
the HCN and DCN transitions are similar to each other and increase from
$\sim15$ K for \tkin =75~K and \nhh = $5\times10^5$ cm$^{-3}$ to
almost 40 K for \tkin = 200~K and \nhh = $5\times10^6$ cm$^{-3}$. 

\begin{table*}
\caption{Estimated HCO$^+$ column densities towards the source centre.}
\label{t:HCOobs}
\begin{tabular}{llccccccc}
\hline \hline \noalign{\smallskip}
Offset\tablefootmark{a} & Transition & Velocity & \tmb\ & FWHM & \txc\ & $\tau$ & $N$ & $N$(HCO$^+$)\tablefootmark{b} \\
 ($'',''$) & & (km~s$^{-1}$) & (K)& (km~s$^{-1}$)& (K) & & (10$^{12}$ cm$^{-2}$) & (10$^{14}$ cm$^{-2}$)\\
\noalign{\smallskip} \hline \noalign{\smallskip}
15.0,--7.5 &H$^{13}$CO$^+$ J=4--3&  5.0 &  1.2 & 10.0 & 37.6 & 0.04 & 4.5 & 3.5 \\
15.0,0.0   &H$^{13}$CO$^+$ J=4--3&  4.0 &  2.6 &  8.4 & 38.3 & 0.08 & 8.3 & 6.4  \\
  15.0,7.5 &H$^{13}$CO$^+$ J=4--3&  4.0 &  2.2 & 11.1 & 38.1 & 0.07 & 9.5 & 7.3  \\
 15.0,15.0 &H$^{13}$CO$^+$ J=4--3&  8.8 &  1.4 & 14.2 & 37.7 & 0.04 & 7.3 & 5.6 \\
7.5,--15.0 &H$^{13}$CO$^+$ J=4--3&  4.2 &  0.6 & 10.4 & 37.3 & 0.02 & 2.4 & 1.9  \\
 7.5,--7.5 &H$^{13}$CO$^+$ J=4--3&  4.0 &  3.0 &  8.7 & 38.5 & 0.10 & 9.9 & 7.7  \\
   7.5,0.0 &HC$^{18}$O$^+$ J=4--3&  4.2 &  0.5 &  7.4 & 38.6 & 0.01 & 1.3 & 7.2 \\
   7.5,7.5 &HC$^{18}$O$^+$ J=4--3&  4.3 &  0.4 &  8.1 & 38.6 & 0.01 & 1.3 & 7.3  \\
  7.5,15.0 &H$^{13}$CO$^+$ J=4--3&  5.5 &  1.9 & 10.6 & 38.0 & 0.06 & 7.7 & 6.0  \\
0.0,--15.0 &H$^{13}$CO$^+$ J=4--3&  3.8 &  0.7 &  8.2 & 37.3 & 0.02 & 2.1 & 1.7 \\
 0.0,--7.5 &HC$^{18}$O$^+$ J=4--3&  4.4 &  0.4 &  7.8 & 38.6 & 0.01 & 1.1 & 6.0  \\
  0.0,0.0 & HC$^{18}$O$^+$ J=4--3&  6.2 &  0.5 &  9.7 & 38.7 & 0.02 & 2.0 & 11.  \\
   0.0,7.5 &HC$^{18}$O$^+$ J=4--3&  6.9 &  0.5 & 10.2 & 38.6 & 0.02 & 1.9 & 11.  \\
  0.0,15.0 &H$^{13}$CO$^+$ J=4--3&  5.5 &  1.7 & 12.3 & 37.9 & 0.06 & 8.0 & 6.2  \\
--7.5,--15.0 &H$^{13}$CO$^+$ J=3--2&  6.4 &  1.7 & 27.3 & 86.2 & 0.02 & 20. & 16.\\
--7.5,--7.5 &H$^{13}$CO$^+$ J=4--3&  8.8 &  1.2 & 12.7 & 37.6 & 0.04 & 5.7 & 4.4 \\
--7.5,0.0 &H$^{13}$CO$^+$ J=4--3&  8.3 &  2.8 & 11.5 & 38.4 & 0.09 & 12. & 9.4  \\
--7.5,7.5 &HC$^{18}$O$^+$ J=4--3&  8.9 &  0.4 &  8.6 & 38.6 & 0.01 & 1.2 & 7.0  \\
--7.5,15.0 &H$^{13}$CO$^+$ J=4--3&  6.1 &  1.0 & 13.3 & 37.5 & 0.03 & 4.9 & 3.8  \\
--15.0,--15.0 &H$^{13}$CO$^+$ J=4--3&  9.6 &  0.5 & 5.6 & 37.2 & 0.02 & 1.0 & 0.8\\
--15.0,--7.5 &H$^{13}$CO$^+$ J=4--3&  9.4 &  0.6 &  9.6 & 37.3 & 0.02 & 2.2 & 1.7\\
--15.0,0.0 &H$^{13}$CO$^+$ J=4--3&  9.4 &  1.0 & 11.1 & 37.5 & 0.03 & 4.0 & 3.1  \\
--15.0,7.5 &H$^{13}$CO$^+$ J=4--3&  7.4 &  1.0 & 12.0 & 37.5 & 0.03 & 4.4 & 3.4 \\
--15.0,15.0 &H$^{13}$CO$^+$ J=4--3&  6.4 &  0.6 &  8.4 & 37.3 & 0.02 & 1.8 & 1.4 \\
\noalign{\smallskip} \hline \noalign{\smallskip}
\end{tabular}
\tablefoot{HCO$^+$ column densities are calculated with RADEX assuming \tkin=100~K and \nhh= 2$\times$10$^6$~cm$^{-3}$. For the offset positions where HC$^{18}$O$^+$ was not detected, the data for H$^{13}$CO$^+$ J=4--3 were used, except for (--7.5$''$,--15.0$''$), where H$^{13}$CO$^+$ J=4--3 was not detected and H$^{13}$CO$^+$ J=3--2 was used. For the other positions, where all transitions were detected, data for the HC$^{18}$O$^+$ J=4--3 transition are listed and we note that the column density determinations from the H$^{13}$CO$^+$ J=4--3 line agree with these to within $\pm$50\%. \\
\tablefoottext{a}{Offsets from R.A.=19:10:13.4, Dec.=09:06:14} \\
\tablefoottext{b}{based on $^{12}$C/$^{13}$C = 77; $^{16}$O/$^{18}$O = 560 \citep{wilson_rood}.}
}
\end{table*}

\begin{table*}
\caption{Estimated HCN and DCN column densities towards the source centre.}
\label{t:DCNobs}
\begin{tabular}{llccccccccc}
\hline \hline \noalign{\smallskip}
Offset\tablefootmark{a} & Transition & Velocity & \tmb\ & FWHM & \txc\ & $\tau$ & $N$ & $N$(HCN)\tablefootmark{b} & $\frac{N({\rm DCN})}{N({\rm HCN})}$ & $\frac{N({\rm HCN})}{N({\rm HCO}^+)}$ \\
 ($'',''$) & & (km~s$^{-1}$) & (K)& (km~s$^{-1}$)& (K) & & (10$^{12}$ cm$^{-2}$) &  (10$^{15}$ cm$^{-2}$) & ($\times$10$^3$) & \\
\hline \noalign{\smallskip}
15.0,7.5  &HC$^{15}$N J=4--3 & 8.8 &  0.4 & 18.1 & 16.3 & 0.04 & 7.8 & 3.5 && 4.8\\
15.0,15.0 &HC$^{15}$N J=4--3 &10.7 &  0.4 & 14.7 & 16.3 & 0.04 & 7.3 & 3.3 && 5.8\\
7.5,0.0 &HC$^{15}$N J=4--3 & 8.3 &  0.9 & 14.6 & 16.4 & 0.09 & 16.0 & 7.2 && 10. \\
7.5,7.5 &HC$^{15}$N J=4--3 & 7.1 &  0.8 & 14.1 & 16.4 & 0.08 & 13.6 & 6.1 && 8.4\\
7.5,15.0 &H$^{13}$CN J=4--3&  7.1 &  1.6 & 15.6 & 16.5 & 0.17 & 31.8 & 2.4 && 4.1\\
0.0,0.0 &HC$^{15}$N J=4--3 &10.4 &  1.1 & 11.7 & 16.4 & 0.11 & 15.1 & 6.8 && 6.2\\
0.0,7.5 &HC$^{15}$N J=4--3 & 9.1 &  0.9 & 12.2 & 16.4 & 0.09 & 13.0 & 5.9 && 5.4\\
0.0,15.0 &H$^{13}$CN J=4--3&  6.5 &  1.7 & 13.7 & 16.5 & 0.18 & 28.7 & 2.2 && 3.6\\
--7.5,0.0 &H$^{13}$CN J=4--3& 9.3 &  2.0 & 15.0 & 16.6 & 0.22 & 38.4 & 3.0 && 3.1\\
--7.5,7.5 &H$^{13}$CN J=4--3& 8.1 &  2.0 & 15.1 & 16.6 & 0.21 & 37.8 & 2.9 && 4.2\\
\noalign{\smallskip} \hline \noalign{\smallskip}
  15.0,7.5 & DCN J=5--4 & 5.0 &  0.1 & 14.0 & 18.3 & 0.01 & 4.5 && 1.3 \\
 15.0,15.0 & DCN J=5--4 &10.1 &  0.2 & 14.2 & 18.3 & 0.02 & 9.1 && 2.8 \\
   7.5,0.0 & DCN J=5--4 & 4.2 &  0.2 &  7.8 & 18.3 & 0.02 & 3.2 && 0.4 \\
   7.5,7.5 & DCN J=5--4 & 5.2 &  0.3 &  9.5 & 18.3 & 0.03 & 8.3 && 1.4 \\
  7.5,15.0 & DCN J=5--4 & 6.8 &  0.3 & 14.5 & 18.3 & 0.03 & 10.0 && 4.1 \\
   0.0,0.0 & DCN J=5--4 & 9.3 &  0.2 & 14.4 & 18.3 & 0.02 & 7.9 && 1.2 \\
   0.0,7.5 & DCN J=5--4 & 7.2 &  0.3 & 13.1 & 18.3 & 0.03 & 10.8 && 1.9 \\
  0.0,15.0 & DCN J=5--4 & 5.8 &  0.2 & 11.3 & 18.3 & 0.02 & 6.8 && 3.1 \\
 --7.5,0.0 & DCN J=5--4 &11.5 &  0.3 &  7.5 & 18.3 & 0.03 & 5.2 && 1.8 \\
 --7.5,7.5 & DCN J=5--4 & 9.4 &  0.2 & 10.4 & 18.3 & 0.02 & 5.8 && 2.0 \\
\noalign{\smallskip} \hline \noalign{\smallskip}
\end{tabular}
\tablefoot{DCN and HCN column densities are calculated with RADEX assuming \tkin=100~K and \nhh= 2$\times$10$^6$~cm$^{-3}$. For the offset positions where HC$^{15}$N was not detected, the data for H$^{13}$CN are shown; for the other positions, where both were detected, data for the HC$^{15}$N J=4--3 transition are listed and we note that the column density determinations from the H$^{13}$CN J=4--3 line agree with these to within $\pm$35\%. \\
\tablefoottext{a}{Offsets from R.A.=19:10:13.4, Dec.=09:06:14} \\
\tablefoottext{b}{based on $^{12}$C/$^{13}$C = 77; $^{14}$N/$^{15}$N = 450 \citep{wilson_rood}.}
}
\end{table*}

Tables~\ref{t:HCOobs} and \ref{t:DCNobs} list fit parameters and resulting column densities, optical depths and excitation temperatures calculated by RADEX for the central pixels. As we expect, optical depths are generally higher for the transitions of the $^{13}$C-substituted molecules than for those containing rarer isotopes.  The excitation temperatures are $\sim$40~K for the H$^{13}$CO$^+$ and HC$^{18}$O$^+$ J=4--3 lines ($\sim$86~K for the H$^{13}$CO$^+$ J=3--2 lines), $\sim$16~K for the H$^{13}$CN and HC$^{15}$N J=4--3 lines, and $\sim$18~K for the DCN J=5--4 lines. We estimate $N$(HCO$^+$) $\sim$ 10$^{14}$--10$^{15}$~cm$^{-2}$, $N$(HCN) $\sim$ a few times 10$^{15}$~cm$^{-2}$, DCN/HCN = 4$\times$10$^{-4}$--3$\times$10$^{-3}$, and HCN/HCO$^+$ = 3--10. Based on 3~$\sigma$ r.m.s.\ noise levels in the DCO$^+$ J=5--4 spectra, we calculate upper limits on the DCO$^+$/HCO$^+$ ratios towards the centre of W49A to be 3--4$\times$10$^{-4}$ for \txc\ = 20--40~K.  

We have also estimated the HNC column density towards the centre of W49A, from the J=4--3 transition. For \tkin=100~K and \nhh=2$\times$10$^6$\,\ccm, we find $N$(HNC) = 10$^{14}$~cm$^{-2}$ so the HCN/HNC ratio is $\sim$50. The result that $N$(HCN)/$N$(HCO$^+$) ratios are $>$1 is contrary that found by naively taking line ratios of the main $J$=4--3 lines (Fig~\ref{f:ratios}; Table~\ref{t:ratios}), which shows the importance of taking excitation effects into account when estimating column densities or abundance ratios.

\subsubsection{The South-West clump}\label{s:SWclump}

\begin{figure}
\rotatebox{270}{\resizebox{!}{4.5cm}{\includegraphics*[2cm,0.5cm][21cm,26cm]{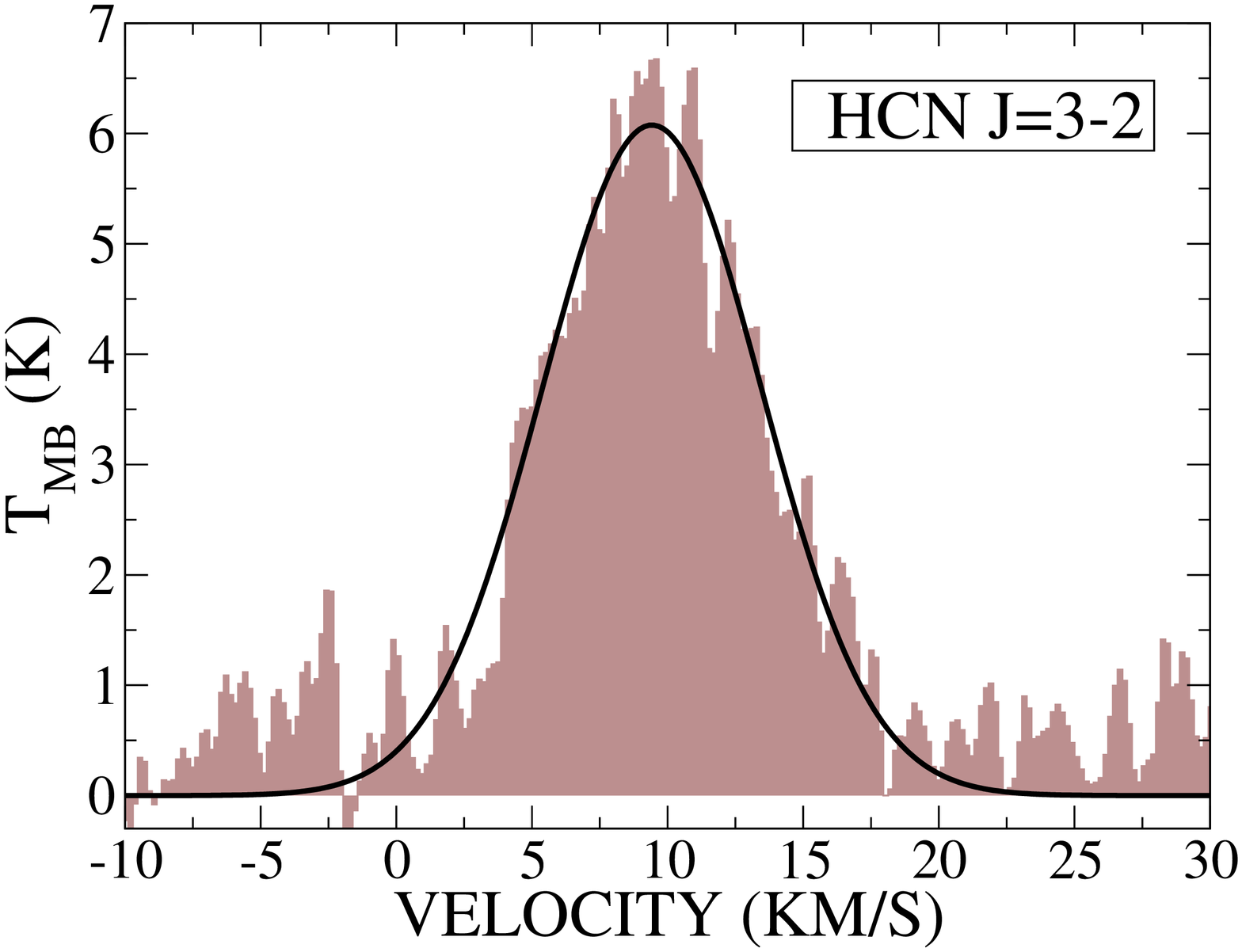}}}\rotatebox{270}{\resizebox{!}{4.5cm}{\includegraphics*[2cm,0.5cm][21cm,26cm]{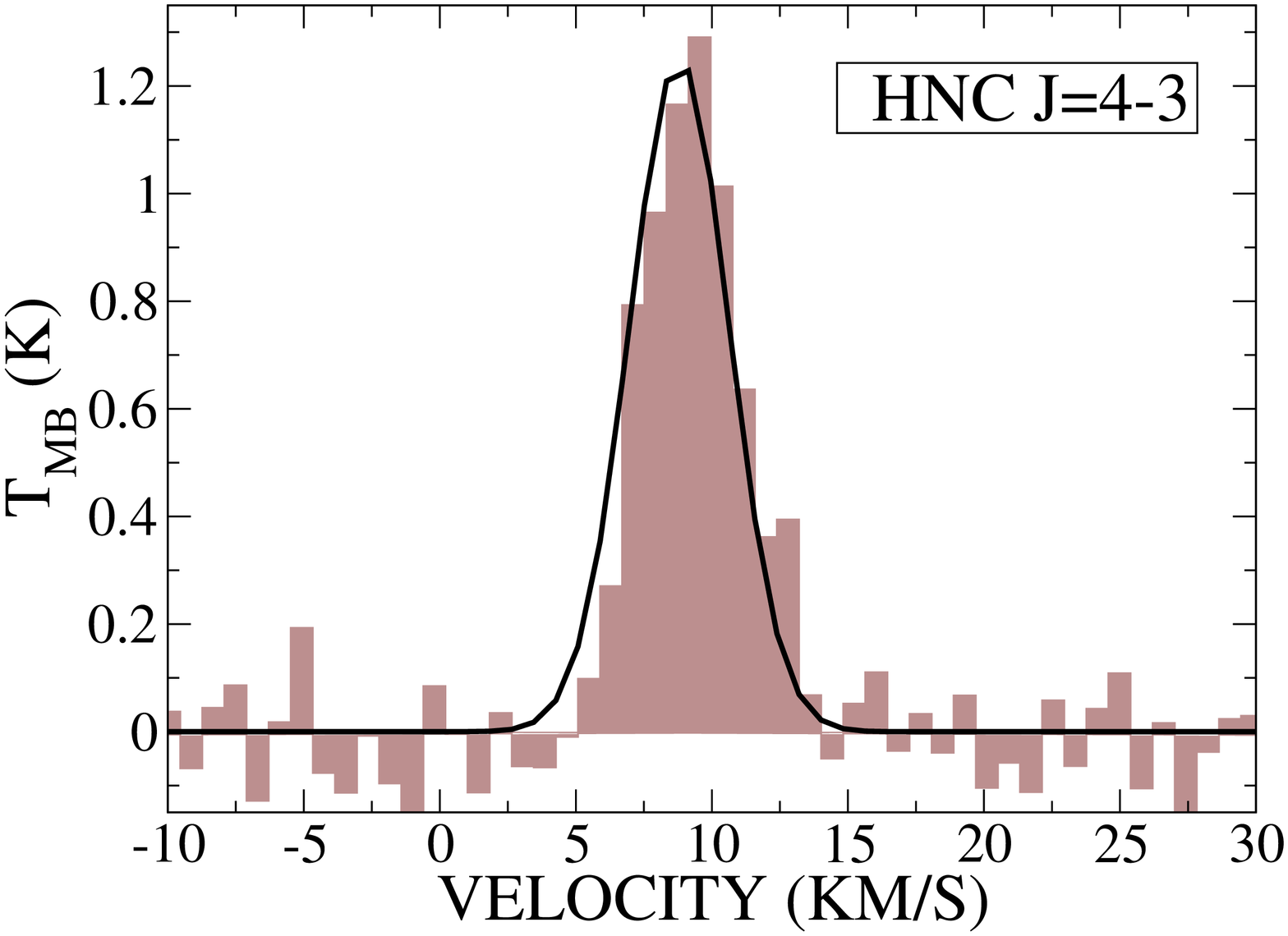}}}

\rotatebox{270}{\resizebox{!}{4.5cm}{\includegraphics*[2cm,0.5cm][21cm,26cm]{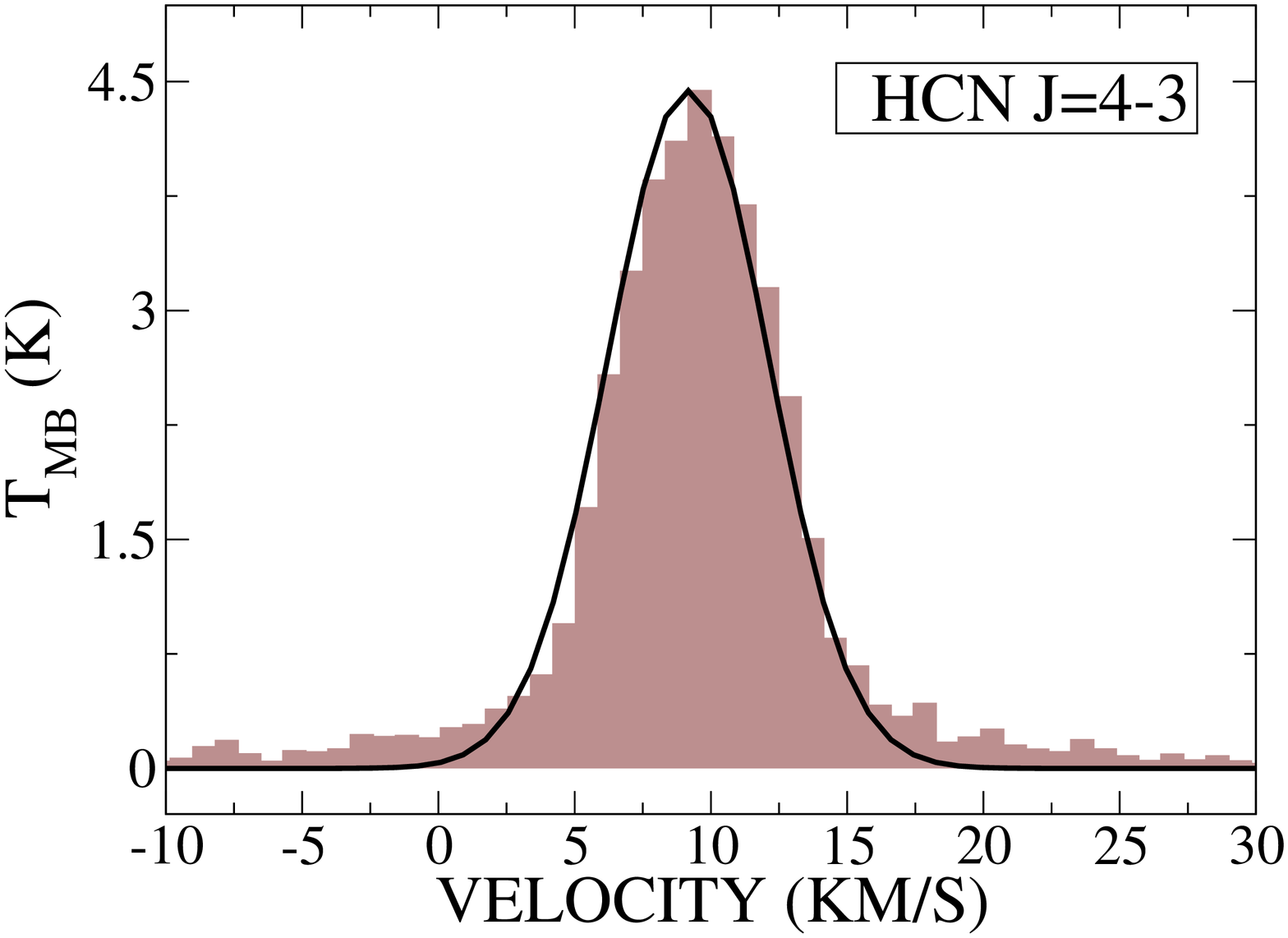}}}\rotatebox{270}{\resizebox{!}{4.5cm}{\includegraphics*[2cm,0.5cm][21cm,26cm]{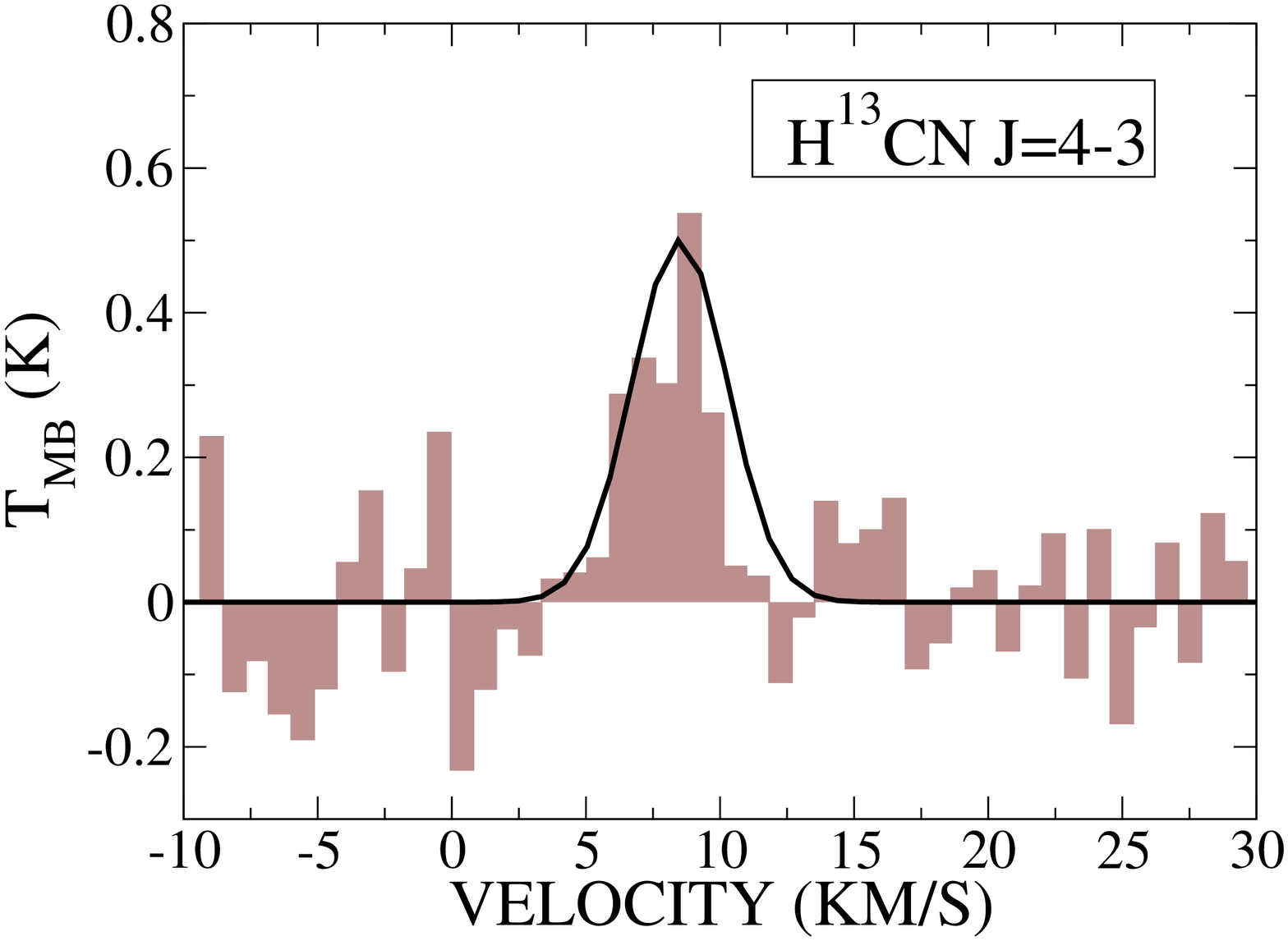}}}

\rotatebox{270}{\resizebox{!}{4.5cm}{\includegraphics*[2cm,0.5cm][21cm,26cm]{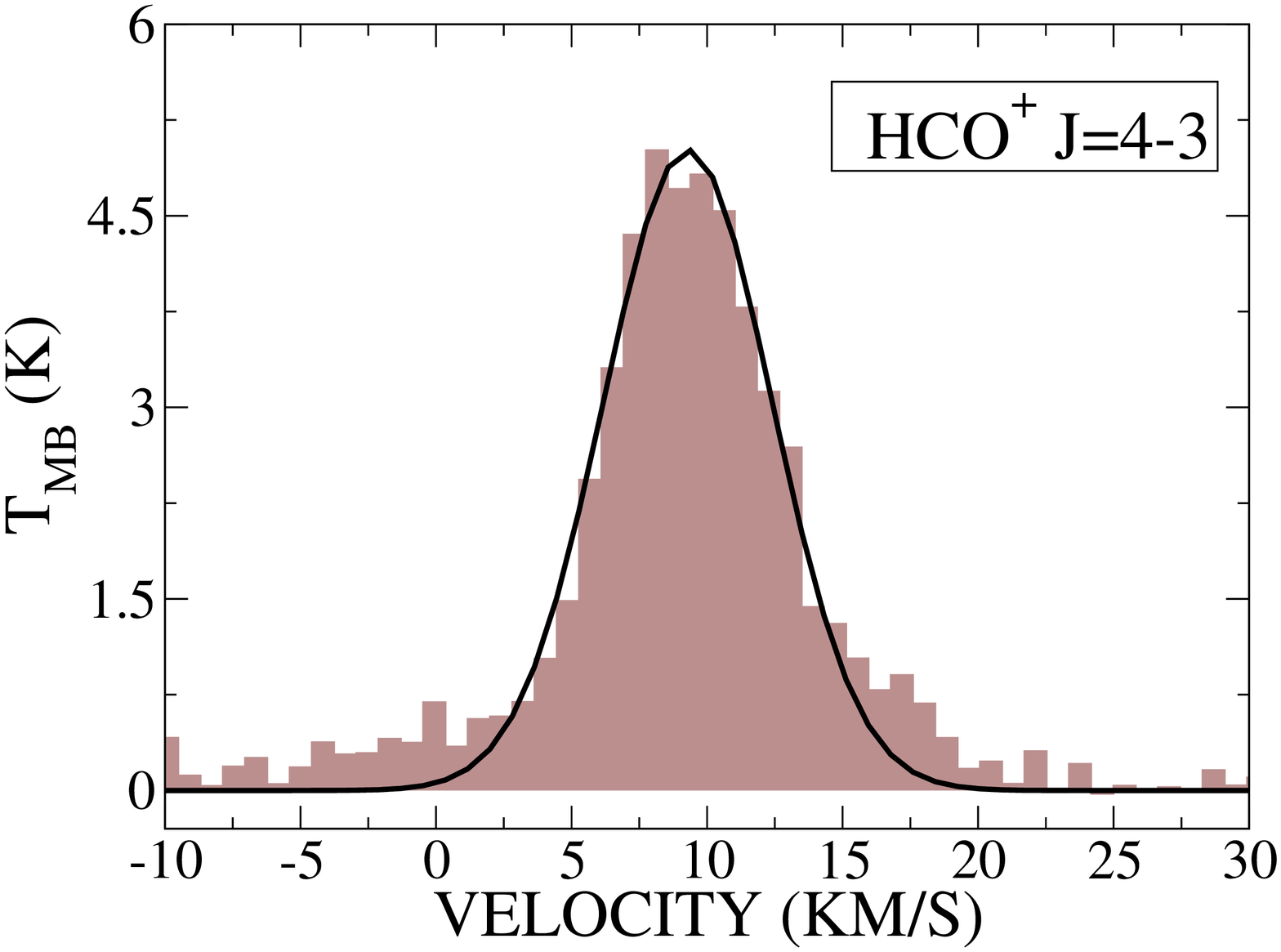}}}\rotatebox{270}{\resizebox{!}{4.5cm}{\includegraphics*[2cm,0.5cm][21cm,26cm]{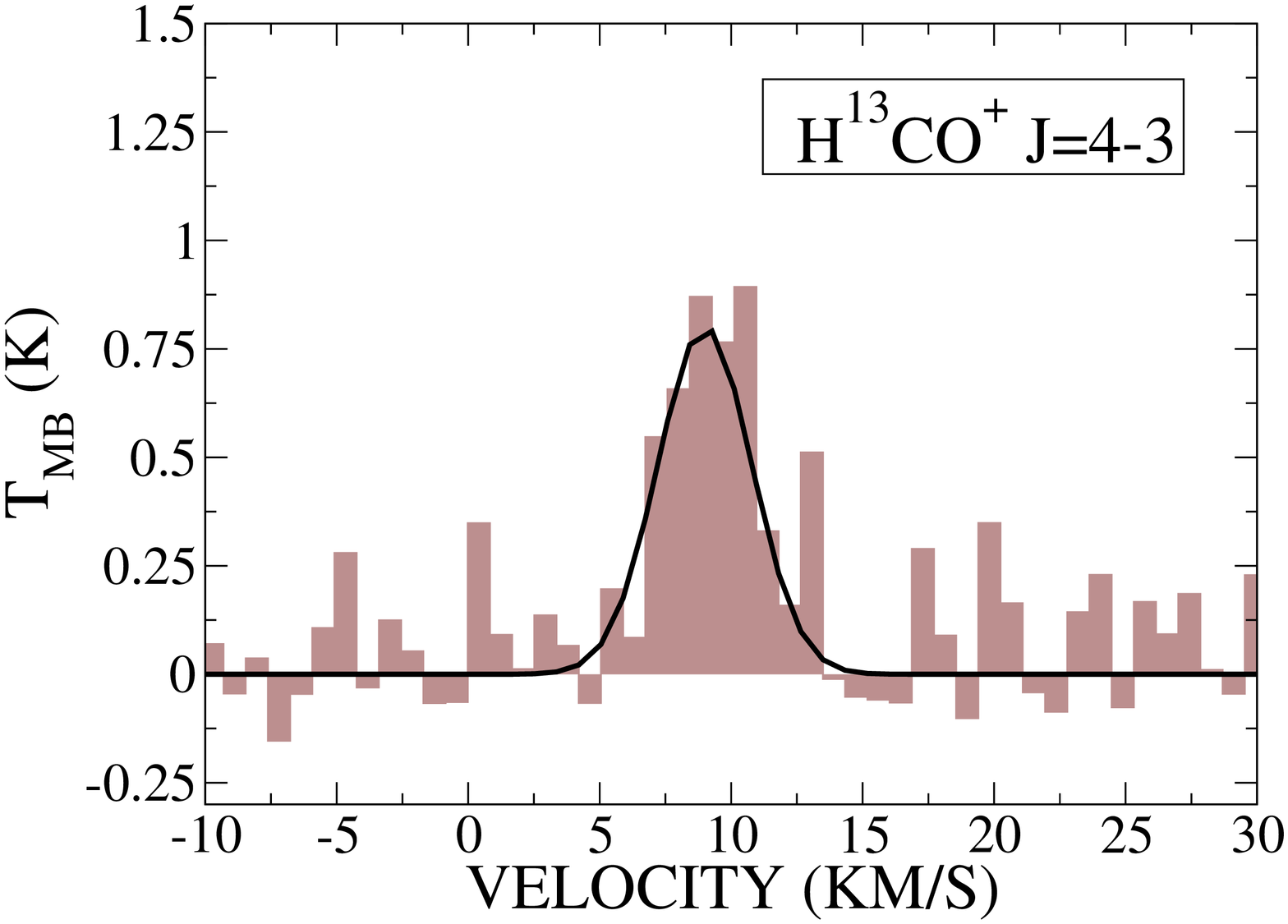}}}
\caption{Observed line profiles (shaded grey) and Gaussian fits (black lines) for the transitions observed towards the ``south-west clump'' position: R.A.=19:10:10.6; Dec.=09:05:18}\label{f:W49clump}
\end{figure}

As Figs.~\ref{f:W49_map_centre_HCOiso} and \ref{f:W49_map_centre_HCNiso} show, the strength of the isotope-substituted transitions falls rapidly away from the source centre. Only the $^{13}$C-substituted J=4--3 lines are observed towards the SW-clump. Fig.~\ref{f:W49clump} shows the detected lines along with Gaussian fit profiles. The lines peak close to the expected source velocity and are well-fitted by single Gaussian profiles.  Again, we have estimated column densities using RADEX, this time varying the physical parameters over the ranges \tkin\ =20--100~K and \nhh\ = 5$\times$10$^3$--10$^6$~cm$^{-3}$. We find that, unless \nhh\ $>$ 10$^5$~cm$^{-3}$, the HCN and HCO$^+$ column densities towards the clump are at least 10~times higher than towards the source centre, due to the high critical densities of these transitions. As this would imply fractional abundances at least 2 orders of magnitude higher at the clump position, it seems more likely that this gas also has a relatively high density of 10$^5$--10$^6$~cm$^{-3}$. 

We have tried to constrain the kinetic temperature by looking at the upper limits on the DCN/HCN and DCO$^+$/HCO$^+$ ratios and at the HCN/HNC ratio. The upper level energies of the DCN and DCO$^+$ 5--4 transitions are $\sim$50~K, though, so we would not expect these lines to be very strong at lower temperatures.  For \tkin = 20~K, the upper limit on the DCN/HCN ratio from the observations is 0.01 whereas steady-state chemical models predict DCN/HCN $\sim$ 0.02 at 20~K (see Section~\ref{s:chem}), slight evidence that the kinetic temperature in the SW clump is higher than 20~K. The HCN and HNC column densities predicted by RADEX 
\newnewold{are fairly insensitive to the adopted physical conditions, and}
give a HCN/HNC ratio $\sim$7, which suggests that this region has a lower temperature than the source centre. 

\begin{table*}
\caption{Estimated column densities towards the South-West clump.}
\label{t:clump}
\begin{tabular}{lcccccccl}
\hline \hline \noalign{\smallskip}
Line            & \vlsr & \tmb\ & FWHM & r.m.s. & \txc\ & $\tau$ & $N$  \\
& (km~s$^{-1}$) & (K)& (km~s$^{-1}$)& (K)& (K) & & (10$^{13}$\,cm$^{-2}$)   \\
\noalign{\smallskip} \hline \noalign{\smallskip}
HCN J=3--2            & 9.4 & 6.1  & 9.5  & 0.52 & 11  & \newnew{4.9} & 33  \\ 
HCN J=4--3            & 9.2 & 4.4  & 7.0  & 0.06 & 11  & \newnew{5.7} & 45  \\ 
H$^{13}$CN J=4--3     & 8.5 & 0.5  & 4.2  & 0.07 & 8.3 & \newnew{0.2} & 2.1  \\ 
DCN J=5--4            & --- & ---  & ---  & 0.04 & 10  & ---  & \newnew{$<$0.5} & \\ 
HNC J=4--3            & 8.8 & 1.25 & 4.3  & 0.05 & 8.3 & \newnew{0.7} & 6.6  & \\ 
HCO$^+$ J=4--3        & 9.3 & 5.0  & 7.3  & 0.12 & 13  & \newnew{1.4} & 6.6  & \\ 
H$^{13}$CO$^+$ J=4--3 & 9.0 & 0.8  & 4.2  & 0.08 & 12  & \newnew{0.2} &\newnew{0.5} & \\ 
DCO$^+$ J=5--4        &---  & ---  & ---  & 0.07 & 10  & ---  & \newnew{$<$0.5} & \\ 
\noalign{\smallskip} \hline
\end{tabular}
\tablefoot{Column densities are calculated with RADEX assuming \tkin=40~K and \nhh= 5$\times$10$^5$~cm$^{-3}$ towards the South-West clump at R.A.=19:10:10.6; Dec.=09:05:18. We have based the column density ratios on the main lines of HCN and HCO$^+$ since the $^{13}$C-substituted spectra are rather noisy.}
\end{table*}

Table~\ref{t:clump}, therefore, lists the line parameters and calculated column densities assuming \tkin = 40~K and \nhh = 5$\times$10$^5$~cm$^{-3}$. This temperature is consistent with that determined by \citet{jaffe} for the outer parts of the cloud (see \S~\ref{s:intro}), although the density is ten times higher than estimated by \citet{welch87}.  We note that the HCN and HCO$^+$ column densities based on the $^{13}$C substituted isotopologues, assuming $^{12}$C/$^{13}$C = 77, are 4--6 times higher than those based on observations of the main lines. The upper limits on the column densities of the deuterated species do not put very strong constraints on the deuterium fractionation: DCN/HCN $<$0.01 and DCO$^+$/HCO$^+$ $<$ 0.07. Based on the main isotopic J=4--3 transitions, the HCN/HCO$^+$ column density ratio towards the SW clump is $\sim$7 ($\sim$4 based on the $^{13}$C lines), similar to the values calculated for the source centre.

\begin{figure}
\rotatebox{270}{\resizebox{!}{4.5cm}{\includegraphics*[2cm,0.5cm][21cm,26cm]{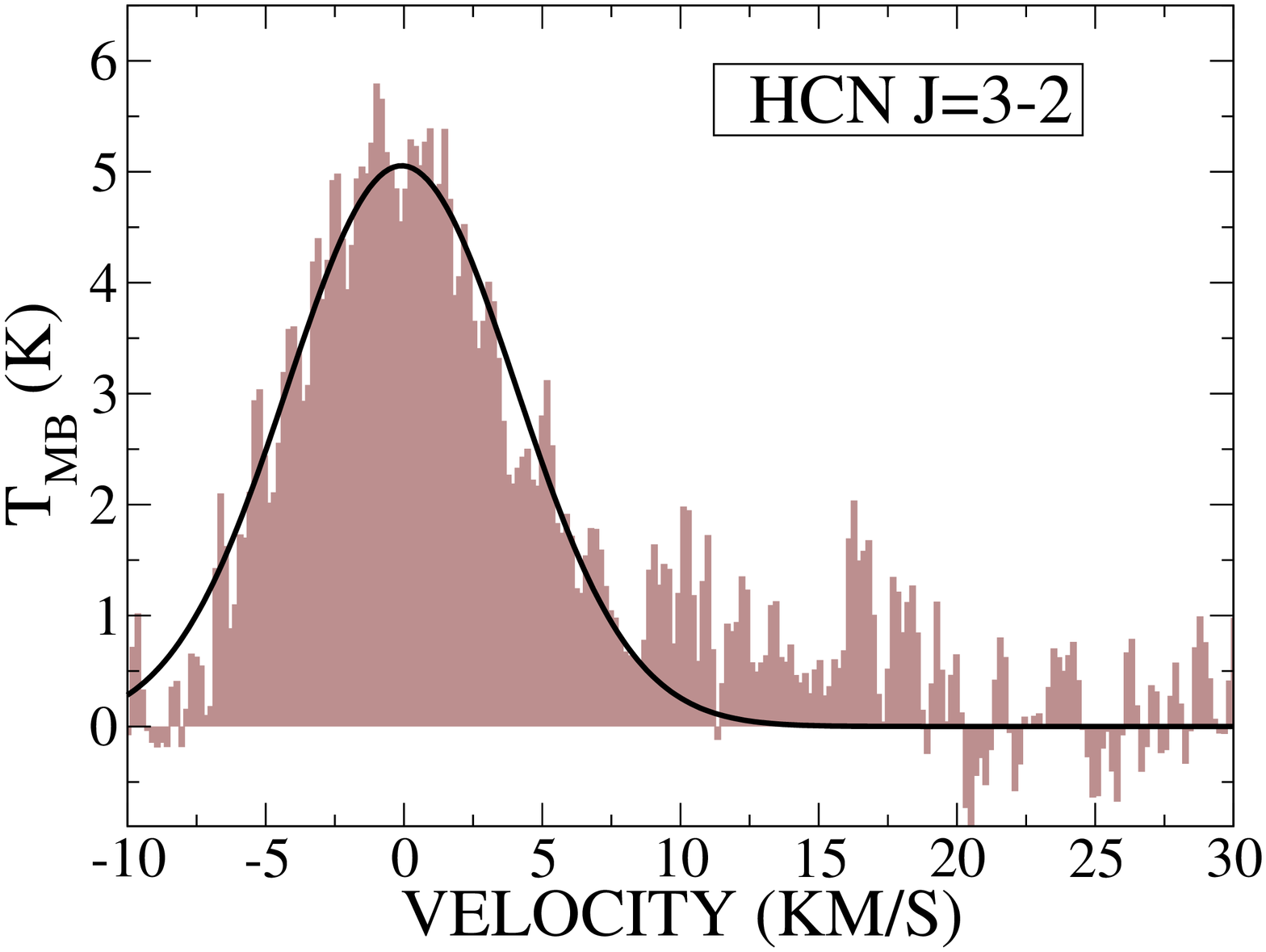}}}\rotatebox{270}{\resizebox{!}{4.5cm}{\includegraphics*[2cm,0.5cm][21cm,26cm]{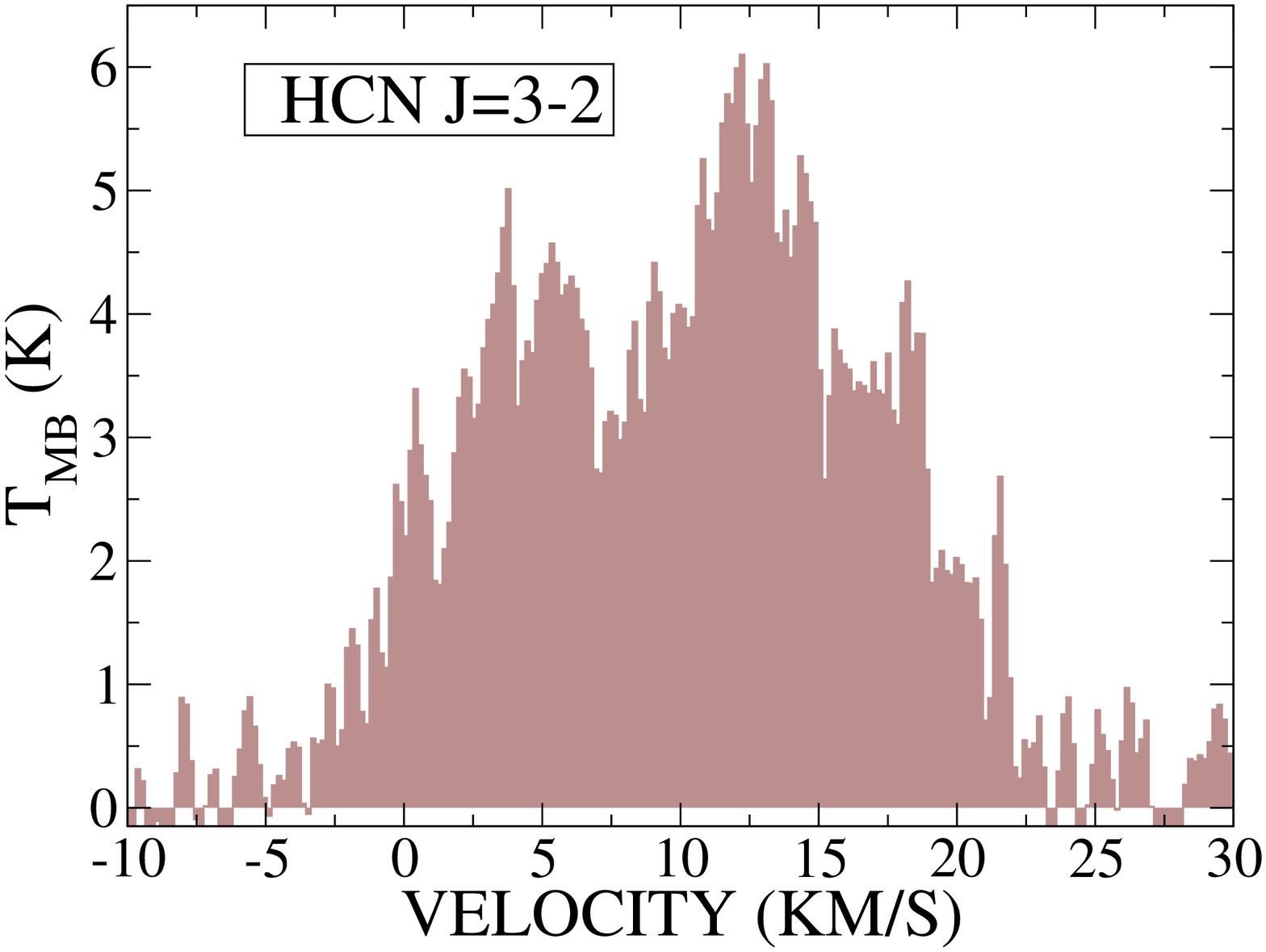}}}

\rotatebox{270}{\resizebox{!}{4.5cm}{\includegraphics*[2cm,0.5cm][21cm,26cm]{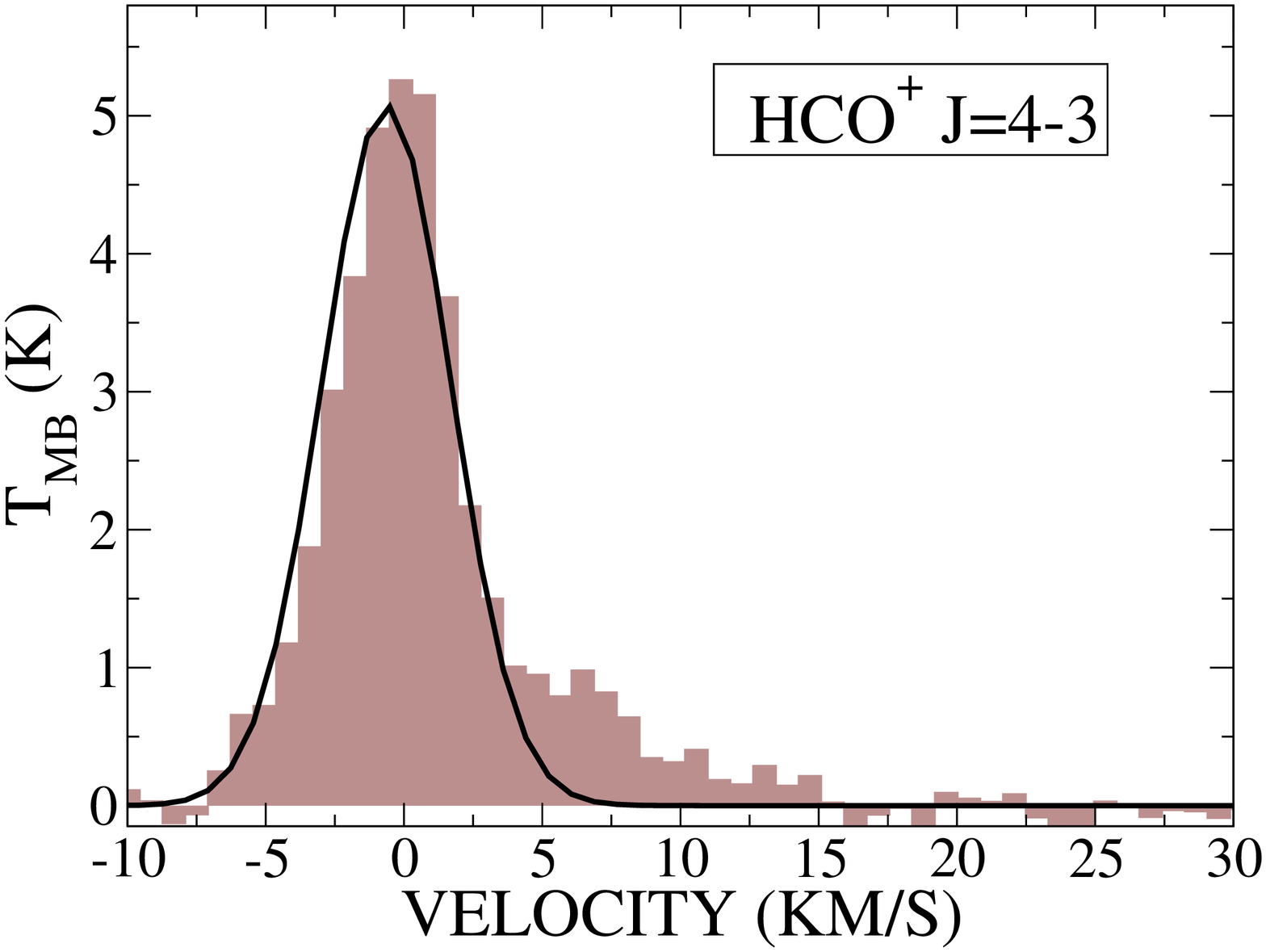}}}\rotatebox{270}{\resizebox{!}{4.5cm}{\includegraphics*[2cm,0.5cm][21cm,26cm]{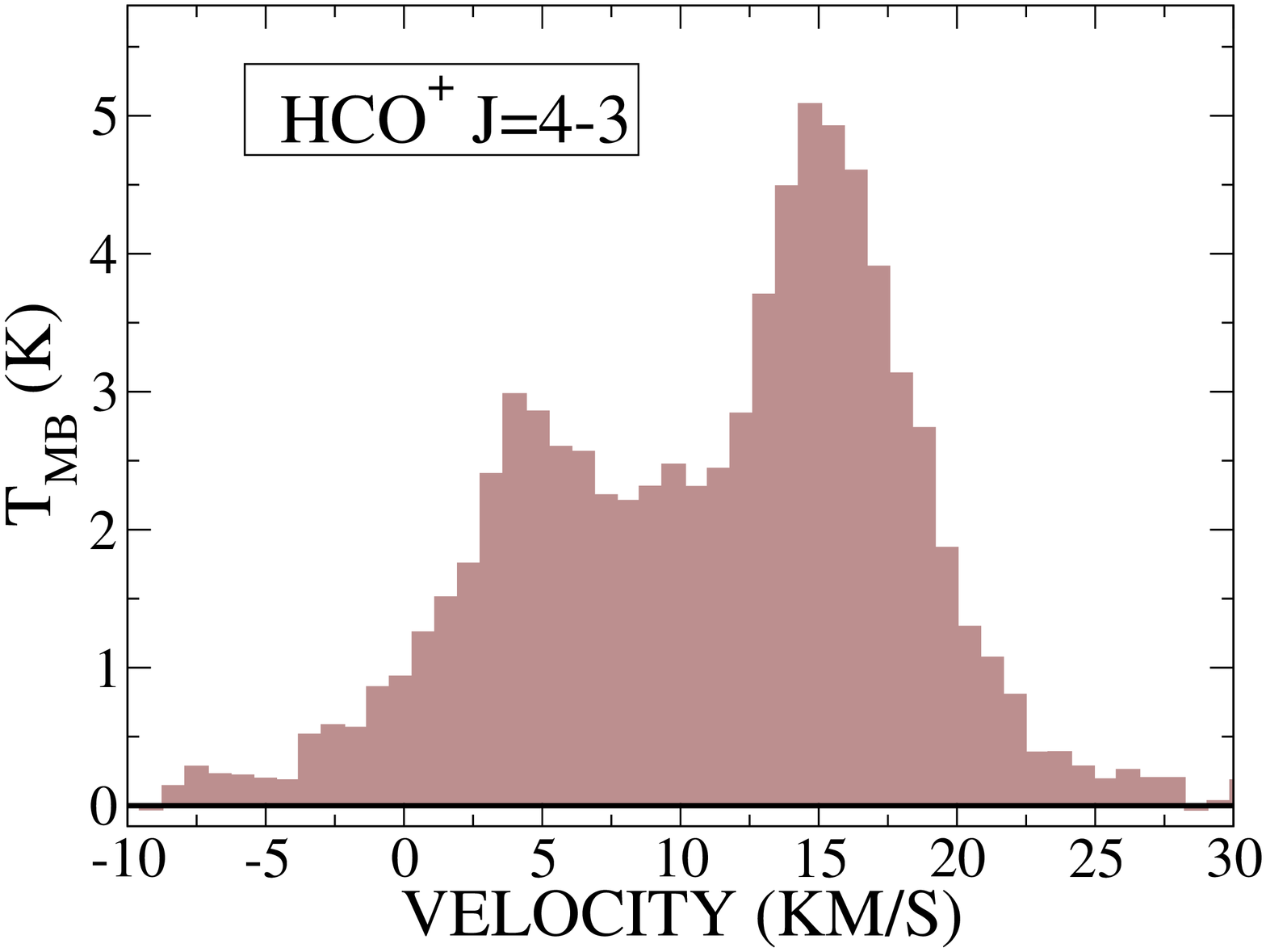}}}

\rotatebox{270}{\resizebox{!}{4.5cm}{\includegraphics*[2cm,0.5cm][21cm,26cm]{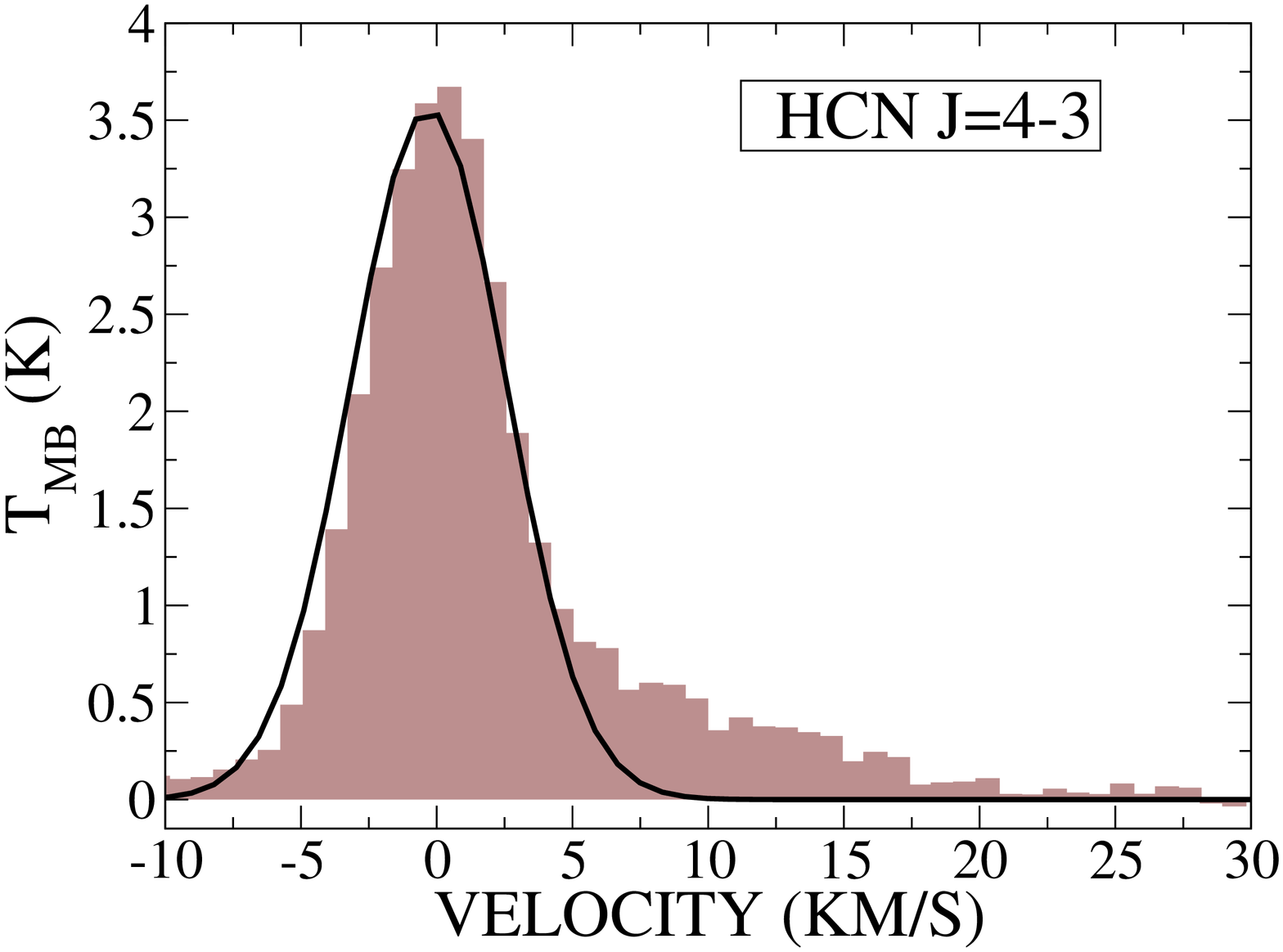}}}\rotatebox{270}{\resizebox{!}{4.5cm}{\includegraphics*[2cm,0.5cm][21cm,26cm]{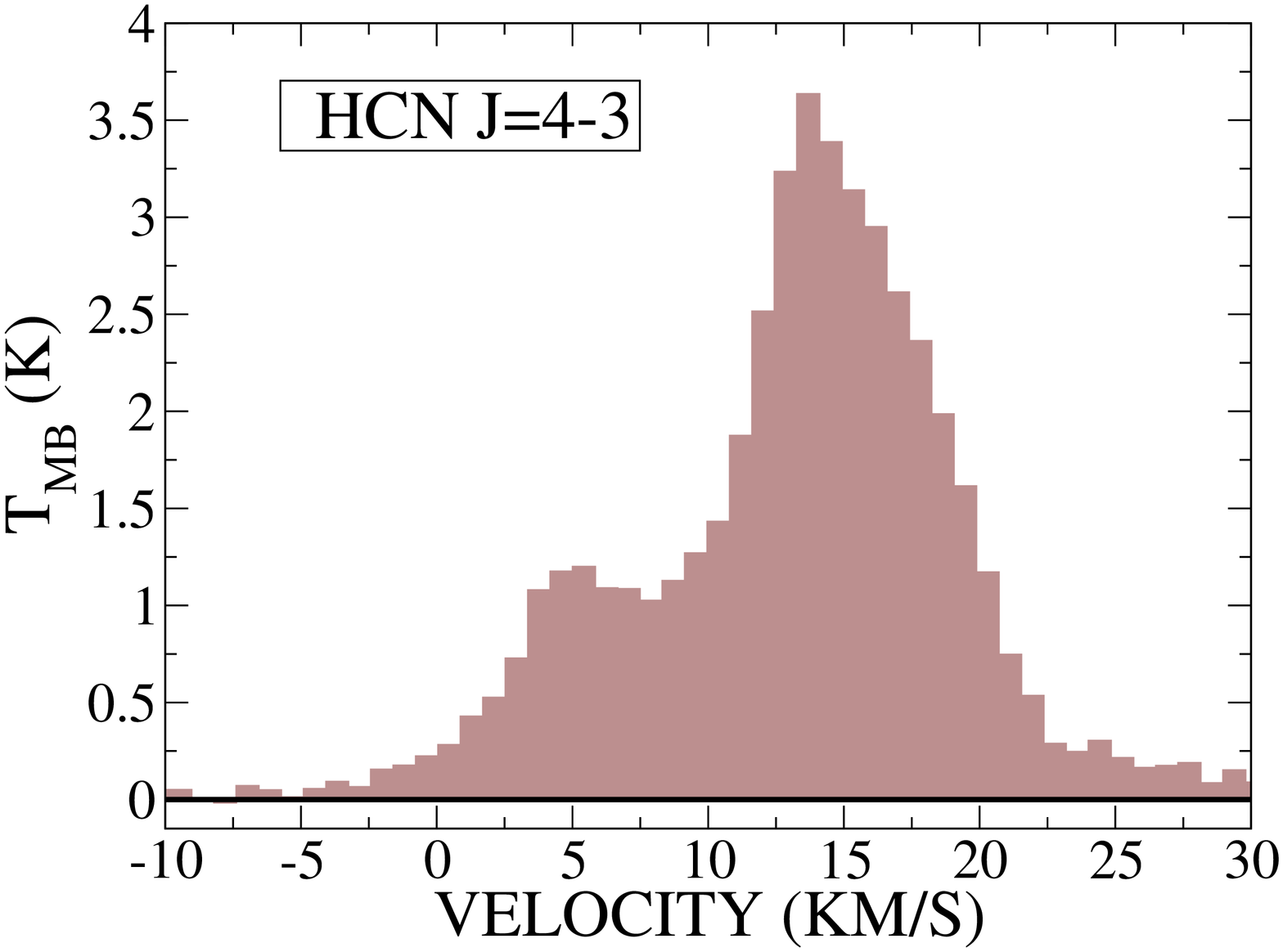}}}

\rotatebox{270}{\resizebox{!}{4.5cm}{\includegraphics*[2cm,0.5cm][21cm,26cm]{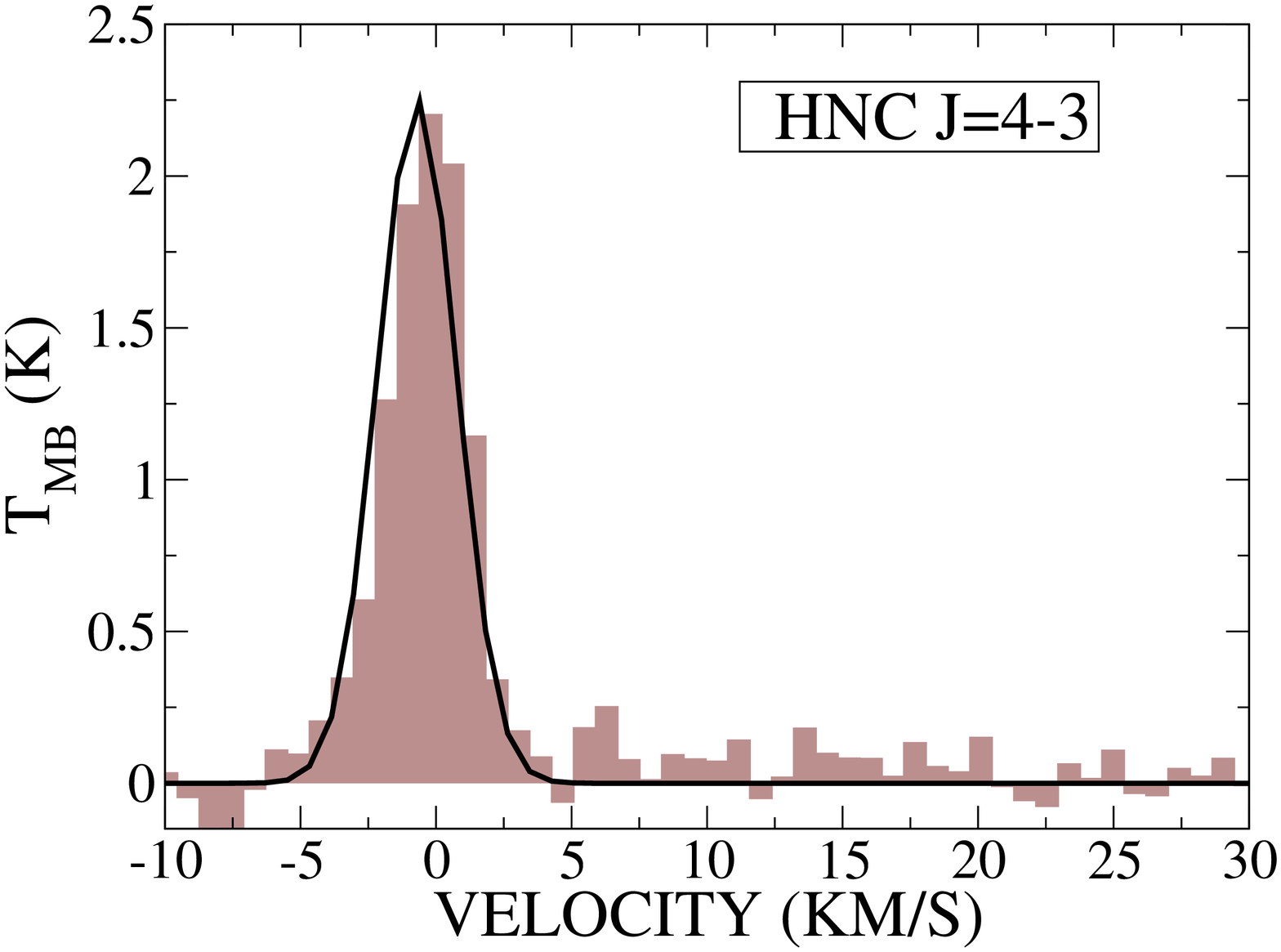}}}\rotatebox{270}{\resizebox{!}{4.5cm}{\includegraphics*[2cm,0.5cm][21cm,26cm]{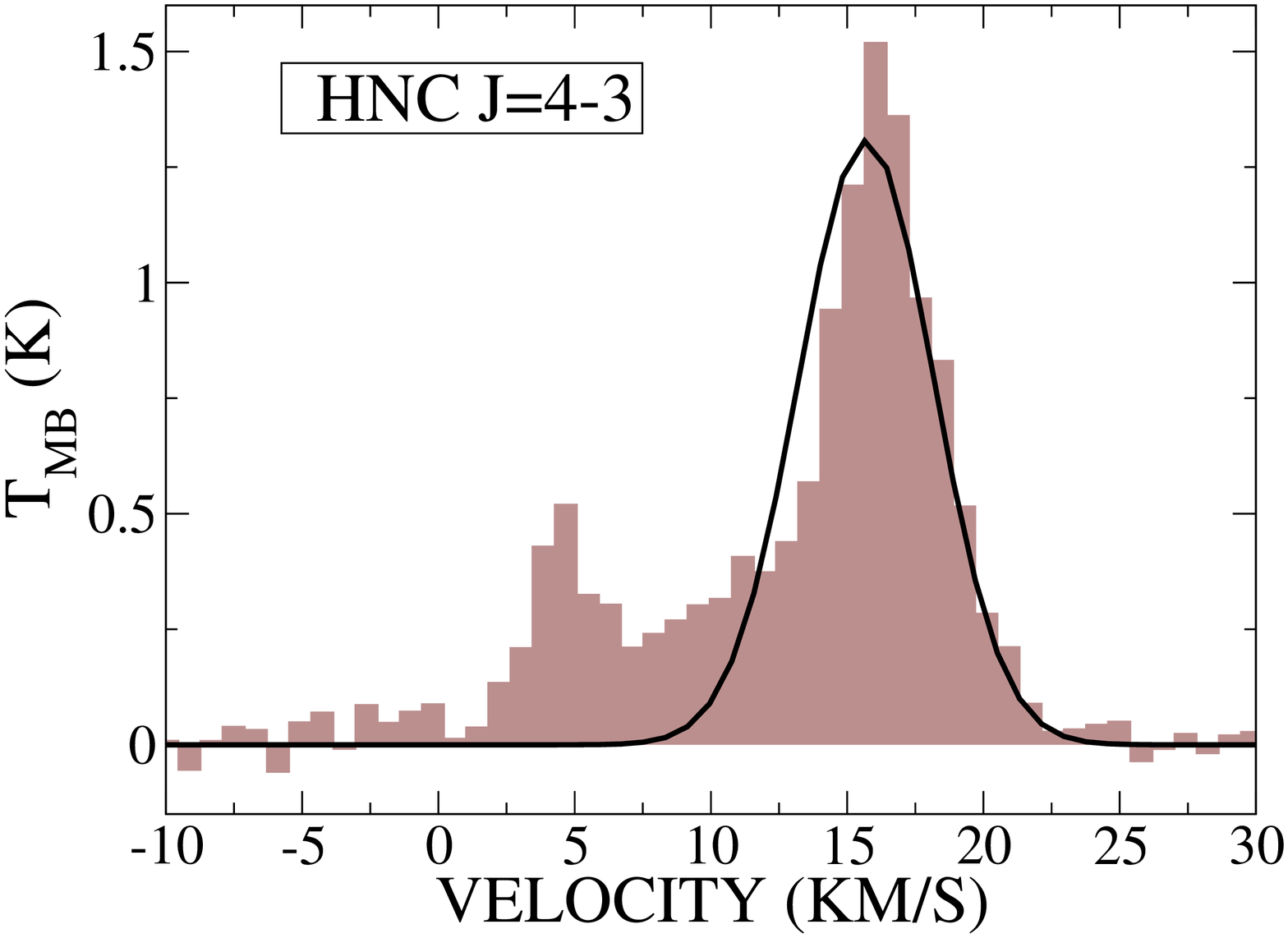}}}

\rotatebox{270}{\resizebox{!}{4.5cm}{\includegraphics*[2cm,0.5cm][21cm,26cm]{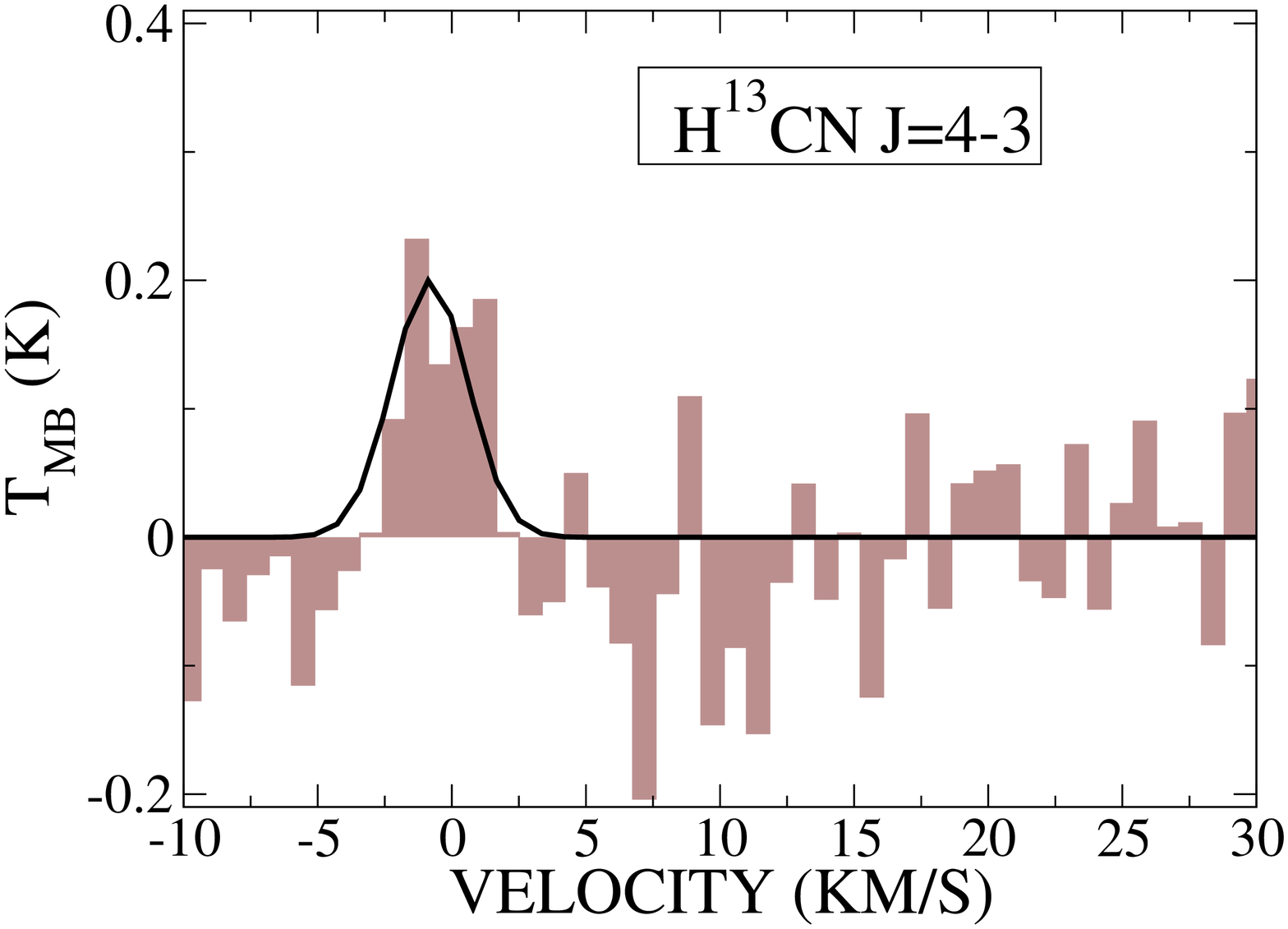}}}\rotatebox{270}{\resizebox{!}{4.5cm}{\includegraphics*[2cm,0.5cm][21cm,26cm]{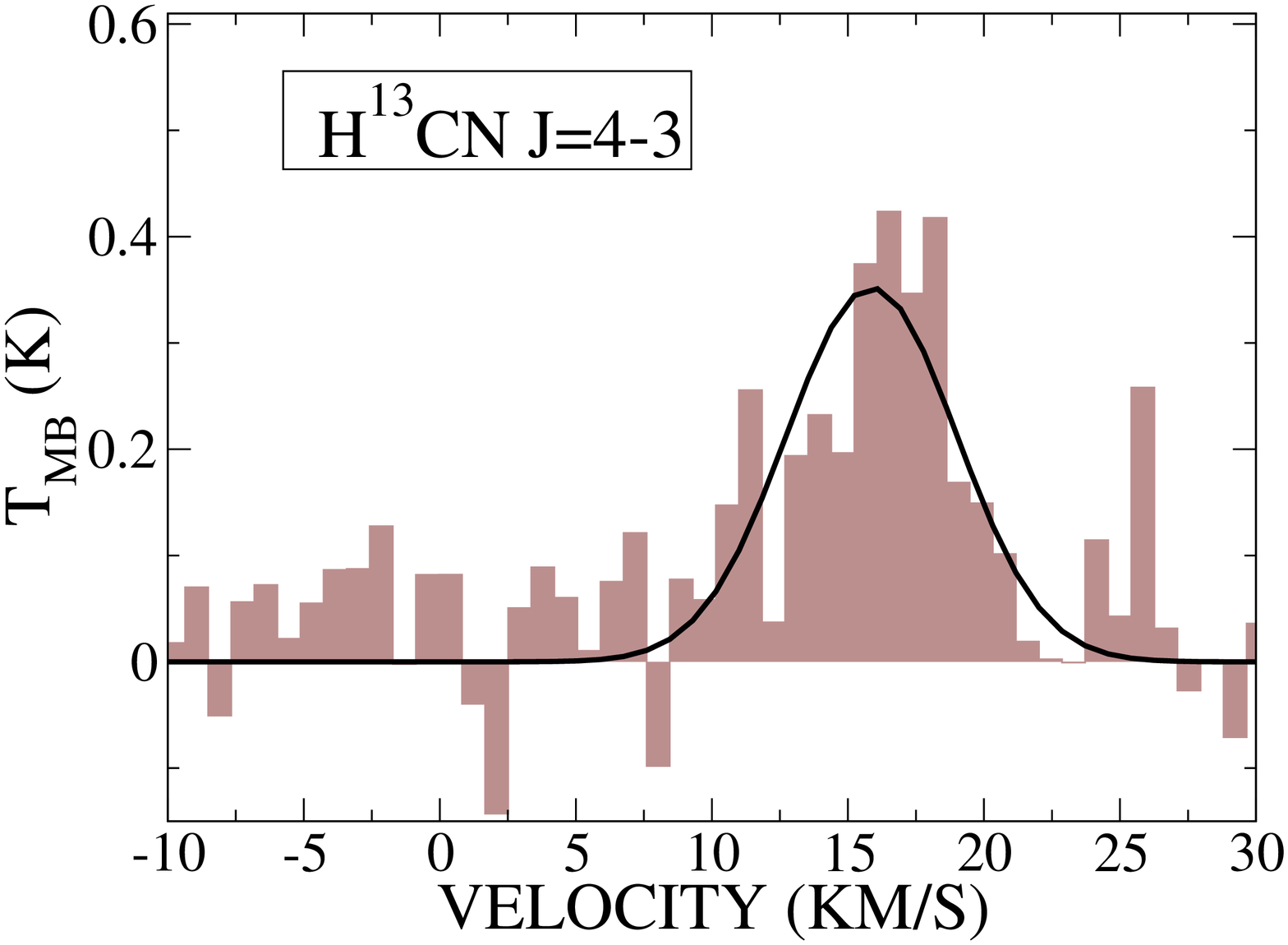}}}

\rotatebox{270}{\resizebox{!}{4.5cm}{\includegraphics*[2cm,0.5cm][21cm,26cm]{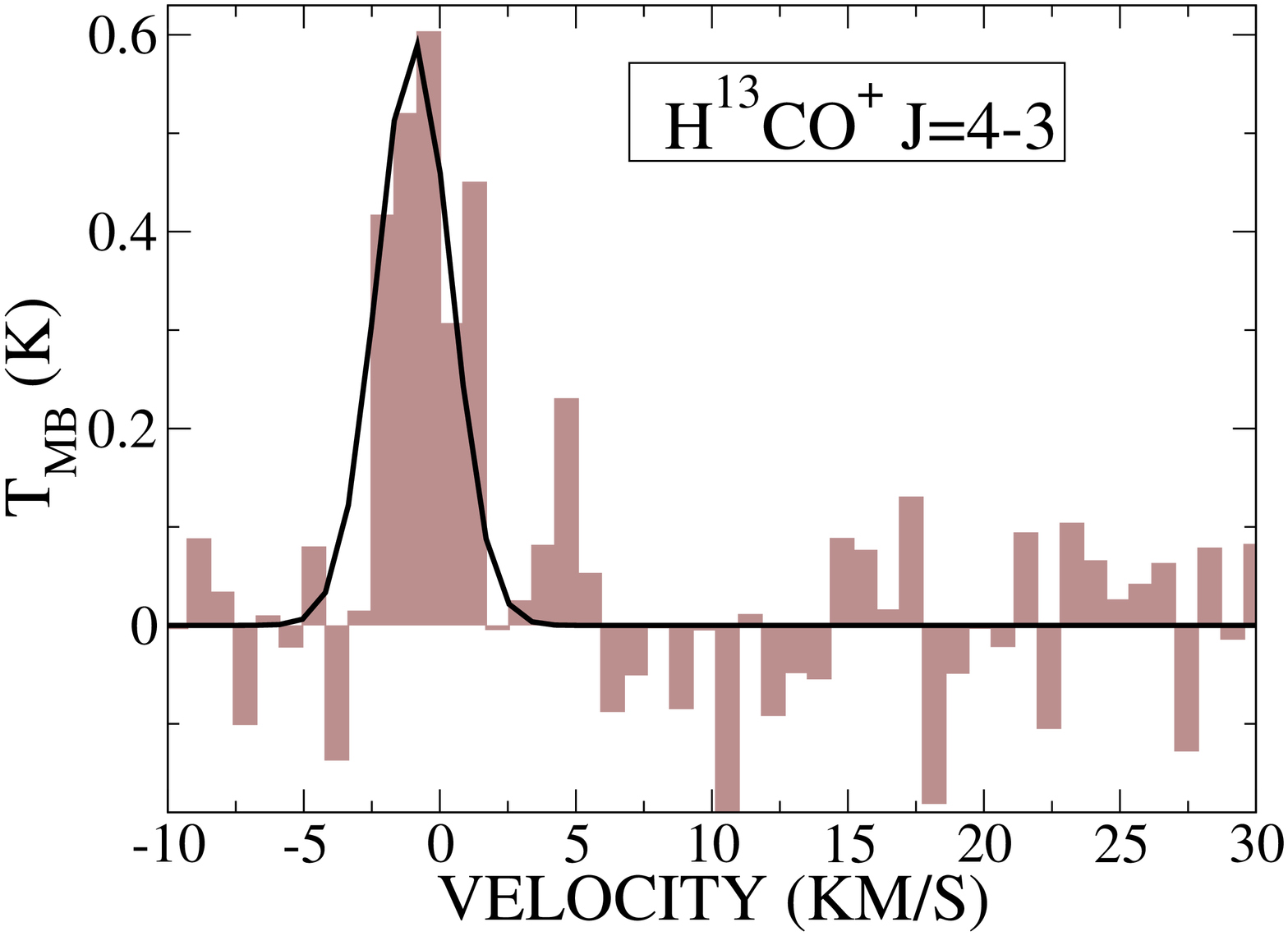}}}\rotatebox{270}{\resizebox{!}{4.5cm}{\includegraphics*[2cm,0.5cm][21cm,26cm]{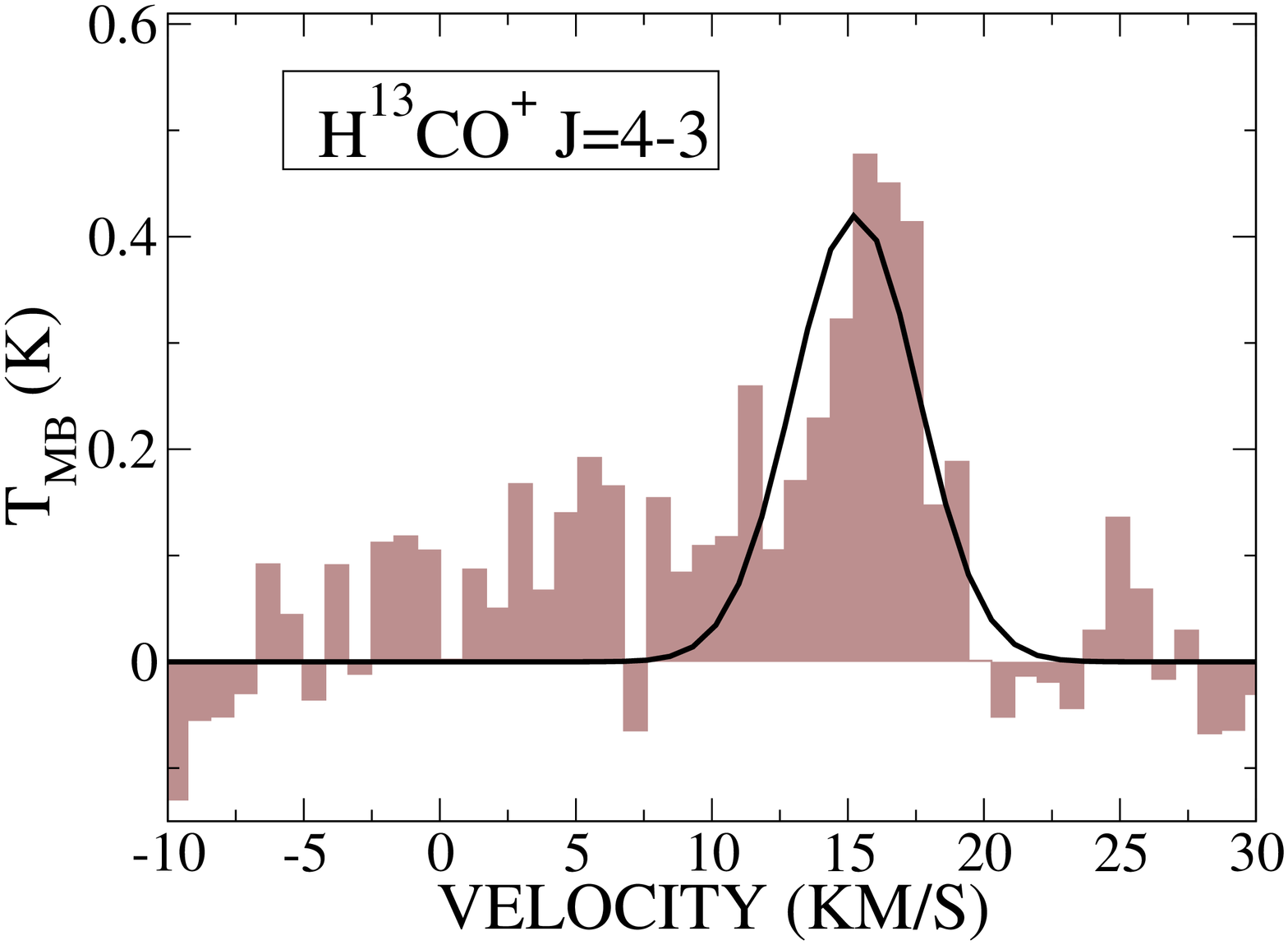}}}

\rotatebox{270}{\resizebox{!}{4.5cm}{\includegraphics*[2cm,0.5cm][21cm,26cm]{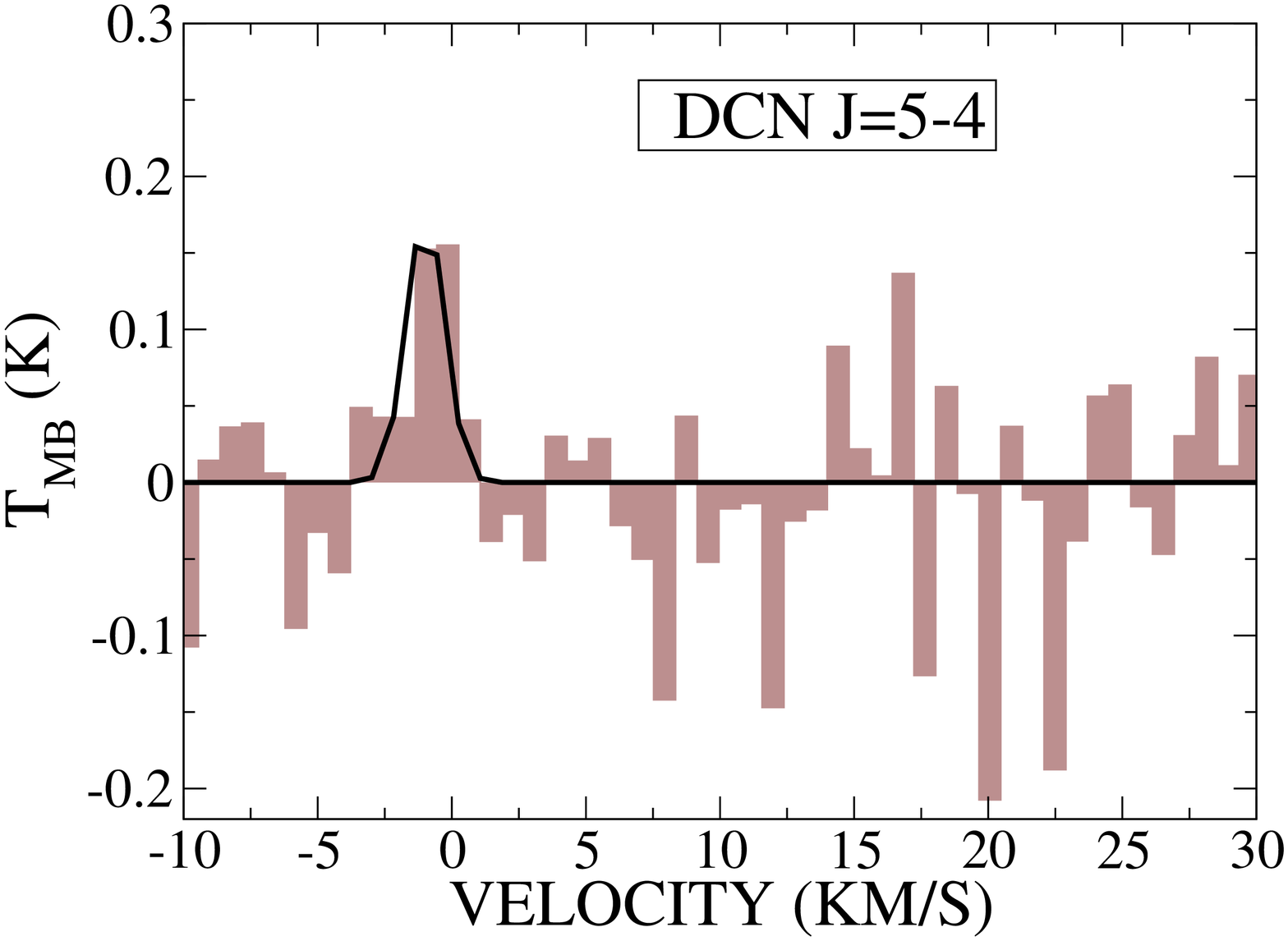}}}\rotatebox{270}{\resizebox{!}{4.5cm}{\includegraphics*[2cm,0.5cm][21cm,26cm]{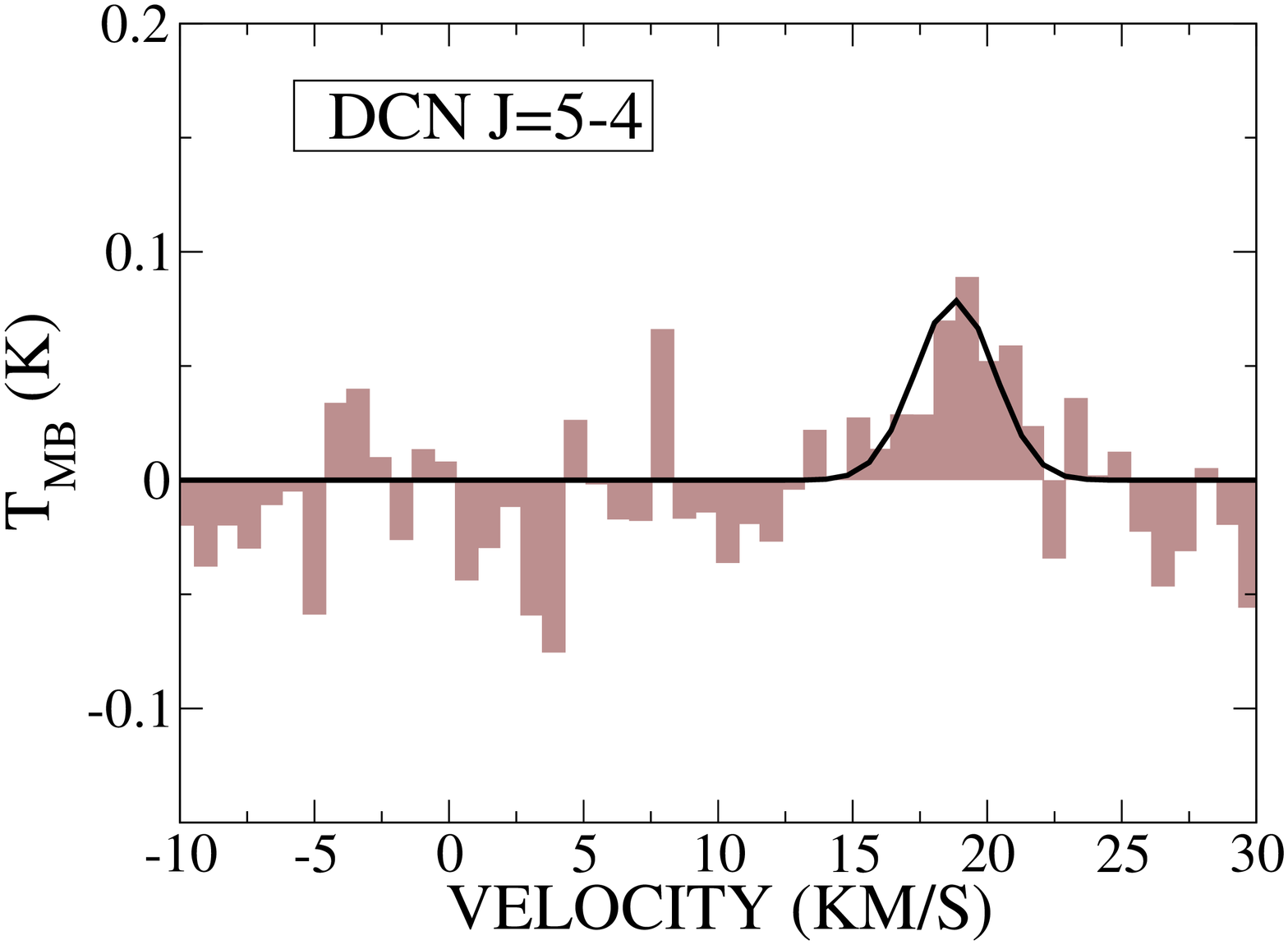}}}
\caption{Observed line profiles (shaded grey) and Gaussian fits (black lines) for the transitions observed towards the ``East tail'' (left) and the ``North clump'' (right) positions (R.A.=19:10:16.6; Dec.=09:05:48 and R.A.=19:10:13.6; Dec=09:06:48, respectively)}\label{f:W49tail}
\end{figure}

\subsubsection{The Eastern tail}\label{s:Etail}

Figure~\ref{f:W49tail} (left column) shows the observed line profiles towards the E-tail. Again, the lines are well-fitted by single Gaussian profiles, but the peak velocity of the lines are close to 0~km~s$^{-1}$, blue-shifted from the central source velocity. As discussed for the SW-clump above, we consider a range of physical conditions and find that column densities based on the HCN J=4--3 compared with the J=3--2 transition and the calculated HCN/HNC ratio suggest that this gas is also at a lower temperature than the source centre.

\begin{table*}
\caption{Estimated column densities towards the Eastern tail.}
\label{t:Etail}
\begin{tabular}{lcccccccl}
\hline \hline \noalign{\smallskip}
Line & \vlsr & \tmb\ & FWHM & r.m.s. & \txc & $\tau$ & $N$  \\
& (km~s$^{-1}$) & (K)& (km~s$^{-1}$)& (K)& (K) & & (10$^{13}$\,cm$^{-2}$)   \\
\noalign{\smallskip} \hline \noalign{\smallskip}
HCN J=3--2         & -0.1 & 5.05 & 9.7  & 0.52 &  10 & 3.85 & 25.8 \\ 
HCN J=4--3         & -0.3 & 3.55 & 6.7  & 0.04 &  9.5 & 4.01 & 33.0 \\ 
H$^{13}$CN J=4--3  & -0.8 & 0.20 & 3.4  & 0.08 &  8.5  & 0.07 & 0.7 \\
DCN J=5--4\tablefootmark{a} & -1.0 & 0.18 & 1.7  & 0.07 &  12 & 0.04 & 1.2 & \\     
HNC J=4--3         & -0.7 &  2.25 & 3.4 & 0.08 & 8.6 & 1.7 & 9.9& \\ 
HCO$^+$ J=4--3     & -0.6 & 5.07 & 5.5  & 0.16 & 13 & 1.4 & 5.0 & \\ 
H$^{13}$CO$^+$ J=4--3 & -0.9 & 0.59 & 3.2 &0.10 & 12 & 0.1 & 0.3 & \\ 
DCO$^+$ J=5--4      &--- & --- & ---     & 0.06 &  $\geq$10. & --- & $<$0.2 & \\ 
\noalign{\smallskip} \hline \noalign{\smallskip}
\end{tabular}
\tablefoot{Column densities are calculated with RADEX assuming \tkin = 40~K and \nhh = 5$\times$10$^5$~cm$^{-3}$ towards the Eastern tail at R.A.=19:10:16.6; Dec.=09:05:48. We show column density ratios based on both the main lines of HCN and HCO$^+$ and the $^{13}$C-substituted lines, since there is some excess emission at higher velocities in the main HCN and HCO$^+$ lines that has not been fitted (see Fig.~\ref{f:W49tail}). \\
\tablefoottext{a}{Tentative detection based on the r.m.s.\ noise in the spectrum, $N$(DCN)$<$4$\times$10$^{12}$~cm$^{-2}$ for \txc$\geq$ 10~K and DCN/HCN $<$0.002.}
}
\end{table*}

We therefore use the same physical conditions for the E-tail as for the SW-clump: \tkin = 40~K and \nhh = 5$\times$10$^5$~cm$^{-3}$. Table~\ref{t:Etail} lists fit parameters and resulting column densities. The HCN/HCO$^+$ and HCN/HNC ratios are very similar to those found for the SW-clump. The upper limit on the DCO$^+$/HCO$^+$ ratio is more stringent and constrains \tkin\ to $>$25~K. The DCN/HCN ratio may be 0.02--0.04, but the DCN J=5--4 detection is uncertain (see Fig.~\ref{f:W49tail}) as the line only covers two spectral channels and the signal-to-noise ratio is $<$3. Based on the r.m.s. noise in the DCN spectrum, \new{using the integrated intensity for HCN and assuming a line width of 3~km~s$^{-1}$ for DCN as measured for \hthcn\ and \hthcop}, the 3$\sigma$ upper limit on the DCN/HCN ratio is $\sim$3$\times$10$^{-3}$.

\subsubsection{The Northern clump}\label{s:Nclump}

The line profiles of the strongest lines towards the N-clump are double peaked, but in contrast to the source centre, the red-shifted peaks are stronger than the blue-shifted peaks (see Figure~\ref{f:W49tail}; right column). 
\new{These line profiles are not consistent with spherical infall motions. Since the isotopic species peak closely in velocity to the redshifted emission, the bulk of the gas appears to be at this velocity. The blueshifted peak is probably due to a bipolar outflow or other unresolved kinematic substructure.}

We have fitted single Gaussian profiles to the detected lines that are not strongly self-absorbed (HNC, H$^{13}$CN, and H$^{13}$CO$^+$ J=4-3 and DCN J=5--4; though the signal-to-noise ratio of the DCN line is $<$2) and find that these peak at 15-16~km~s$^{-1}$ ($\sim$19~km~s$^{-1}$ for DCN J=5--4). The calculated $N$(HCN)/$N$(HNC) ratio, based on the H$^{13}$CN observation, is $\sim$20, which is between the values for the source centre ($\sim$50) and the two clumps (5--7). 
\newnewnew{We therefore adopt an intermediate value for the kinetic temperature (see also \S\,\ref{s:chem}).}
Note that our fit to the HNC line profile is dominated by the red-shifted peak and thus underestimates the HNC column density, but only by $\sim$10\%, which does not change our result.

\begin{table*}
\caption{Estimated column densities towards the Northern clump.}
\label{t:Nclump}
\begin{tabular}{lcccccccl}
\hline \hline \noalign{\smallskip}
Molecule & \vlsr\ & \tmb\ & FWHM & r.m.s. & \txc\ & $\tau$ & $N$  \\
& (km~s$^{-1}$) & (K)& (km~s$^{-1}$)& (K)& (K) & & (10$^{12}$ cm$^{-2}$)   \\
\noalign{\smallskip} \hline \noalign{\smallskip}
H$^{13}$CN J=4--3     & 15.9 & 0.35 & 7.4 & 0.06 &  10 &  0.09 & 13.8 \\ 
DCN J=5--4\tablefootmark{a} & 18.8 & 0.08 & 3.5 & 0.05 &  14 & 0.01 & 4.3 & \\ 
HNC J=4--3           & 15.7 & 1.31 & 5.8 & 0.05 & 10 & 0.38 & 46.8  & \\ 
H$^{13}$CO$^+$ J=4--3 & 15.3 & 0.42 & 5.4 & 0.07 & 15 &  0.04 & 0.9  & \\ 
DCO$^+$ J=5--4        & --- & --- & ---   & 0.06 & 15 & --- & $<$0.7  & \\  
\noalign{\smallskip} \hline \noalign{\smallskip}
\end{tabular}
\tablefoot{Column densities are calculated with RADEX assuming \tkin = 75~K and \nhh = 5$\times$10$^5$~cm$^{-3}$ towards R.A.=19:10:13.6; Dec.=09:06:48. \\
\tablefoottext{a}{Tentative detection based on the r.m.s.\ noise in the spectrum, $N$(DCN) $<$3.7$\times$10$^{12}$~cm$^{-2}$ for \txc$\geq$10~K and DCN/HCN $<$0.008}
}
\end{table*}

Table~\ref{t:Nclump} lists the fit parameters and resulting column densities for \tkin\ = 75~K and \nhh\ = 5$\times$10$^5$~cm$^{-3}$. The calculated DCN/HCN ratio is an order of magnitude lower than towards the E-tail, while the HCN/HCO$^+$ ratio is $\geq$1.5 times higher than towards the other regions we looked at. 

\section{Discussion}\label{s:disc}

We have presented velocity-resolved maps of the W49A region in HCN, HNC, \hcop, and isotopic line emission over a 2$'$ (6.6\,pc) field at 15$''$ (0.83\,pc) resolution. The HCN and \hcop\ emission extends over 60--100$''$, while HNC, \hthcn\ and \hthcop\ emission is confined to $\sim$40$''$ and rarer isotopes are only seen at the source centre. The maps show a molecular 'core', elongated East-West, with 'tails' to the North and East, and a separate 'clump' to the South-West.
While most of the rare isotopic species show single-peaked velocity profiles, the main HCN and \hcop\ species show double-peaked, asymmetric profiles suggestive of infall around the peak of the emission.

Table~\ref{t:summ} summarizes our adopted physical conditions and derived column densities for the various gas components in the W49A region. The HCN/\hcop\ column density ratio is seen to vary from $\approx$3 in the SW clump and E~tail to $\approx$10 in the core and the N~clump. The HCN/HNC column density ratio is generally higher and varies from $\approx$6 in the E~tail to $\approx$20 in the other gas components.
The HCN/\hcop\ column density ratios are quite different from the observed line ratios which range from 0.3 to 1.4, while the isotopic line ratio \hthcn/\hthcop\ of 0.3--5.3 is closer to the column density ratio. 

\begin{table}
\caption{Summary of adopted physical conditions and derived column densities for the various gas components in the W49A region}\label{t:summ}
\begin{tabular}{lccccc} \hline \hline \noalign{\smallskip}
Component & $T_{\rm kin}$ & $n$(H$_2$) & \multicolumn{3}{c}{$N$(10$^{14}$\,\scm)} \\
 & & & \multicolumn{3}{c}{\hrulefill} \\
          & (K)         & (10$^5$\,\ccm)    & HCN  & \hcop & HNC \\ 
\noalign{\smallskip} \hline \noalign{\smallskip}
Core      & 100         & 20              & 20--70 & 2--10 & $\approx$1 \\
SW clump  &  40         &  5              & 10     & 4     & 0.7 \\
E tail    &  40         &  5              &  6     & 2     & 1 \\
N \newnewnew{clump}    &  75         &  5              & 10     & 0.7   & 0.5 \\
\noalign{\smallskip} \hline
\end{tabular} 
\end{table}

The HNC line has a similar central velocity and line width as the \hthcop\ and \hthcn\ lines, which is considerably different from the main HCN and \hcop\ lines. The HNC line thus seems to trace the inner core regions more than HCN and \hcop. This difference may be due to a low HNC abundance in the warm central gas, but \newnewold{accurate collisional rate coefficients for HNC are needed to quantify} a possible difference in critical density between HCN and \hcop\ on one hand and HNC on the other \newnewnew{may also play a role \citep{sarrasin:hcn-he}}.

Since the HCN, HNC and \hcop\ main isotopic lines are optically thick, the observed line intensity is not a direct measure of column density, but is rather the product of surface area, filling factor and temperature. 
\new{The cloud structure of W49 is not homogeneous, as shown by the difference in line width between the main and rare isotopic lines, and by the main beam brightness of the lines which is well below the estimated kinetic temperature.}
A possible geometry, commonly invoked to explain the large-scale CO emission from Giant Molecular Clouds, is an ensemble of small gas clumps in the beam.
\new{The inhomogeneous source structure does not directly affect our  estimates of the beam-averaged column density, which are based on rare isotopic lines.}
\newnewnew{The column densities are indirectly affected through our estimate of the volume density, which influenced the fraction of HCN in the $J$=4 state.}

We now discuss our derived column densities toward W49A by comparing them to predictions from physical and chemical models (\S~\ref{s:pdrxdr} and ~\ref{s:chem}) and to the results of extragalactic observations (\S~\ref{s:xgal}).

\subsection{Comparison to PDR/XDR models}
\label{s:pdrxdr}

One interesting comparison which these observations allow are with the models
by \citet*{MSI} who predict the molecular line emission from irradiated
dense gas in galaxies using models of PDRs (photon dominated regions) and XDRs
(X-ray dominated regions). Specifically \citeauthor{MSI} predict the intensity
ratios in the $J$=1--0 and $J$=4--3 transitions of HCN, HCO$^+$ and HNC. The
measured \new{velocity-integrated} values of these ratios towards W49A are summarised in
Table~\ref{t:ratios} where the values for the $J$=1--0 transition come from
maps made with the IRAM 30m telescope by \citet{peng} and Peng et al (in prep).

The observed HCN/\hcop\ ratios in both the J=1--0 and J=4--3 transitions 
are inconsistent with all the \citeauthor{MSI} XDR models, but are
consistent with emission from a PDR. The J=1--0 line ratio implies a well
constrained number density of $n\sim4\times10^4$ cm$^{-3}$, 
consistent with the ratio in the J=4--3 transitions. 
However, the PDR model has difficulties explaining our observed HNC/HCN ratios.
Our observed HNC/HCN ratio in the J=1--0 line would imply
number densities $>3\times 10^5$ \ccm, while the J=4--3 transitions imply
$n<2\times10^4$ \ccm. The HNC/HCN line ratios are better explained by
the XDR model, where both transitions are consistent with densities between $10^4$ \ccm\ and $\sim10^5$ \ccm\ but require X-ray fields with fluxes $>5$ erg cm$^{-2}$ s$^{-1}$ at the lowest densities up to $>100$ erg cm$^{-2}$ s$^{-1}$ at the highest densities.
We conclude that no single PDR or XDR model is able to reproduce our observations, so that a more complex, multi-component model must be invoked. 
\newnewnew{One caveat is the Meijerink et al models should be updated with the new collision data for HCN and HNC \citep{sarrasin:hcn-he}.}

\begin{figure}
\rotatebox{270}{
\resizebox{!}{\hsize}{\includegraphics*[2cm,-3cm][21cm,30cm]{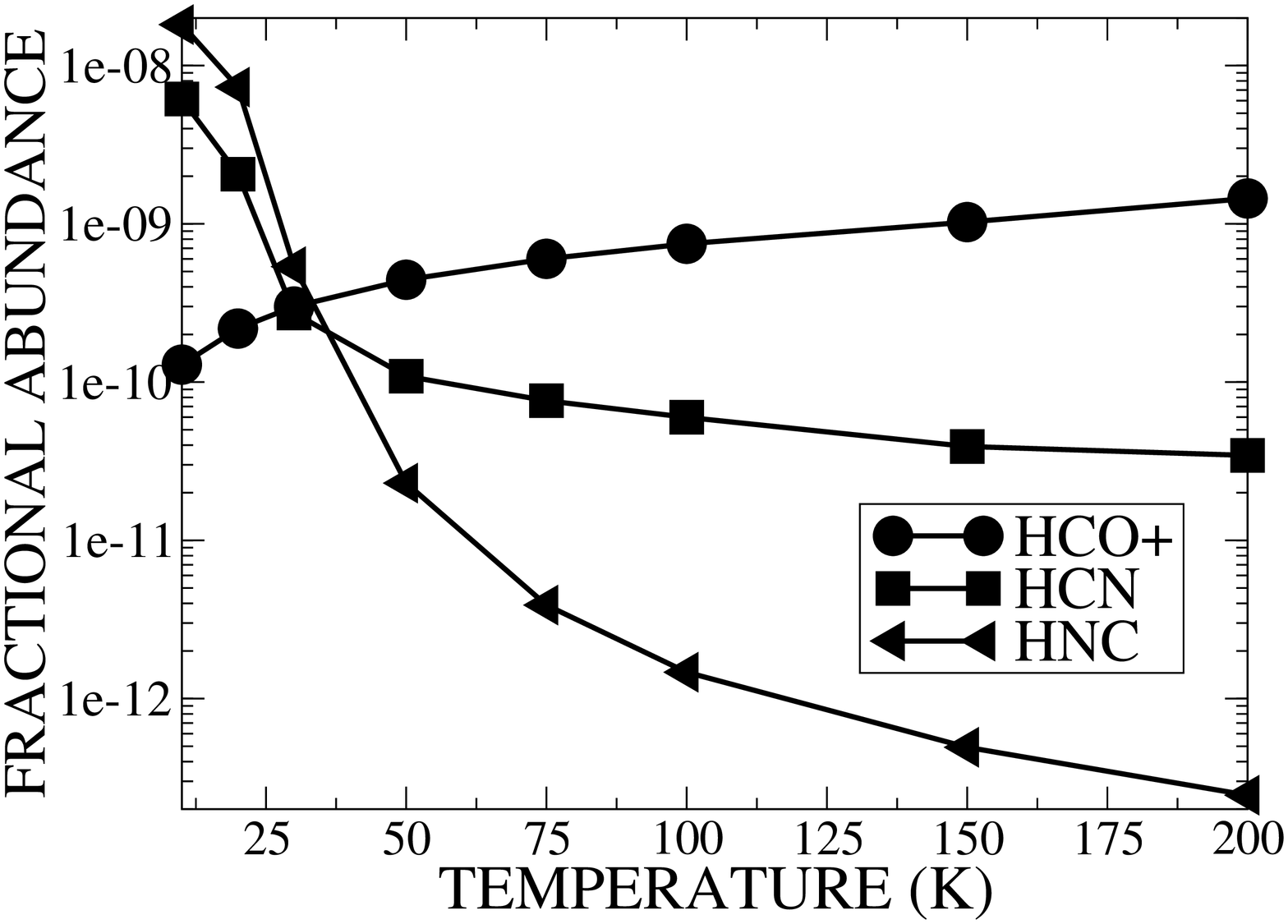}}}
\rotatebox{270}{
\resizebox{!}{\hsize}{\includegraphics*[2cm,-3cm][21cm,30cm]{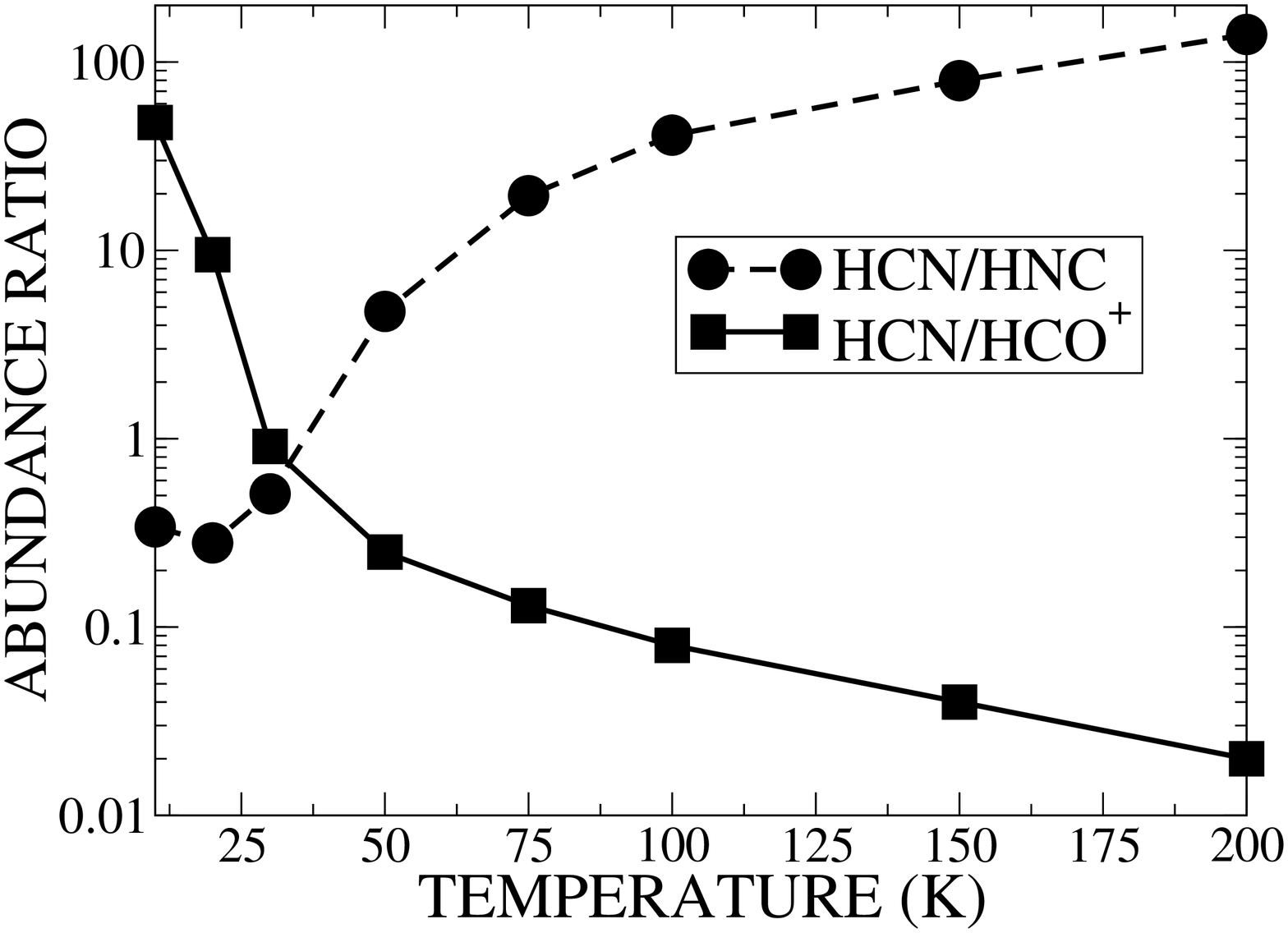}}}
\rotatebox{270}{
\resizebox{!}{\hsize}{\includegraphics*[2cm,-3cm][21cm,30cm]{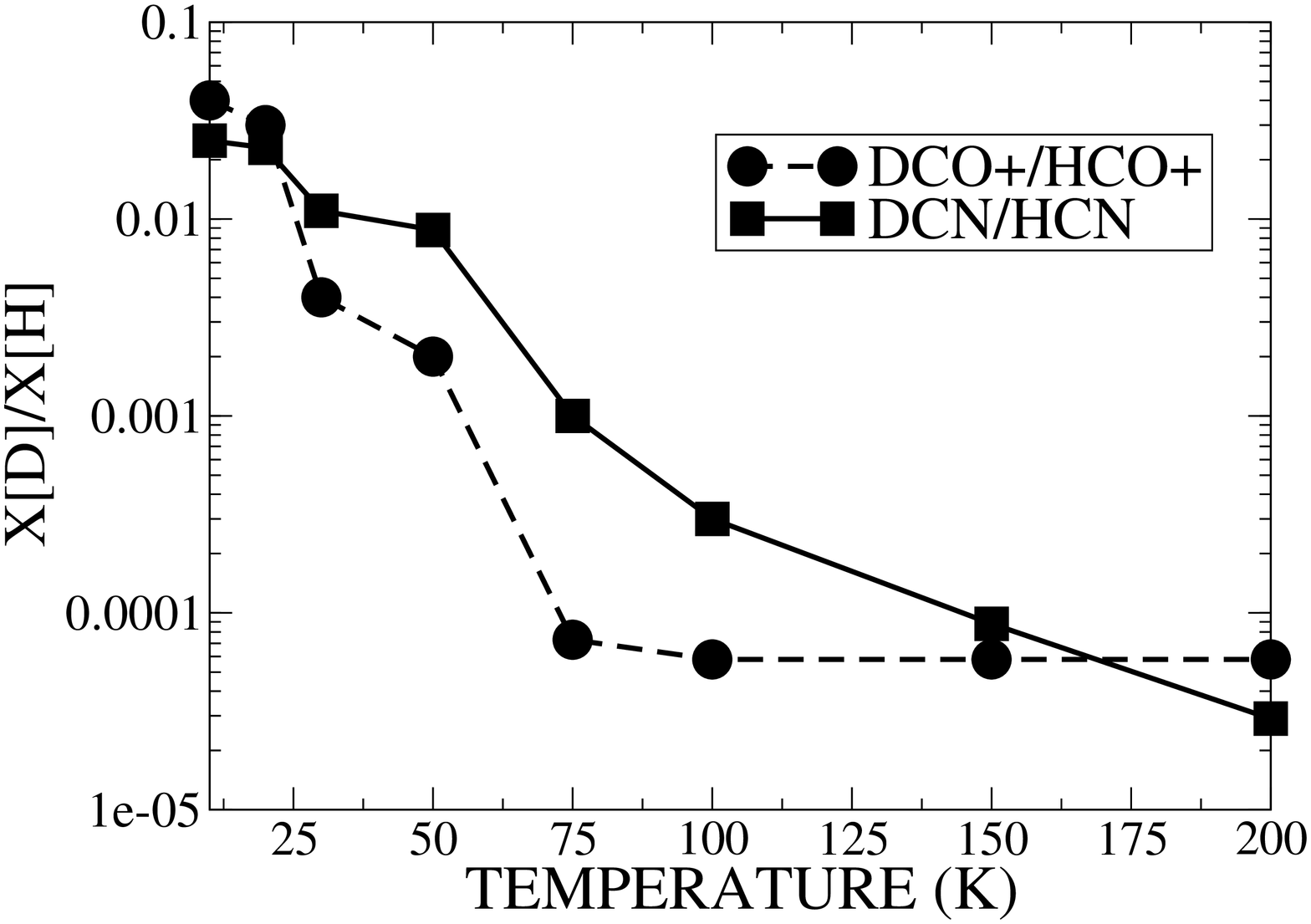}}}
\caption{Steady-state fractional abundances and abundance ratios predicted by gas-phase chemical models run at different temperatures; n(H$_2$)~=~10$^6$~cm$^{-3}$. }\label{f:model1}
\end{figure}

\subsection{Comparison to chemical models}
\label{s:chem}

We also compare the observed column density ratios with predictions from kinetic models of interstellar chemistry. The models consider 340 species linked by $>$10,000 reactions with rate co-efficients based on the latest release of the UMIST Database for Astrochemistry, RATE06\footnote{http://www.udfa.net} \citep{woodall07}. The chemical reaction set has been extended to include deuterated analogues of all hydrogen-containing species \citep[see e.g.][]{roberts2000}. The models are ``pseudo-time-dependent'', in that the physical conditions remain constant over time and they consider only gas-phase chemical reactions (with the exception of H$_2$, HD, and D$_2$ formation on grain-surfaces).  Fig.~\ref{f:model1} shows steady-state fractional abundances and abundance ratios predicted by a set of models run with \nhh\ =~10$^6$~cm$^{-3}$ and values for \tkin\ between 10 and 200~K.

As discussed in \S~\ref{s:intro}, the HCN/HNC ratio increases and the molecular D/H ratios decrease as \tkin\ increases.  The fractional abundances of HCN and HNC fall with increasing \tkin, while the abundance of HCO$^+$ increases. We expect, therefore, that the HCN/HCO$^+$ ratio will also trace the gas temperature. 
 
The HCN/HNC ratios we measure in W49A vary between 50 at the source centre to $\sim$20 towards the N-clump and 6--7 in the SW-clump and E-tail. It is these ratios that we have mainly used to constrain the temperatures of these regions as 100, 75, and 40~K, respectively. \newnewnew{In the future, we plan to use the SLS data to put more direct constraints on the gas temperature and density across the W49A region.}

The HCN/HCO$^+$ ratios we observe, however, are between 3 and 15. The steady-state chemical models predict HCN$>$HCO$^+$ only for \tkin\ $<$ 25~K. Presumably, though, the source formed from a molecular cloud where the gas-phase abundances had equilibrated at a much lower temperature. If we use the steady-state abundances at 10~K as the initial abundances for the models at higher temperatures it takes $>$10$^5$ years for the chemistry to reach a new steady-state. This is larger than the expected lifetimes of hot-cores in high mass star forming regions \newnewnew{(3--5 $\times$ 10$^4$\,yr: \citealt{wilner01})}.

\begin{figure}
\rotatebox{270}{
\resizebox{!}{\hsize}{\includegraphics*[0cm,-3cm][21cm,30cm]{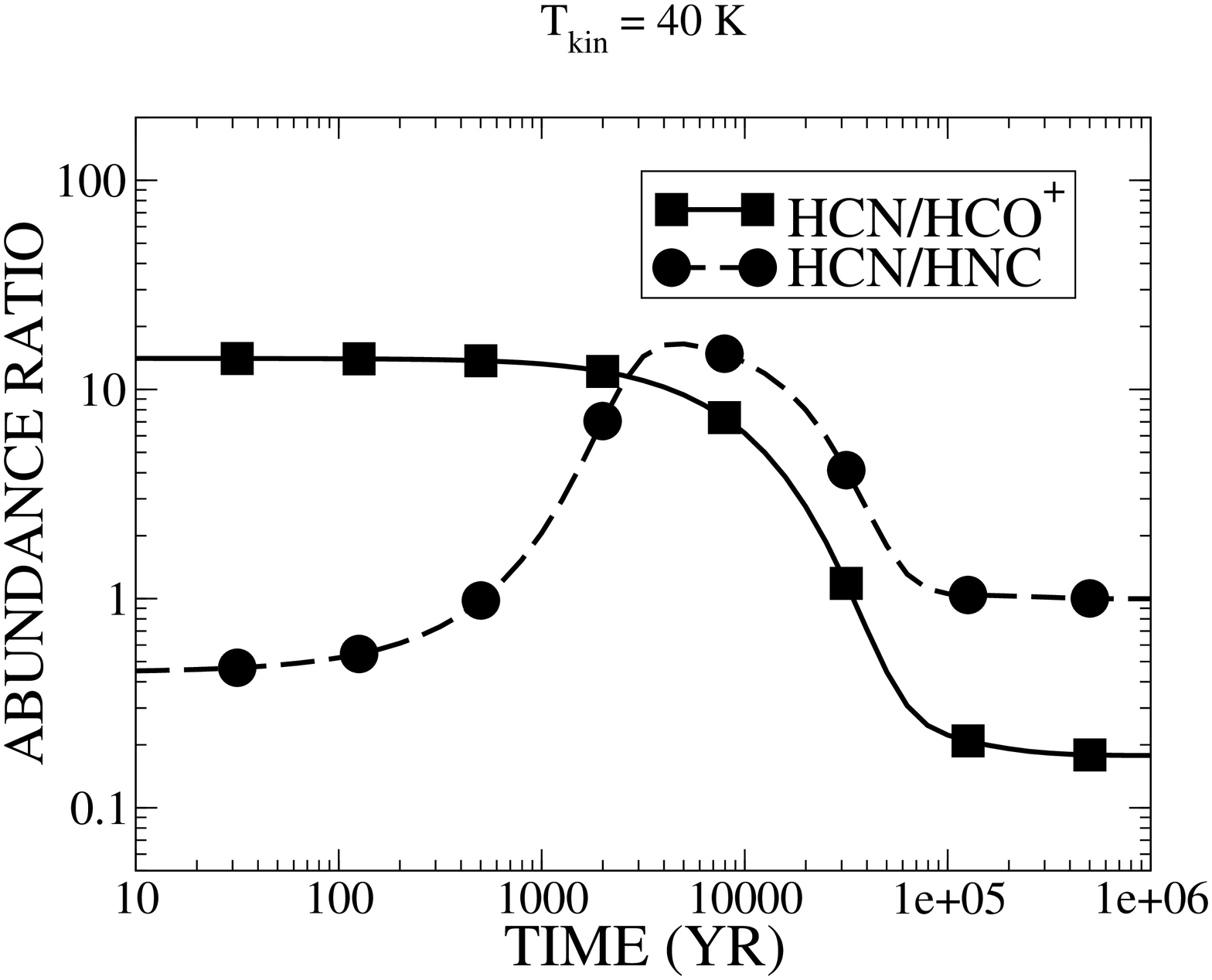}}}
\rotatebox{270}{
\resizebox{!}{\hsize}{\includegraphics*[0cm,-3cm][21cm,30cm]{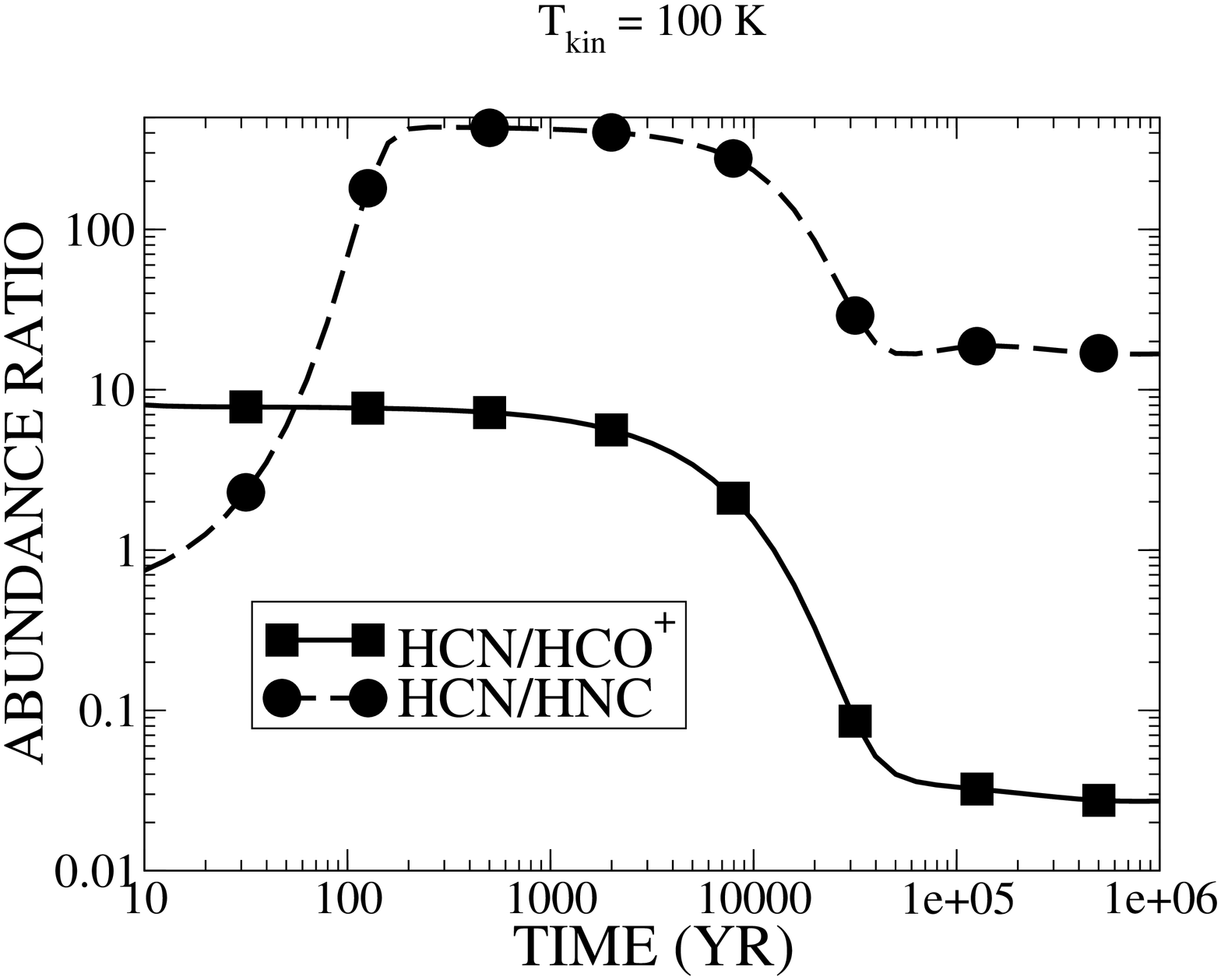}}}
\caption{Fractional abundance ratios varying over time from gas-phase chemical models assuming n(H$_2$)~=~10$^6$~cm$^{-3}$. The initial abundances come from a steady-state model at 10~K; the kinetic temperatures are 40~K (top) and 100~K (bottom). }\label{f:model2}
\end{figure}

Fig.~\ref{f:model2} shows the evolution of the HCN/HCO$^+$ and HCN/HNC abundance ratios from gas which has reached steady state at 10~K at time=0~yr and is then heated to 40 or 100~K. While the HCN/HCO$^+$ ratio maintains its ``low-temperature'' value, $>$1, for $\geq$10$^4$~yr, the HCN/HNC ratio varies much more with time, increasing to $\sim$10$\times$ its ``high temperature'' value before decreasing to steady-state. It is not really possible from these simple models to put an absolute age on the W49A core, as the timescales do depend on the density and on the rate of warming of the gas, but they could explain why the observed HCN/HNC ratios better probe the current gas temperature, while the HCN/HCO$^+$ ratios are closer to what we expect in colder gas. 

\begin{figure}
\rotatebox{270}{
\resizebox{!}{\hsize}{\includegraphics*[2cm,-3cm][21cm,30cm]{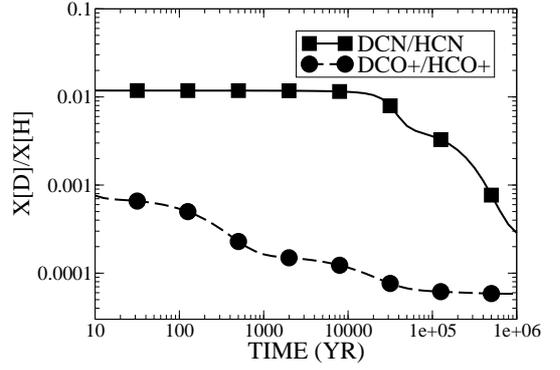}}}
\caption{Fractional abundance ratios varying over time from a gas-phase chemical models assuming $n$(H$_2$)~=~10$^6$~cm$^{-3}$ and \tkin = 100~K. The initial abundances come from a steady-state model at 10~K. }\label{f:model3}
\end{figure}

Similarly, for the DCN/HCN ratio, Fig.~\ref{f:model3} shows that it also takes $>$10$^4$~yr at 100~K for the fractionation set at low temperatures to begin to fall (see \S\ref{s:intro}). In contrast, the DCO$^+$/HCO$^+$ ratio resets to a ``high-temperature'' value within 1000~yr. Thus, the DCN/HCN ratios of 4$\times$10$^{-4}$--10$^{-3}$ observed towards the centre of W49A (and the tentative estimate of DCN/HCN = 0.02--0.04 in the E-tail at 40~K) are not inconsistent with current higher kinetic temperatures as, like the HCN/HCO$^+$ ratio, the fractionation may be preserved from an earlier, colder phase of the core's evolution.  

Of course, the chemistry in W49A is unlikely to be this simple. For each line of sight we may be looking through multiple gas-components with different physical conditions (certainly, at the source centre the JCMT beam covers several embedded hot-core sources; see Fig.~\ref{f:spatial}). The gas-phase chemical models also assume that the density and, therefore, the visual extinction are high enough that photo-reactions are unimportant. \newnewnew{This is supported by the IRAC 8\,$\mu$m image of W49A \citep{peng:shells} not showing maxima at any of our 'clumps' and 'tails', which would be indicative of a PDR.}

\subsection{Comparison to extragalactic regions}\label{s:xgal}

Both HCN and \hcop\ have been advocated as tracers of dense gas in external
galaxies \citep{baan08,jgc06}, but to date the majority of extragalactic
observations have focussed on the J=1--0 transitions of these species. However,
due to their higher critial densities and higher upper energy levels, the higher J transitions are likely to be more efficient probes of
the high density, warm gas associated with star formation. In addition, the 
comparison with lower energy transitions provides a probe of the (range of)
excitation conditions in the gas \citep{papad07}. A survey of nearby active
galaxies in the J=1--0 to J=3--2 transitions of both HCN and \hcop\ shows that
the line intensity from higher J transitions relative to lower transitions
increases in increasingly starburst-dominated galaxies, as does the ratio of
\hcop\ to HCN emission in a particular J transition \citep{krips08}.  The
temperature ($50-100$\,K), density ($(0.5 - 2)\times10^6$\,cm$^{-3}$) and column
density ($10^{14}-10^{15}$\,cm$^{-2}$) properties of W49A \new{on the scales observed here} are similar to
those derived by \citet{krips08}, indicating that W49A may be a valuable analogue
for understanding the emission from other galaxies. Consistent with the
observations of \citet{krips08}, the strength of the \hcop\ J=4--3
transition in W49A suggests that this line should be a strong and robust
tracer of the dense gas in extragalactic star forming regions.
\newnewnew{One advantage of the 4--3 line over the 1--0 line is that it is much less affected by absorption in low-density foreground layers.}

\begin{table*}
\caption{Observed line ratios in selected extragalactic objects and comparison to W49A}
\label{t:xgal}
\begin{tabular}{lcccccc} \hline \hline \noalign{\smallskip}
Ratio                              & APM 08279 &  NGC 253 & \multicolumn{4}{c}{W49A} \\
 & & & Centre & North & East & Southwest \\
\noalign{\smallskip} \hline \noalign{\smallskip}
HCN/HCO$^+$            4--3$^a$    & 0.98 &  1.03          & 0.5 & 0.6 & 0.9 & 0.8 \\
H$^{13}$CN/H$^{13}$CO$^+$   2--1   & --   &  1.66--3.72$^b$ & $\approx$5$^c$ & $\approx$5$^c$ & $\approx$1.4$^c$ & $\approx$1.4$^c$ \\
HCN 4--3$^a$ / HNC 1--0            & 1.57 &  0.03--0.14$^b$ & 2.0 & 1.0 & 0.9 & 1.4 \\
H$^{13}$CN/HN$^{13}$C       2--1   & --   &  2.29       \\
HCN 4--3 / H$^{13}$CN 2--1         & --   &  0.42--0.56$^b$ \\
HCO$^+$ 4--3 / H$^{13}$CO$^+$ 2--1 & --   &  0.90--1.51$^b$ \\
HCO$^+$ 4--3 / HC$^{18}$O$^+$ 2--1 & --   &  2.60       \\
DNC 2--1 / HNC 1--0                & --   &  0.01--0.05$^b$ \\
\noalign{\smallskip} \hline \noalign{\smallskip} 
\multicolumn{3}{l}{$^a$: $J$=5--4 for APM 08279} \\
\multicolumn{3}{l}{$^b$: Values for the two observed velocity components} \\
\multicolumn{3}{l}{$^c$: Estimated from the 4--3 ratio (see text)} \\
\end{tabular}
\end{table*}

The $J$=5-4 lines of HCN, HCO$^{+}$ and HNC have been observed for the high-z galaxy APM 08279+5255 \citep{guelin07}.
In NGC 253, several isotopologues of HCN, HNC and HCO$^{+}$ have also been observed \citep{huettemeister,wagg05,sgb06,martin06}.
Estimates of ratios derived from these studies are presented in Table \ref{t:xgal}. These studies make these two very active sources particularly relevant to compare to the properties derived for each component identified in W49A (\S\S~\ref{s:centre}--\ref{s:Nclump}). 

The comparison between APM 08279+5255 and the components identified in W49A is limited by the number of ratios available for the high-z source. In contrast, the data for NGC 253 are much more complete than those for W49A. Concentrating on those ratios where data exist for all sources, we note the following from the data in Table~\ref{t:xgal}. First, the HCN/\hcop\ 4--3 ratios in the East and Southwest regions are similar to those in APM 08279 and NGC 253, whereas the values in the Centre and North regions are lower. The minimum value of this ratio measured in the entire W49A region is $\approx$0.3. Second, the HCN 4--3 / HNC 1--0 ratios at all positions in W49A are much higher than measured toward NGC 253, while the values toward the Centre and North regions bracket the ratio measured toward APM 08279. Third, the \hthcn / \hthcop\ 2--1 ratios, which are based on the observed 4--3 ratios, using RADEX at the temperature and density of the region to convert the 4--3 ratio to the 2--1 ratio, bracket the value measured toward NGC 253.
Interestingly, there is little evidence that the fact that APM 08279 is hosting a quasar is making any impact on the emission in our observed lines.

Based on the similarity of the line ratios and the column densities, we conclude that W49A appears to be a reasonable model for extragalactic star-forming regions. 

\section{Conclusions}\label{s:concl}

We have mapped HCN, HNC, HCO$^+$, and related isotopologues in the luminous star-forming region W49A. Line ratios of these dense gas tracers are often used to probe the chemical state and ionisation processes in external galaxies.
We find that the HCN/HCO$^+$ J=4--3 line ratios are $<$1 across most of the source, but that these lines, particularly HCO$^+$, are highly optically thick; the H$^{13}$CN/H$^{13}$CO$^+$ J=4--3 line ratios are $\geq$1 towards the source centre and slightly higher just to the north-west; the HCN/HCO$^+$ column density ratios are 3--10. 
The HCN/HNC column density ratio varies from $\approx$6 in the Eastern tail to $\approx$20 in the other gas components.

We also find that, although there are embedded hot cores towards the centre of W49A which are not resolved in these observations, column densities towards this region are consistent with an average gas temperature of $\sim$100~K and density $>$10$^6$~cm$^{-3}$. There appear to be several cooler (\tkin\ $\sim$ 40~K) but still relatively dense (\nhh\ $>$ 10$^5$~cm$^{-3}$) clumps of gas surrounding, and possibly infalling onto, the source centre. 

Comparison of our observations with the PDR and XDR models of \citet{MSI} shows
contradictory results. While the HCN and HCO$^+$ emission could be
consistent with a PDR, this is inconsistent with the HNC to HCN ratio
which points to emisssion from a XDR. 
We interpret this discrepancy as evidence that irradiation plays only a minor part in the chemistry of the molecular gas in W49A.

We also find that regions of W49A have significant similarities to
observations of the starburst galaxy NGC 253 and the quasar host APM 08279.
This agreement seems to imply that over scales of a few 100 pc, the line emission from external galaxies is not dominated by hot core physics but rather by extended warm gas-phase chemistry. The W49A region thus may be a good template for starburst galaxies, 
and the current data indicate that irradiation by an AGN has a limited effect on the chemistry of the surrounding molecular gas.

The observations discussed here highlight the potential problems in the
interpretation of extragalactic systems on the basis of a few transitions of a
limited set of molecular species. The extragalatic star forming regions which such observations probe are likely at least as complex and structured as
W49A.  A better understanding of the origin of the emission towards W49A and
the internal physical and kinematic structure of the source will require
velocity resolved, high angular resolution observations with instruments such as the SMA and ALMA interferometers.  These observations will be able to address the origin of the velocity structure seen in the JCMT observations as
well as determine whether different components of the emission are spatially
segregated due to differences in excitation or chemistry.  Together with the
analysis of the more complete set of molecular species probing other physical
and chemical environments which will be available in the full JCMT
SLS-observations of W49A and the Orion Bar \citep{vanderwiel09}, such
studies of W49A can provide more insight into the relationship between the
molecular emission, the structure and the star formation in star forming
regions in other galaxies.

\begin{acknowledgements}
The authors would like to thank all members of the SLS consortium for carrying out the JCMT Legacy survey observations, Libby Miller (Belfast) for reducing the JCMT 230\,GHz data, Jenny Williams (Manchester) for reducing the SCUBA data, Zs\'ofia Nagy (Groningen) for reducing the HNC data, and Tzu-Cheng Peng \& Friedrich Wyrowski (Bonn) for making their IRAM 30m maps available to us ahead of publication.

The James Clerk Maxwell Telescope is operated by the Joint Astronomy Centre on behalf of the Science and Technology Facilities Council of the United Kingdom, the Netherlands Organisation for Scientific Research, and the National Research Council of Canada. Astrophysics at Queens University Belfast and The Jodrell Bank Centre for Astrophysics is supported by STFC. EB acknowledges support from an STFC grant.

\newnewnew{We thank the referee, Paul Ho, and the editor, Malcolm Walmsley, for their useful comments on the manuscript.}

\end{acknowledgements}

\bibliography{w49a}
\bibliographystyle{aa}

\clearpage 

\end{document}